\documentclass[12pt]{article}

\usepackage{ulem}

\usepackage{amssymb,amsmath,bm}
\usepackage{mathabx, mathtools}
\usepackage{amsfonts}
\usepackage{url,graphicx,subfig} 
\usepackage{enumerate} 
\usepackage{color}
\usepackage[margin=1.5cm]{geometry}
\usepackage{makecell}
\usepackage{tikz}
\usetikzlibrary{decorations.pathreplacing,calligraphy}

\usepackage{algorithm}
\usepackage{algpseudocode}

\usepackage{tikz-qtree}

\usepackage{float}

\newtheorem{Lemma}{Lemma}
\newtheorem{Remark}{Remark}
\newtheorem{Example}{Example}
\newtheorem{Problem}{Problem}
\newtheorem{Definition}{Definition}

\newcommand{\balpha}{\mbox{\boldmath$\alpha$}}

\title{Matrix-analytic methods for the evolution of \\species trees, gene trees, and their reconciliation
	\thanks{
		We thank the Australian Research Council for funding this research through Discovery Project DP180100352.
	}
}

\author{
	Albert C. Soewongsono
	\thanks{University of Tasmania, Hobart TAS 7001, Australia, email: albert.soewongsono@utas.edu.au.}
	\and
	Jiahao Diao
	\thanks{University of Tasmania, Hobart TAS 7001, Australia, email: jiahao.diao@utas.edu.au.}	
	\and
	Tristan Stark
	\thanks{University of Tasmania, Hobart TAS 7001, Australia, email: tristanstark@gmail.com.}
	\and
	Amanda E. Wilson
	\thanks{Temple University, Philadelphia, PA 19122, USA, email: amandaewilson@temple.edu.}
	\and
	David A. Liberles 
	\thanks{Temple University, Philadelphia, PA 19122, USA, email: daliberles@temple.edu.}
	\and
	Barbara R. Holland
	\thanks{Australian Research Council Centre of Excellence for Plant Success, University of Tasmania, Hobart TAS 7001, Australia, email: barbara.holland@utas.edu.au.}	
	\and	
	Ma{\l}gorzata M. O'Reilly
	\thanks{University of Tasmania, Hobart TAS 7001, Australia, email: malgorzata.oreilly@utas.edu.au.}
}

\begin{document}
	
	\maketitle
	
\begin{abstract}
We consider the reconciliation problem, in which the task is to find a mapping of a gene tree into a species tree, so as to maximize the likelihood of such fitting, given the available data. We describe a model for the evolution of the species tree, a subfunctionalisation model for the evolution of the gene tree, and provide an algorithm to compute the likelihood of the  reconciliation. We derive our results using the theory of matrix-analytic methods and describe efficient algorithms for the computation of a range of useful metrics. We illustrate the theory with examples and provide the physical interpretations of the discussed quantities, with a focus on the practical applications of the theory to incomplete data.
\end{abstract}

	
\section{Introduction}\label{sec:intro}
	
In the areas of comparative genomics and of environmental sequencing, it is common to collect samples of genes that come from the genomes of different species \cite{hermansen2017adaptive,walker2022structural}. From such sets of genes, gene families are typically constructed for all genes showing evidence of descent from common ancestral genes. These gene families can include multiple instances of genes from the same species that have been derived through gene duplication events and also can include homologous genes from other species. To understand lineage-specific genome evolution and organismal adaptation, it is common to apply data analysis pipelines that include gene tree/species tree reconciliation \cite{anisimova2013state}  and many approaches for this either use the assumption of parsimony or maximize likelihoods based upon overly simple models that neglect underlying expectations at different levels of biological organisation \cite{roth2007evolution}.
		
For data that comes from closely related species, there are several biological processes that can be modeled with increasing levels of mechanistic realism. Speciation is one such process. While there are many different species concepts that have been described by Wheeler et al. \cite{wheeler2000species},  a core principle in speciation is the cessation of gene flow between the two populations that are speciating, although other mechanisms that enable differentiation are possible \cite{nosil2008speciation}. The simplest scheme to describe this is as a clock-like Poisson process with a simple rate~\cite{neeetal,neeetal2}.
		
In addition to generating a species model that is independent of or coupled to an underlying genome evolution model for the analysis of lineage-specific gene content, the species evolution model can be used to address fundamental questions in organismal evolutionary and conservation biology. For example, it has been argued that the ape clade is less speciose than would be expected by extrapolation from primate evolutionary speciation rates \cite{peng2009primate}. Similar arguments have been made elsewhere across the tree of life in characterizing speciation rate heterogeneity \cite{singhal2022no}. The modeling framework generated here enables addressing such questions with model-based statistical formalism.
		
With the independence of speciation and underlying gene content evolution, gene trees are then initially described by species trees, subject to lineage-specific changes in gene content. Gene duplication can occur through relatively infrequent whole genome duplication events and with more frequent smaller scale duplication events that typically affect one gene at a time. Duplicate genes are initially redundant and the initial period of redundancy is when they are most likely to be lost. This assertion is supported both by empirical data fitting of comparative genomic data and by theoretical expectations from the retention process, given retention mechanisms that include neofunctionalisation and subfunctionalisation, both resulting in gene copy-specific changes in function and selection relative to the redundant state~\cite{2020DSLOH}.
		
It should be recognized that species and genomes are described as single individuals, but this simplicity is consistent with population level modeling when mutation is infrequent (weak) and selection is strong, such that only one relevant mutation is segregating at a time \cite{stark2021characterizing}. In this case, the mutation-selection process can be reasonably approximated by single rates, either as a Poisson process or a more complex process to describe time-dependence \cite{stark2021characterizing,teufel2018using}. It is in this spirit that a multi-scale model for speciation and gene content evolution coupled to their reconciliation is presented here.

 One important aim of this paper is to provide a bridge between the phylogenetics literature and an area of stochastic modeling known as matrix-analytic methods (MAMs)~\cite{latouche2012matrix,neuts1981matrix}. MAMs encompass several different important classes of models including quasi-birth-and-death processes (QBDs)~\cite{asmussen2003applied,latouche2010level,latouche1999introduction,latouchet1998invariant} and Markovian Binary Trees (MBTs)~\cite{2015S,2014HF,2009HLR,2011H,Kontoleon}. Both of these classes of models have wide application in phylogenetics (indeed many commonly used models turn out to be special cases). We hope that demonstrating these links will be useful to the phylogenetics community as a strength of the MAMs literature is the existence of algorithms for efficiently computing many metrics of interest.

The rest of the paper is structured as follows. In Section~\ref{sec:speciestree} we introduce a general MBT and describe some variations that are of relevance to generating species trees. We consider the MuSSE model presented by FitzJohn in~\cite{fitzjohn2010quantitative} and the BiSSE model presented by Maddison et al. in~\cite{2007MMO}. We give an extended example that describes how the MuSSE (and hence BiSSE) class of models can be expressed as a special case of an MBT.

In Section~\ref{sec:recspetree}, we discuss the important distinction between a phylogenetic tree connecting taxa that have all survived to the present day (known as a reconstructed tree) and the true tree which includes species that have gone extinct and are hence unobserved in the current day. We perform our mathematical analysis to account for this. We derive a novel expression for the likelihood of a reconstructed tree, in which we consider the probability of observing an internal branch such that all potential speciation events that occurred within that branch, resulted in extinctions by the present time (and so beyond the end of the internal branch). To formulate this likelihood expression, we define new probability matrices and derive related differential equations for the computation of the likelihood.

This analysis is performed within the theory of the MBTs, and so is also applicable to the MuSSE and the BiSSE models, since they are special cases of the MBT model. We demonstrate the results for the BiSSE model in~\cite{2007MMO} and for the MuSSE model in~\cite{fitzjohn2010quantitative} are the special case of our general results for the MBTs. We also note that the results in~\cite{fitzjohn2010quantitative,2007MMO}  do not consider the probability of a lineage going extinct after the end of an  internal branch but before the present day, which is an interesting problem that we address here.

Next, we derive the conditional probability for computing the tree balance. We give an extended example showing how we can compute expected tree balance under the MBT models (and hence also the MuSSE and the BiSSE models). We conclude Section~\ref{sec:recspetree} by showing that the special case of the MBT, the MuSSE model, can be equivalently expressed as a QBD, and explain why this will sometimes have computational advantages for the evaluation of distribution of the time until species extinction.

In Section~\ref{Sec:GenTree} we give several examples of how the process of gene trees evolving under a subfunctionalisation model can be approximated using level-dependent QBDs. These QBD-based models extend those given in \cite{2020DSLOH} to also record the size of the subtrees which means they are useful for assessing tree balance. We describe and extend some useful algorithms from the MAMs literature so that they are applicable for this application.

In Section~\ref{sec:reconciliation}, we consider the reconciliation problem and derive the recursive expression for the likelihood of a given mapping of a gene tree into a species tree. We illustrate our methodology through a numerical example in which we compute the likelihoods of the three species-gene trees based on an example in G{\'o}recki~\cite[Figure 1]{gorecki2014drml}.

\section{Species trees}\label{sec:speciestree}

For the species tree we apply a class of models in which each branch from the moment of birth evolves independently of all other branches, and only its initial state is drawn from a distribution that depends on the state of the parent branch at the time of giving birth.

Consider the following Markovian binary tree (MBT) model discussed in~\cite{2014HF,2009HLR,2011H,Kontoleon}, which models the lifetime of an individual and counts the number of times the individual gives birth to children until it eventually dies. Note that in our application of this class of models later on, an individual will correspond to one species.

Each individual is assumed to follow a replica of the same independent process. Let $\{(M(t),\varphi(t)):t\geq 0\}$ be a continuous-time Markovian process with state space $\mathbb{N}\times \{0,1,\ldots,n\}$, where the level process $M(t)\in\mathbb{N}$ counts the total number of children of the individual that were born in $[0,t]$ and the phase process $\varphi(t)\in\{0,1,\ldots,n\}$ is a continuous-time Markov chain that drives the evolution of the individual, with absorbing state $0$. The model has the following parameters:
	\begin{itemize}
		\item $\balpha=[\alpha_j]_{j=1,\ldots,n}$ is some initial distribution vector of phase observed at the time of the birth of the individual, 
		\item ${\bf d}=[d_i]_{i=1,\ldots,n}$ is vector of transition rates $d_i$ from $(\ell,i)$ to $(\ell,0)$ at which the individual's life is terminated when in phase $i$ (the number of children $\ell$ does not change), 
		\item ${\bf D}_0=[(D_0)_{ij}]_{i,j=1,\ldots,n}$ is a matrix of transition rates $(D_0)_{ij}$ from $(\ell,i)$ to $(\ell,j)$ at which the individual changes phase from $i$ to $j$ (the number of children $\ell$ does not change), 
		\item ${\bf D}_1=[(D_1)_{ij}]_{i,j=1,\ldots,n}$ is a matrix of transition rates $(D_1)_{ij}$ from $(\ell,i)$ to $(\ell+1,j)$ at which the individual gives birth and simultaneously transitions to phase $j$ when in phase $i$ (the number of children increases from $\ell$ to $\ell +1$),  
		\item ${\bf P}=[P_{j,ik}]_{i,j,k=1,\ldots,n}$ is is an $n^2\times n$ matrix which records conditional probabilities $P_{j,ik}$ that a child starts in phase $j$ given its parent made a transition from phase $i$ to phase $k$ at the time of giving birth, and
		\item  ${\bf B}=[B_{i,jk}]_{i,j,k=1,\ldots,n}$ is a matrix of transition rates $B_{i,jk}=(D_1)_{ik}P_{j,ik}$ at which a parent in phase $i$ makes a transition to phase $k$ and simultaneously gives birth to child in phase $j$.
	\end{itemize}

We apply the MBT model as follows. Suppose that a species tree $T$ starts with a root that grows into a branch. Then the lifetime (length) of the initial branch of the species tree $T$ is a random variable that records the time to absorption in a continuous-time Markov chain (CTMC) $\{X(t):t\geq 0\}$ with state space $\mathcal{A}=\{1,\ldots,n,S,E\}$, some initial distribution vector $\balpha=[\alpha_i]_{i=1,\ldots,n}$, and generator matrix
	\begin{eqnarray}
		{\bf Q}^*
		&=&
		\left[
		\begin{array}{c|c|c}
			{\bf D}_0& {\bf D}_1 {\bf 1}&{\bf d}\\
			\hline
			{\bf O}&{\bf 0}&{\bf 0}
		\end{array}
		\right]
		,
	\end{eqnarray}
	partitioned according to $\{1,\ldots,n\}\cup\{S\}\cup\{E\}\times \{1,\ldots,n\}\cup\{S\}\cup\{E\}$, so that ${\bf D}_0=[Q^*_{ij}]_{i,j=1,\ldots,n}$, ${\bf D}_1 {\bf 1}=[Q^*_{i,S}]_{i=1,\ldots,n}$, ${\bf d}=[Q^*_{i,E}]_{i=1,\ldots,n}$, and ${\bf O}$ and ${\bf 0}$ is a matrix and a vector of zeros of appropriate sizes. States $1,\ldots,n$ are transient states containing some information about the branch and/or underlying environment which drives the evolution, $S$ and $E$ are absorbing states corresponding to speciation and extinction respectively.

Further, whenever absorption to $E$ occurs (according to rate vector ${\bf d}$), the branch becomes extinct, while if absorption to $S$ occurs, a speciation event occurs (according to rate matrix ${\bf D}_1$), and so the branch gives birth to two new branches, each evolving according to an absorbing CTMC, eventually either becoming extinct or giving birth to two new branches, and the process continues in the same manner from then on. The evolution of the two child branches is conditionally independent, given their initial phase.

\begin{figure}[H]
\begin{center}
\begin{tikzpicture}[>=stealth,redarr/.style={->}]	
			\draw [dashed] (0,10) -- (10,10);
			
			\draw [dashed] (0,9.5) -- (10,9.5);
			\draw (-0.5,9.5) node[anchor=north, below=-0.2cm] {\scriptsize{\color{black} S }};
			\draw [black,dashed] (0,9) -- (10,9);
			\draw (-0.5,9) node[anchor=north, below=-0.2cm] {\scriptsize{\color{black} S }};
			\draw [red,dashed] (0,8.5) -- (10,8.5);
			\draw (-0.5,8.5) node[anchor=north, below=-0.2cm] {\scriptsize{\color{red} E }};
			\draw [black,dashed] (0,8) -- (10,8);
			\draw (-0.5,8) node[anchor=north, below=-0.2cm] {\scriptsize{\color{black} S }};
			\draw [dashed] (0,7.5) -- (10,7.5);
			\draw (-0.5,7.5) node[anchor=north, below=-0.2cm] {\scriptsize{\color{black} S }};
			\draw [red,dashed] (0,7) -- (10,7);
			\draw (-0.5,7) node[anchor=north, below=-0.2cm] {\scriptsize{\color{red} E }};
			\draw [dashed] (0,6.5) -- (10,6.5);
			
			\draw (2.5,10.4) node[anchor=north, below=-0.17cm] {{\color{black} $T^{(1)}$ }};
			\draw [green,very thick] (2.5,10) -- (2.5,9.5);
			\draw [green,very thick] (2.5,9.5) -- (1.5,9.5);
			\draw [green,very thick] (1.5,9.5) -- (1.5,9);
			\draw [black,thick] (1.5,9) -- (1,9);
			\draw [black,thick] (1,9) -- (1,8.5);			
			\draw [green,very thick] (1.5,9) -- (2.5,9);
			\draw [green,very thick] (2.5,9) -- (2.5,8);
			\draw [green,very thick] (2.5,8) -- (2.5,7.5);
			\draw [green,very thick] (2.5,8) -- (2.5,8.5);
			\draw [green,very thick] (2.5,7.5) -- (2,7.5);
			\draw [green,very thick] (2,7.5) -- (2,6.5);
			\draw [green,very thick] (2.5,7.5) -- (3,7.5);
			\draw [green,very thick] (3,7.5) -- (3,6.5);
			\draw [green,very thick] (2.5,9.5) -- (4,9.5);
			\draw [green,very thick] (4,9.5) -- (4,8);
			\draw [green,very thick] (4,8) -- (3.5,8);
			\draw [green,very thick] (3.5,8) -- (3.5,6.5);
			\draw [black,thick] (4,8) -- (4.5,8);
			\draw [black,thick] (4.5,8) -- (4.5,7);

			\draw (6,10.4) node[anchor=north, below=-0.17cm] {{\color{black} $T^{(2)}$ }};
   \draw [green,thick] (6,10) -- (6,9);
			\draw [black,thick] (6,9) -- (6,8.5);
			\draw [green,very thick] (6,9.5) -- (8,9.5);
			\draw [green,very thick] (8,9.5) -- (8,6.5);
			\draw [black,thick] (8,8) -- (8.5,8);
			\draw [black,thick] (8.5,8) -- (8.5,7);
			\draw [green,very thick] (6,9) -- (6.5,9);	
			\draw [green,very thick] (6.5,9) -- (6.5,6.5);
			\draw [green,very thick] (6.5,7.5) -- (7,7.5);
			\draw [green,very thick] (7,7.5) -- (7,6.5);

			\end{tikzpicture}
			\caption{Tree $T^{(1)}$ may evolve according to variant I or II of the MBT model. Each of the new branches either restarts in a phase according to some fixed distribution (I); or is a continuation of the parent (II). Tree $T^{(2)}$ evolves according to variant III of the MBT model. The left branch on each subtree is a continuation of the parent, the right branch is the new species. Branches in black are extinct.}
			\label{MBTbehaviours}
		\end{center}
	\end{figure}
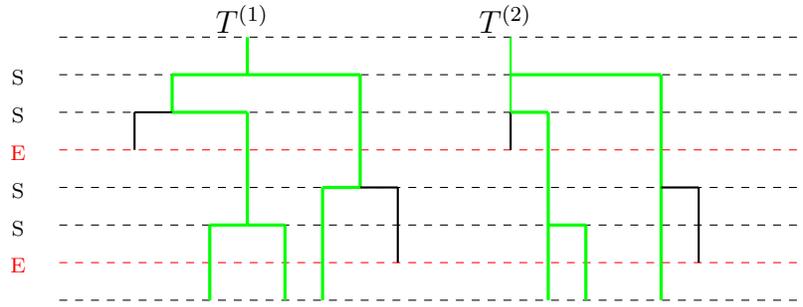

The above model is capable of describing a wide range of possible evolutionary behaviours, including those depicted in Figure~\ref{MBTbehaviours}, by choosing suitable parameters. For example, we describe the following three variants of the model (amongst many other possibilities). Suppose that the matrix ${\bf D}_1=[(D_1)_{ij}]_{i,j=1,\ldots,n}$ is of the form 
	\begin{eqnarray}
		(D_1)_{ij}&=&\beta_{i,S}\times p_{ij}
	\end{eqnarray}
	where $\beta_{i,S}>0$ is a rate of speciation and $p_{ij}$ is the probability that the parent transitions from phase $i$ to phase~$j$ at the time of giving birth (at the time of speciation event), with $0\leq p_{ij}\leq 1$, $\sum_j p_{ij}=1$. We consider the following three possibilities which are special cases of the model above.
\begin{enumerate}[I.]
\item A change in phase of the parent triggers speciation. Each of the two new branches starts in some phase according to the same distribution of phases as the root branch. See tree $T^{(1)}$ in Figure~\ref{MBTbehaviours}.
		
To model this we assume that $p_{ij}=\alpha_j$ for all $i$ so that the parent transitions to phase $j$ with probability $\alpha_j$ at the time of giving birth, and $P_{j,ik}=\alpha_j$ for all $i$ and $k$, so that the child starts from phase $j$ with probability $\alpha_j$ at birth. 
		
\item Each of two new branches is a continuation of the parent branch, that is, it starts in the same phase as the parent just before giving birth. See tree $T^{(1)}$ in Figure~\ref{MBTbehaviours}. 
		
To model this, we assume that $p_{ii}=1$ for all $i$ so that the parent does not change the phase at the time of giving birth, and $P_{i,ii}=1$ for all $i$, so that the starting phase of the child is the same as the phase of the parent at the time of birth.	
		
\item The left branch follows model II while the right branch follows model I. That is, the left branch is the continuation of the parent branch (the existing species) while the right branch is the child branch (new species). See tree $T^{(2)}$ in Figure~\ref{MBTbehaviours}.
		
To model this we assume that $p_{ii}=1$ for all $i$ so that the parent does not change the phase at the time of giving birth, and $P_{j,ik}=\alpha_j$ for all $i$ and $k$, so that the child starts from phase~$j$ with probability $\alpha_j$ at birth.
\end{enumerate}
	
Variants I and III of the above model are useful for modelling evolution of species where speciation and extinction rates depend on an inherited trait. Variant III can be used to model a situation in which speciation event occurs after some individuals became isolated from their parent population due to geographical factors, such as mountain range or waterway. As Duchen et al. stated in~\cite{duchen2021effect}, such geographic isolation is likely to generate descendent species with traits different from its parent. An example of this is the evolution of Darwin's finches in the Gal$\acute{\text{a}}$pagos island studied by Grant in~\cite{grant1999ecology}. This type of speciation occurs due to the fact that when a group of individuals separates from parent population by some geographical barriers, then over generations this group will develop some unique trait different from their parent population, as influenced by their unique habitat. When the separated group is large, it is called allopatric speciation, while if only small groups of individuals get separated, it is called peripatric speciation~\cite{mayr1982speciation,mayr1999systematics}.
	
Furthermore, variant I can be used to model a situation in which a change in the trait of parent species leads to a speciation event in which the descending lineages of the parent do not inherit the parent's trait, but both evolve independently. An example of such independent evolution of children is asymmetrical trait inheritance described by Duchen et al. in~\cite{duchen2021effect}. We note that many real-life examples of trait evolution examples show evidence of asymmetrical trait inheritance, these include the evolution of floral shape and size in flowering plants studied by Serrano-Serrano et al. in~\cite{serrano2015decoupled}, and the evolution of insect genitalia considered by Schilthuizen in~\cite{schilthuizen2003shape}.
	
On the other hand, variant II of the model can be used to model a situation in which both descending lineages inherit their parent's trait following a speciation event, leading to  symmetrical trait inheritance, this is assumed in most phylogenetic models of trait evolution ~\cite{felsenstein1973maximum}. Another example of variant II can also be seen in a model described by Cavalli-Sforza and Edwards in~\cite{cavalli1967phylogenetic}  for evolution of gene frequencies between populations. In this model, the authors assume that only genetic drift and selection affect changes in gene frequencies, and that at each splitting event, the two daughter populations are identical to their parent population.

	
\begin{Example}
\label{ex:BiSSE} 
Consider the following BiSSE model proposed by Maddison, Midford and Otto in~\cite{2007MMO}. Suppose $\{J_t:t\geq 0\}$ is a continuous-time Markov chain with state space $\{0,1\}$ and transition rates $q_{01}$, $q_{10}$, which drives the speciation and extinction events so that when $J_t=i\in\{0,1\}$ then the speciation and extinction rates are $\lambda_i$ and $\mu_i$, respectively. In order to derive the likelihood of observing a particular reconstructed tree under such model, the authors wrote differential equations for the speciation and extinction processes and proposed numerical integration as a method to solve them.

We note that the BiSSE model can be represented as an MBT $\{(M(t),\varphi(t)):t\geq 0\}$ with state space $\mathbb{N}\times \{E,0,1\}$ where $E$ is an absorbing state corresponding to extinction event, and parameters 
		\begin{eqnarray}
			\balpha&=&
			\left[
			\begin{array}{cc}
				\alpha_0&\alpha_1\\
			\end{array}
			\right]
			,\
			{\bf d}=
			\left[
			\begin{array}{c}
				\mu_0\\
				\mu_1
			\end{array}
			\right]
			,\
			{\bf D}_0=
			\left[
			\begin{array}{cc}
				q_{00}&q_{01}\\
				q_{10}&q_{11}
			\end{array}
			\right]
			,\
			{\bf D}_1=
			\left[
			\begin{array}{cc}
				\lambda_0&0\\
				0&\lambda_1
			\end{array}
			\right]
			,
			\nonumber\\
			{\bf P}&=&
			\left[
			\begin{array}{cc}
				P_{0,00}&P_{1,00}\\
				P_{0,01}&P_{1,01}\\
				P_{0,10}&P_{1,10}\\
				P_{0,11}&P_{1,11}
			\end{array}
			\right]
			=
			\left[
			\begin{array}{cc}
				1&0\\
				0&0\\
				0&0\\
				0&1
			\end{array}
			\right],\ 
			{\bf B}
			=
			\begin{bmatrix}
				B_{0,00} & B_{0,01} & B_{1,00} & B_{1,01}\\
				B_{0,10} & B_{0,11} & B_{1,10} & B_{1,11}
			\end{bmatrix}
			=
			\begin{bmatrix}
				\lambda_{0} & 0 & 0 & 0 \\
				0 & 0 & 0 & \lambda_{1}
			\end{bmatrix}.
		\end{eqnarray}
		
A similar observation applies to the MuSSE model discussed by FitzJohn in~\cite{2012F}, which is a generalisation of the BiSSE model to include more than two states $0$ and $1$, that is, to $\{J_t:t\geq 0\}$ with state space $\{1,2,\ldots ,n\}$. The MuSSE model can be represented as an MBT $\{(M(t),\varphi(t)):t\geq 0\}$ with state space $\mathbb{N}\times \{E,1,2,\ldots,n\}$ where $E$ is an absorbing state corresponding to extinction event, and parameters 
		\begin{eqnarray}
			\balpha&=&[\alpha_i]_{i=1,2,\ldots,n}
			,\
			{\bf d}=\left([\mu_i]_{i=1,2,\ldots,n}\right)^T
			,\ 
			{\bf D}_0=[q_{ij}]_{i=1,2,\ldots,n}
			,\ 
			{\bf D}_1=
			diag(\lambda_i)_{i=1,2,\ldots,n}
			,
			\nonumber\\
			{\bf P}&=&[P_{j,ik}]_{i,j,k=1,\ldots,n} \mbox{ is such that } P_{j,ik}=1 \mbox{ for }i=j=k \mbox{ and } P_{j,ik}=0 \mbox{ otherwise},
			\nonumber\\
			{\bf B}&=&[B_{i,jk}]_{i,j,k=1,\ldots,n}
			\mbox{ is such that } B_{i,jk}=\lambda_i \mbox{ for }i=j=k \mbox{ and } B_{i,jk}=0 \mbox{ otherwise}.
			\label{eq7}
		\end{eqnarray}
		
The MBT model is more general than the MuSSE model, and so the existing theory of the MBTs, as well as the new results and algorithms presented here, can be applied to the analysis of the MuSSE model. \hfill$\Box$
	\end{Example}
	
	\begin{Example}
	 \label{ex:CopesRule}    
  Cope's rule proposes that animal lineages evolve towards larger body sizes over time. Investigations of Cope's Rule are complicated by the fact that there might be conflicting forces at micro and macro scales of evolution~\cite{10.1111/evo.12653}. For example, it is possible that within a population larger individuals have an advantage over smaller individuals, but at the species level, large species may be more vulnerable to extinction compared to smaller species. A discretized version of this proposed scenario could be modeled within the MBT framework, with
  \begin{eqnarray}
			\balpha&=&
			\left[
			\begin{array}{ccc}
				\alpha_0&\alpha_1&\alpha_2\\
			\end{array}
			\right]
			,\
			{\bf d}=
			\left[
			\begin{array}{c}
				\mu_0\\
				\mu_1\\
                \mu_2
			\end{array}
			\right]
			,\
			{\bf D}_0=
			\left[
			\begin{array}{ccc}
				q_{00}&q_{01}&0\\
				0&q_{11}&q_{12}\\
                0&0&0
			\end{array}
			\right]
			,\
			{\bf D}_1=
			\left[
			\begin{array}{ccc}
				\lambda_0&0&0\\
				0&\lambda_1&0\\
                0&0&\lambda_2
			\end{array}
			\right]
			,
			\nonumber\\
			{\bf P}&=&
			\left[
			\begin{array}{ccc}
				P_{0,00}&P_{1,00}&P_{2,00}\\
				&\cdots\\
				P_{0,11}&P_{1,11}&P_{2,11}\\
               & \cdots\\
				P_{0,22}&P_{1,22}&P_{2,22}\\
                &\cdots
			\end{array}
			\right]
			=
			\left[
			\begin{array}{ccc}
				1&0&0\\
                &\cdots\\
				0&1&0\\
               & \cdots\\
				0&0&1\\
				&\cdots
			\end{array}
			\right].\ 
		\end{eqnarray}
  Here the states 0,1 and 2 correspond to species being small, medium-sized, or large. The ${\bf D}_0$ matrix encodes the fact that species can transition to larger sizes (or stay the same) but cannot get smaller. Choosing $\mu_0 < \mu_1 < \mu_2$ would impose that larger species are more likely to go extinct. The ${\bf D}_1$ and ${\bf P}$ matrices imply that speciation does not change a species' size and that child species start out out the same size as their parents (only rows of the ${\bf P}$ matrix with non-zero entries are shown).
	\end{Example}
	
	\begin{Example}
	 \label{ex:Allopatry}    
  MBTs could be used to represent an allopatric model of speciation. For example, the state space could represent possible locations $1, 2, \ldots, k$ that species can inhabit, and so
  \begin{eqnarray}
			\balpha&=&
			\left[
			\begin{array}{ccc}
				\alpha_1&\alpha_2&\cdots\\
			\end{array}
			\right]
			,\
			{\bf d}=
			\left[
			\begin{array}{c}
				\mu_1\\
				\mu_2\\
                \vdots
			\end{array}
			\right]
			,\
			{\bf D}_0=
			{\bf 0}
			,\
			{\bf D}_1=
			\left[
			\begin{array}{ccc}
				\lambda_1&0&\cdots\\
				0&\lambda_2&\\
                \vdots&&\ddots
			\end{array}
			\right]
			,
   \nonumber\\
			{\bf P}
			&=&
			\left[
			\begin{array}{cccc}
				0&\frac{1}{k-1}&\frac{1}{k-1}&\cdots\\
                &\cdots\\
				\frac{1}{k-1}&0&\frac{1}{k-1}&\cdots\\
               & \cdots\\
				\frac{1}{k-1}&\frac{1}{k-1}&0&\cdots\\
				&\cdots
			\end{array}
			\right].\ 
		\end{eqnarray}
  
  The ${\bf D}_0$ and ${\bf D}_1$ matrices encode the fact that parent species stay in the same location before and after speciating. The ${\bf P}$ matrix imposes that a child species must move to a new location from its parent (only rows of the ${\bf P}$ matrix with non-zero entries are shown). Here the structure of ${\bf P}$ means that any new location is equally likely, but a more realistic geographical structure could be imposed. Note that this model of speciation is asymmetric.
	\end{Example}

 \section{Analysis of a reconstructed species tree}\label{sec:recspetree}

\begin{figure}[h]
		\begin{center}
			\begin{tikzpicture}[>=stealth,redarr/.style={->}]	
			\draw [dashed] (0,10) -- (10,10);
			
			\draw [dashed] (0,9.5) -- (10,9.5);
			\draw (-0.5,9.5) node[anchor=north, below=-0.2cm] {\scriptsize{\color{black} S }};
			\draw [red,dashed] (0,9) -- (10,9);
			\draw (-0.5,9) node[anchor=north, below=-0.2cm] {\scriptsize{\color{red} S }};
			\draw [red,dashed] (0,8.5) -- (10,8.5);
			\draw (-0.5,8.5) node[anchor=north, below=-0.2cm] {\scriptsize{\color{red} E }};
			\draw [red,dashed] (0,8) -- (10,8);
			\draw (-0.5,8) node[anchor=north, below=-0.2cm] {\scriptsize{\color{red} S }};
			\draw [dashed] (0,7.5) -- (10,7.5);
			\draw (-0.5,7.5) node[anchor=north, below=-0.2cm] {\scriptsize{\color{black} S }};
			\draw [red,dashed] (0,7) -- (10,7);
			\draw (-0.5,7) node[anchor=north, below=-0.2cm] {\scriptsize{\color{red} E }};
			\draw [dashed] (0,6.5) -- (10,6.5);
			
			\draw (2.5,10.4) node[anchor=north, below=-0.17cm] {{\color{black} $T$ }};
			\draw [green,very thick] (2.5,10) -- (2.5,9.5);
			\draw [green,very thick] (2.5,9.5) -- (1.5,9.5);
			\draw [green,very thick] (1.5,9.5) -- (1.5,9);
			\draw [black,thick] (1.5,9) -- (1,9);
			\draw [black,thick] (1,9) -- (1,8.5);			
			\draw [green,very thick] (1.5,9) -- (2.5,9);
			\draw [green,very thick] (2.5,9) -- (2.5,8);
			\draw [green,very thick] (2.5,8) -- (2.5,7.5);
			\draw [green,very thick] (2.5,8) -- (2.5,8.5);
			\draw [green,very thick] (2.5,7.5) -- (2,7.5);
			\draw [green,very thick] (2,7.5) -- (2,6.5);
			\draw [green,very thick] (2.5,7.5) -- (3,7.5);
			\draw [green,very thick] (3,7.5) -- (3,6.5);
			\draw [green,very thick] (2.5,9.5) -- (4,9.5);
			\draw [green,very thick] (4,9.5) -- (4,8);
			\draw [green,very thick] (4,8) -- (3.5,8);
			\draw [green,very thick] (3.5,8) -- (3.5,6.5);
			\draw [black,thick] (4,8) -- (4.5,8);
			\draw [black,thick] (4.5,8) -- (4.5,7);

			\draw (7.5,10.4) node[anchor=north, below=-0.17cm] {{\color{black} $T^*$ }};
			\draw [green,very thick] (7.5,10) -- (7.5,9.5);
			\draw [green,very thick] (7.5,9.5) -- (6.5,9.5);
			\draw [green,very thick] (6.5,9.5) -- (6.5,7.5);
			\draw [green,very thick] (6.5,7.5) -- (6,7.5);
			\draw [green,very thick] (6,7.5) -- (6,6.5);	
			\draw [green,very thick] (6.5,7.5) -- (7,7.5);
			\draw [green,very thick] (7,7.5) -- (7,6.5);
			\draw [green,very thick] (7.5,9.5) -- (8,9.5);
			\draw [green,very thick] (8,9.5) -- (8,6.5);			
			
			\end{tikzpicture}
			\caption{An example of a true tree $T$ and its reconstructed tree $T^*$. The information about speciation events followed by extinction events is missing in the reconstructed tree $T^*$ due to lack of data.}
			\label{fig:TTstar}
		\end{center}
	\end{figure}
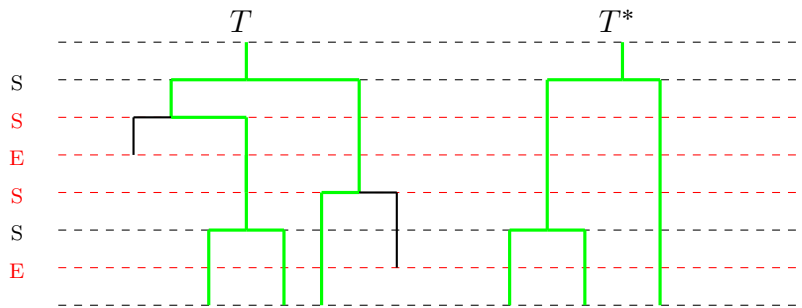

The above theoretical model describes the evolution of a species tree~$T$. Fitting the parameters of the model to data is a difficult task since the data is typically incomplete in that only species at the present day are observed. Therefore, we now make a distinction between a true species tree $T$ and a {\em reconstructed} species tree $T^*$, which may contain incomplete information about~$T$, see Figure~\ref{fig:TTstar}. 
	\begin{Definition}
		Let $T$ be a true (potentially unknown) species tree that consists of internal branches ending with speciation events, internal branches ending with extinction events, and external branches which are yet to be absorbed into either $S$ or $E$. Denote by $T^*$ a corresponding reconstructed tree with some given topology, which contains some partial information about the true species tree $T$.
	\end{Definition}

	Below we discuss the computation of the likelihood of observing a reconstructed tree $T^*$ under an assumption that some information is missing due to lack of data. The results are stated in a convenient matrix form, which are useful for computation using fast algorithms.

First, we introduce some notation. Let $N(t)$ be the total number of individuals at time $t$ in the lineage which started from one individual at time $0$. Define a matrix $\widetilde {\bf D}^{(1)}(t)=[\widetilde D_{ij}^{(1)}(t)]$, a column vector $\mathbf{D}^{(1)}(t)=[D_i^{(1)}(t)]=\widetilde {\bf D}^{(1)}(t){\bf 1}$ and a scalar $D^{(1)}(t)=\balpha\mathbf{D}^{(1)}(t)$ such that
	\begin{eqnarray}\label{defD1t}
		\widetilde D_{ij}^{(1)}(t) &=& 
		\mathbb{P}\left(N(t)= 1, \varphi(t)=j\ |\ N(0) = 1, \varphi(0)=i \right)
	\end{eqnarray}
	is the probability that a lineage on the tree that started at time $0$ in phase $i$ consists of one individual only at time $t$, who is in phase $j$. 
 
    $D_i^{(1)}(t)=\mathbb{P}\left(N(t)= 1\ |\ N(0) = 1, \varphi(0)=i \right)$ is the probability that a lineage on tree that started at time $0$ in phase $i$ consists of one individual only at time~$t$, and $D^{(1)}(t)=\mathbb{P}\left(N(t)= 1 \ |\ N(0) = 1 \right)$ is the probability that a lineage on tree that started at time~$0$ consists of one individual only at time $t$.
	
	Also, define a column vector $\mathbf{E}(t)=[E_i(t)]$ and a scalar $E(t)=\balpha\mathbf{E}(t)$ such that
	\begin{eqnarray}
		E_{i}(t) &=& \mathbb{P}\left(N(t)= 0 \ |\ N(0) = 1, \varphi(0)=i \right)
	\end{eqnarray}
	is the probability that a tree becomes extinct by time $t$ given it started at time $0$ in phase $i$, and $E(t)$ is the probability that a tree becomes extinct by time $t$.

	We note that by~\cite{2009HLR}, it follows that ${\bf E}(t)$ is the solution of the differential equation
	\begin{eqnarray}
		\frac{d {\bf E}(t)}{dt}
		&=& {\bf d} + {\bf D}_0{\bf E}(t) + {\bf B}({\bf E}(t)\otimes {\bf E}(t))
		\label{EtMBT}
	\end{eqnarray}
	with the initial condition ${\bf E}(0)={\bf 0}$, and ${\bf E}=\lim_{t\to\infty}{\bf E}(t)$ is the minimal non-negative solution of
	\begin{eqnarray}
		{\bf E}&=& (-{\bf D}_0)^{-1}{\bf d}  + (-{\bf D}_0)^{-1}{\bf B}({\bf E}\otimes {\bf E}),
		\label{EMBT}
	\end{eqnarray} 
	which can be solved using Newton's method~\cite{Guo2015,2011H}, with ${\bf E}_0={\bf 0}$ and for $k\geq 1$,
	\begin{eqnarray}
		{\bf E}_k&=& ( {\bf I} - (-{\bf D}_0)^{-1}
		{\bf B}({\bf E}_{k-1}\oplus {\bf E}_{k-1}) )
		({\bf E}_{k-1} - (-{\bf D}_0)^{-1}{\bf d} 
		- (-{\bf D}_0)^{-1}({\bf E}_{k-1}\otimes {\bf E}_{k-1}))
	\end{eqnarray}
	until the stopping criterion $|| {\bf E}_{k}-{\bf E}_{k-1}||<\epsilon$ is met for some $\epsilon>0$. Also, denoting by $\lambda$ the maximal eigenvalue of $(-{\bf D}_0)^{-1}{\bf B}({\bf 1}\oplus {\bf 1})$, it follows that ${\bf E}={\bf 1}$ whenever $\lambda\leq 1$, ${\bf E}\not={\bf 1}$ if $\lambda> 1$, and ${\bf E}<{\bf 1}$ if $\lambda> 1$ and $(-{\bf D}_0)^{-1}{\bf B}({\bf 1}\oplus {\bf 1})$ has an inverse~\cite{2011H}.

	\begin{Remark}\label{remBM} By Example~\ref{ex:BiSSE}, applying Equation~\eqref{EtMBT} to the BiSSE model in~\cite{2007MMO} gives,	
		\begin{eqnarray}
			\begin{bmatrix}
				\frac{dE_{0}(t)}{dt}\\
				\frac{dE_{1}(t)}{dt}
			\end{bmatrix} 
			&=&
			\begin{bmatrix}
				\mu_{0} \\
				\mu_{1}
			\end{bmatrix}
			+
			\begin{bmatrix}
				-\left(\mu_{0} + q_{01} + \lambda_{0}\right)E_{0}(t) + q_{01}E_{1}(t)\\
				q_{10}E_{0}(t) -\left(\mu_{1} + q_{10} + \lambda_{1}\right)E_{1}(t)
			\end{bmatrix}
			+
			\begin{bmatrix}
				\lambda_{0}E_{0}(t)^{2}\\
				\lambda_{1}E_{1}(t)^{2}
			\end{bmatrix} 
			,
		\end{eqnarray}
		resulting in the differential equations derived in~\cite{2007MMO}, as expected. Similarly, the differential equations for $E_i(t)$ in the MuSSE model in~\cite{2012F} follow from follow from Equations~\eqref{eq7} and \eqref{EtMBT}. We note that the term ${\bf B}({\bf E}(t)\otimes {\bf E}(t))$ can be computed more efficiently (using smaller-size matrices) with the Hadamard product $\odot$, since here ${\bf B}({\bf E}(t)\otimes {\bf E}(t))=
		{\bf D}_1({\bf E}(t)\odot {\bf E}(t))$.  
	\end{Remark}

	\begin{Lemma}\label{lemDt}
		$\mathbf{D}^{(1)}(t)$ is the solution to the differential equation
		\begin{eqnarray}\label{eq:DE_D1t}
			\frac{d {\bf D}^{(1)}(t)}{dt} = \mathbf{D}_{0}\mathbf{D}^{(1)}(t) +
			\mathbf{B}(\mathbf{E}(t) \otimes \mathbf{D}^{(1)}(t))
			+
			\mathbf{B}(\mathbf{D}^{(1)}(t) \otimes \mathbf{E}(t))
		\end{eqnarray}
		with the initial condition $\mathbf{D}^{(1)}(0)={\bf 1}$. 
		
		Furthermore, 	
		\begin{eqnarray}\label{eq:int_D1t}
			{\bf D}^{(1)}(t) = 
			e^{\mathbf{D}_{0}(t)}{\bf 1}
			+
			\int_{u=0}^t
			e^{\mathbf{D}_{0}(t-u)}
			\mathbf{B}
			\left(
			\mathbf{E}(u) \otimes \mathbf{D}^{(1)}(u)
			+
			\mathbf{D}^{(1)}(u) \otimes \mathbf{E}(u)
			\right)
			du.
		\end{eqnarray}

	\end{Lemma}
	{\bf Proof:} In our proof we present an argument with the aim of highlighting the underlying physical interpretations. An alternative method is to 
 apply the derivatives of the generating function studied in~\cite[Section 3]{2009HLR}, since by standard probability theory (see e.g Ross~\cite{Ross2010}), the $n$th derivative of a probability generating function evaluated at $s=0$ and divided by $n!$, gives the probability mass at $n$.
 
 To show~\eqref{eq:DE_D1t}, we write Kolmogorov differential equations, assuming that the lineage starts from a parent in some phase $\varphi(0)=i\in\{1,\cdots, n\}$, and consider the following two alternatives.
	\begin{enumerate}
		\item The process instantaneously transitions on the parent lineage from phase $i$ to some phase $k \in\{1,\cdots, n\}$ at rate $\left[\mathbf{D}_{0}\right]_{ik}$, and next, given start in phase $\varphi(0)=k$ with $N(0)=1$, we observe $N(t) =1$ with probability $[\mathbf{D}^{(1)}(t)]_{k}$. The total rate of this occurring is,
		\begin{eqnarray}
			\sum_{k}[\mathbf{D}_{0}]_{ik}[\mathbf{D}^{(1)}(t)]_{k}=[\mathbf{D}_{0}\mathbf{D}^{(1)}(t)]_{i}.
			\label{eqalt1}
		\end{eqnarray}
		
		\item The parent simultaneously transitions to some phase $k\in\{1,\cdots, n\}$ and gives birth to a child in some phase $j\in\{1,\cdots, n\}$ at rate $[\mathbf{B}]_{i,jk}$. Conditioning on phases $j$ and $k$ and considering the memoryless property, we have two alternative scenarios observing $N(t)=1$: 
		\begin{enumerate}
			\item Given start in phase $\varphi(0)=k$, the parent lineage becomes extinct by time $t$ with probability $\left[\mathbf{E}(t)\right]_{k}$, and the child at some phase $\varphi(0)=j$ survives until time $t$ with probability $\left[\mathbf{D}^{(1)}(t)\right]_{j}$. The total rate of this occurring is,
			\begin{eqnarray}
				\sum_{j,k=1,\cdots, n} \mathbf{B}_{i,jk}\left([\mathbf{E}(t)]_{k}[\mathbf{D}^{(1)}(t)]_{j}\right)=[\mathbf{B}(\mathbf{E}(t) \otimes \mathbf{D}^{(1)}(t))]_{i}.
				\label{case1a}
			\end{eqnarray}
			
			\item Given start in phase $\varphi(0)=k$, the parent lineage survives until time $t$ with probability $\left[\mathbf{D}^{(1)}(t)\right]_{k}$, and the child at some phase $\varphi_{0}=j$ becomes extinct by time $t$ with probability $\left[\mathbf{E}(t)\right]_{j}$. The total rate of this occurring is,
			\begin{eqnarray}
				\sum_{j,k=1,\cdots, n} \mathbf{B}_{i,jk}\left([\mathbf{D}^{(1)}(t)]_{k}[\mathbf{E}(t)]_{j}\right)=[\mathbf{B}(\mathbf{D}^{(1)}(t) \otimes \mathbf{E}(t))]_{i}
				\label{case1b}
			\end{eqnarray}
		\end{enumerate}
		
	\end{enumerate}
	Therefore, by taking the sum of~\eqref{eqalt1}--\eqref{case1b}, for all $i = 0,1,\cdots,n$, we have
	\begin{eqnarray}
		\left[ \frac{d {\bf D}^{(1)}(t)}{dt}\right]_i
		&=&\frac{dD_i(t)}{dt}	
		=
		[\mathbf{D}_{0}\mathbf{D}^{(1)}(t) + 
		\mathbf{B}(\mathbf{E}(t) \otimes \mathbf{D}^{(1)}(t))
		+
		\mathbf{B}(\mathbf{D}^{(1)}(t) \otimes \mathbf{E}(t))
		]_{i},
	\end{eqnarray}
	and so~\eqref{eq:DE_D1t} follows, where $\mathbf{D}^{(1)}(0)={\bf 1}$ due to $D_i^{(1)}(0)=\mathbb{P}\left(N(0)= 1\ |\ N(0) = 1, \varphi(0)=i \right)=1$ for all $i$ (at time $0$ there is only one branch). 
	
	Next, to show~\eqref{eq:int_D1t}, we consider the evolution of the process in the time interval $(0,t)$ and the following two alternatives.
	\begin{enumerate}
		\item No speciation or extinction events occur in the time interval $(0,t)$. The probability of this alternative is recorded in the vector
		\begin{eqnarray}\label{eq:alt1}
			e^{\mathbf{D}_{0}t}{\bf 1}.
		\end{eqnarray}
		\item The first speciation event occurs at time $(t-u)$ for some $u\in(0,t)$, according to the probability density matrix
		$$e^{\mathbf{D}_{0}(t-u)}
		\mathbf{B}.$$ 
		Next, one of the two new branches becomes extinct while the other ends with one individual by time $t$ (or equivalently, by the remaining time $u$), according to the probability vector
		$$\mathbf{E}(u) \otimes \mathbf{D}^{(1)}(u)
		+
		\mathbf{D}^{(1)}(u) \otimes \mathbf{E}(u).$$ 
		Therefore, by conditioning on $u$, the probability vector of the second alternative is,
		\begin{eqnarray}\label{eq:alt2}
			\int_{u=0}^t
			e^{\mathbf{D}_{0}(t-u)}
			\mathbf{B}
			\left(
			\mathbf{E}(u) \otimes \mathbf{D}^{(1)}(u))
			+
			\mathbf{D}^{(1)}(u) \otimes \mathbf{E}(u)
			\right)
			du.
		\end{eqnarray}
	\end{enumerate}
	By adding the probabilities~\eqref{eq:alt1}-\eqref{eq:alt2} of the two alternatives, we obtain~\eqref{eq:int_D1t}, which completes the proof. Note also that by taking derivatives in~\eqref{eq:int_D1t}, we obtain~\eqref{eq:DE_D1t}, giving another method of showing~\eqref{eq:DE_D1t}. \hfill\rule{9pt}{9pt}

	\begin{Remark}\label{remDt} By Example~\ref{ex:BiSSE}, applying Lemma~\ref{lemDt} to the BiSSE model in~\cite{2007MMO}, gives,	
		\begin{eqnarray}
			\begin{bmatrix}
				\frac{d{D}_{0}^{(1)}(t)}{d(t)}\\
				\frac{d{D}_{1}^{(1)}(t)}{d(t)}
			\end{bmatrix}
			&=&
			\begin{bmatrix}
				-\left(\mu_{0}+q_{01}+\lambda_{0}\right)D_{0}(t) + q_{01}D_{1}(t)\\
				-\left(\mu_{1}+q_{10}+\lambda_{1}\right)D_{1}(t) + q_{10}D_{0}(t)
			\end{bmatrix}
			+
			\begin{bmatrix}
				2\lambda_{0}E_{0}(t)D_{0}(t)\\
				2\lambda_{1}E_{1}(t)D_{1}(t)
			\end{bmatrix},
		\end{eqnarray}
		resulting in the differential equations derived in~\cite{2007MMO} (with $D_{i}^{(1)}(t)$ denoted $D_{Ni}(t)$ there), as expected. Similarly, the differential equations for ${D}_{Ni}(t)$ in the MuSSE model in~\cite{2012F} follow by Lemma~\ref{lemDt} and Equation~\eqref{eq7}. The result in Lemma~\ref{lemDt} is more general (applies to all MBTs) and is also stated in a matrix form, which is convenient for efficient computation. We note that the terms in~\eqref{eq:DE_D1t}-\eqref{eq:int_D1t} can be computed more efficiently with $\mathbf{B}(\mathbf{E}(t) \otimes \mathbf{D}^{(1)}(t))
		+
		\mathbf{B}(\mathbf{D}^{(1)}(t) \otimes \mathbf{E}(t))
		=2	{\bf D}_1\left(\mathbf{E}(t) \odot \mathbf{D}^{(1)}(t)\right)$.
	\end{Remark}

	We are now ready to introduce the following useful conditional probability $D_{n_L\vert n}(t)$, which is the measure of balance of the species tree, given $n$ tips are observed at time $t$.

	First, we extend the definition of $\mathbf{D}^{(1)}(t)$ and, for all $n\geq 0$, define a column vector $\mathbf{D}^{(n)}(t)=[D_i^{(n)}(t)]$ and a scalar $D^{(n)}(t)=\balpha\mathbf{D}^{(n)}(t)$ such that
	\begin{eqnarray}\label{defDnt}
		D_{i}^{(n)}(t) &=& \mathbb{P}\left(N(t)= n \ |\ N(0) = 1, \varphi(0)=i \right)
	\end{eqnarray}
	is the probability that a lineage on a tree that started at time $0$ in phase $i$ consists of $n$ individuals at time $t$, and $D^{(n)}(t)=\mathbb{P}(N(t)=n)$ is the probability that a lineage on a tree that started at time $0$ consists of $n$ individuals at time $t$. Note that  $D^{(0)}(t)\equiv E(t)$ and $\mathbf{D}^{(0)}(t)\equiv {\bf E}(t)$, and so the expressions in the case $n=0$ follow from~\eqref{EtMBT}.

	Next, let $N_L(t)$ be the total number of individuals at time $t$ in the left subtree in the reconstructed tree of the lineage which started from one individual at time $0$. For $n\geq 2$ and $n_L=1,\ldots,n-1$, define scalars
	\begin{eqnarray}\label{eq:defDigivenn}
		D_{n_L\vert n}(t)
		&=&
		\mathbb{P}(N_L(t)=n_L | N(t)=n, N(0) = 1) = \frac{D^{(n_L,n)}(t)}{D^{(n)}(t)},\\
		D^{(n_L,n)}(t)
		&=&
		\mathbb{P}(N_L(t)=n_L , N(t)=n \ |\ N(0) = 1),
	\end{eqnarray}
	where $D_{n_L\vert n}(t)$ is the probability that we observe $n_L$ tips on the left subtree given we have $n$ tips at the present time $t$ on a tree that started at time $0$, and $D^{(n_L,n)}(t)$ is the probability that we observe $n_L$ tips on the left subtree and $n$ tips in total at the present time $t$ on a tree that started at time $0$.

	Denote by $I(\cdot)$ the indicator function which takes value $1$ when the statement inside the brackets is true, and $0$ otherwise. 
	\begin{Lemma}\label{lemDt2}
		For $n\geq 1$, $D^{(n)}(t)=\balpha {\bf D}^{(n)}(t)$ where ${\bf D}^{(n)}(t)$ is the solution of the differential equation
		\begin{eqnarray}\label{iterationDnt}
			\frac{d{\bf D}^{(n)}(t)}{dt} &=& 
			{\bf D}_{0}{\bf D}^{(n)}(t)
			+{\bf B}\left({\bf D}^{(0)}(t) \otimes {\bf D}^{(n)}(t)
			+{\bf D}^{(n)}(t) \otimes {\bf D}^{(0)}(t)\right)
			+\sum_{n_L=1}^{n-1}{\bf B}\left({\bf D}^{(n_L)}(t) \otimes {\bf D}^{(n-n_L)}(t)\right)\nonumber\\
		\end{eqnarray}
		with the initial condition ${\bf D}^{(n)}(0)=I(n=1){\bf 1}$.
		
		For $n\geq 2$ and $n_L=1,\ldots,n-1$, $D^{(n_L,n)}(t)=\balpha{\bf D}^{(n_L,n)}(t)$ where ${\bf D}^{(n_L,n)}(t)$ is a column vector that is the solution of the differential equation
		\begin{eqnarray}\label{eqDint}
			\frac{d{\bf D}^{(n_L,n)}(t)}{dt}={\bf D}_{0}{\bf D}^{(n_L,n)}(t)+{\bf B}
			\left(
			{\bf D}^{(0)}(t)\otimes {\bf D}^{(n_L,n)}(t)
			+{\bf D}^{(n_L,n)}(t)\otimes {\bf D}^{(0)}(t)
			\right)
			+{\bf B}
			\left(
			{\bf D}^{(n_L)}(t)\otimes {\bf D}^{(n-n_L)}(t)\right)\nonumber\\
		\end{eqnarray}
		with the initial condition ${\bf D}^{(n_L,n)}(0)={\bf 0}$.
	\end{Lemma}
	{\bf Proof:} Equation~\eqref{iterationDnt} follows by a straightforward generalisation of the argument in the proof of Lemma~\ref{lemDt}, since in order to observe $n$ tips on a tree, we either have a phase transition and end up with $n$ tips, or there is a branching event and we end up with $n_L$ tips on the left subtree and $n-n_L$ tips on the right subtree at time $t$, for $n_L=0,\ldots,n$. The condition ${\bf D}^{(n)}(0)=I(n=1){\bf 1}$ follows since at time $0$ there is only one branch.

	Equation~\eqref{eqDint} follows since in order to end up with $n_L$ tips on the left subtree in the reconstructed tree and $n$ tips in total on a tree, we must observe
	\begin{itemize}
		\item a phase transition and then end up with $n_L$ tips on the left subtree and $n$ tips in total on a tree, according to ${\bf D}_{0}{\bf D}^{(n_L,n)}(t)$, or
		\item a branching event and then end up with $0$ tips on the left subtree in the true tree due to extinction and $n$ tips on the right subtree, according to ${\bf B}({\bf D}^{(0)}(t)\otimes {\bf D}^{(n_L,n)}(t))$, or
		\item a branching event and then end up with $n$ tips on the left subtree and $0$ tips on the right subtree due to extinction, according to ${\bf B}({\bf D}^{(n_L,n)}(t)\otimes {\bf D}^{(0)}(t))$, or
		\item a branching event and then end up with $n_L$ tips on the left subtree and $n-n_L$ tips on the right subtree in the true tree, according to ${\bf B}({\bf D}^{(n_L)}(t)\otimes {\bf D}^{(n-n_L)}(t))$,
	\end{itemize} 
	and so by summing up these rates, we obtain the total rate on the right-hand side of~\eqref{eqDint}, and ${\bf D}^{(n_L,n)}(0)={\bf 0}$ for $n\geq 2$ follows since at time $0$ there is only one branch. \rule{9pt}{9pt}

	We now consider the following practical problem that is of a significant interest in the analysis of incomplete data.
	\begin{Problem}\label{Prb3}
		Suppose that the only available information are the times of speciation events corresponding to the branches that have not become extinct. That is, no information about the extinct branches is available, and so we do not have any information about potential extinction events that have occurred. Consequently, the times of the speciation events corresponding to the branches that ended with extinction are not known, or how many such events might have occurred.
		\hfill$\Box$
	\end{Problem}
	
\begin{figure}[h!]
		\begin{center}
			\begin{tikzpicture}[>=stealth,redarr/.style={->}]	
			\draw (2,10.4) node[anchor=north, below=-0.17cm] {{\color{black} $T^*$ }};
			\draw [dashed] (-1.0,10) -- (6,10);
			\draw (-1.4,10) node[anchor=north, below=-0.3cm] {\scriptsize{\color{black} $t_{\circ}^{1}$ }};
			\draw [dashed] (-1.0,9) -- (2,9);
			\draw [dashed] (3.5,9) -- (6,9);
			\draw (-2,9) node[anchor=north, below=-0.3cm] {\scriptsize{\color{black} $t_{\bullet}^{1}=t_{\circ}^{2}=t_{\circ}^{3}$ }};
			\draw (7,9) node[anchor=north, below=-0.3cm] {\scriptsize{\color{black} $\widetilde\varphi(1)\in \{0,1\}$ }};
			\draw [dashed] (4.5,8) -- (6,8);
			\draw [dashed] (2.5,8) -- (-1.0,8);
			\draw (-2,8) node[anchor=north, below=-0.3cm] {\scriptsize{\color{black} $t_{\bullet}^{2}=t_{\circ}^{\widehat{3}}=t_{\circ}^{4}$ }};
			\draw (7,8) node[anchor=north, below=-0.3cm] {\scriptsize{\color{black} $\widetilde\varphi(2)\in \{0,1\}$ }};
			\draw [dashed] (1.5,7) -- (6,7);
			\draw [dashed] (-0.5,7) -- (-1.0,7);
			\draw (-2,7) node[anchor=north, below=-0.3cm] {\scriptsize{\color{black} $t_{\bullet}^{3}=t_{\circ}^{\widehat{1}}=t_{\circ}^{\widehat{2}}$ }};
			\draw (7,7) node[anchor=north, below=-0.3cm] {\scriptsize{\color{black} $\widetilde\varphi(3)\in \{0,1\}$ }};
			\draw [dashed] (5.5,6.5) -- (6,6.5);
			\draw [dashed] (3.5,6.5) -- (-1.0,6.5);
			\draw (7,6.5) node[anchor=north, below=-0.3cm] {\scriptsize{\color{black} $\widetilde\varphi(4)\in \{0,1\}$ }};
			\draw (-2,6.5) node[anchor=north, below=-0.3cm] {\scriptsize{\color{black} $t_{\bullet}^{4}=t_{\circ}^{\widehat{4}}=t_{\circ}^{\widehat{5}}$ }};
			\draw [dashed] (-1.0,5.5) -- (6,5.5);
			\draw (-1.4,5.5) node[anchor=north, below=-0.3cm] {\scriptsize{\color{black} $t$ }};
			\draw [green,very thick] (2,10) -- (2,9);
			\draw [green,very thick] (0.5,9) -- (3.5,9);
			\draw (2.3,9.5) node[anchor=north, below=-0.3cm] {\scriptsize{\color{black} $1$ }};
			\draw [green,very thick] (3.5,9) -- (3.5,8);
			\draw (3.8,8.5) node[anchor=north, below=-0.3cm] {\scriptsize{\color{black} $2$ }};
			\draw [green,very thick] (2.5,8) -- (3.5,8);
			\draw [green,very thick] (3.5,8) -- (4.5,8);
			\draw [green,very thick] (2.5,8) -- (2.5,5.5);
			\draw (2.3,6.75) node[anchor=north, below=-0.3cm] {\scriptsize{\color{black} $\widehat{3}$ }};
			\draw [green,very thick] (4.5,8) -- (4.5,6.5);
			\draw (4.8,7.25) node[anchor=north, below=-0.3cm] {\scriptsize{\color{black} $4$ }};
			\draw [green,very thick] (5.5,6.5) -- (3.5,6.5);
			\draw [green,very thick] (3.5,6.5) -- (3.5,5.5);
			\draw (3.3,6) node[anchor=north, below=-0.3cm] {\scriptsize{\color{black} $\widehat{4}$ }};
			\draw [green,very thick] (5.5,6.5) -- (5.5,5.5);
			\draw (5.8,6) node[anchor=north, below=-0.3cm] {\scriptsize{\color{black} $\widehat{5}$ }};
			%
			\draw [green,very thick] (0.5,9) -- (0.5,7);
			\draw (0.3,8.5) node[anchor=north, below=-0.3cm] {\scriptsize{\color{black} $3$ }};
			\draw [green,very thick] (-0.5,7) -- (1.5,7);
			\draw [green,very thick] (1.5,7) -- (1.5,5.5);
			\draw (1.8,6) node[anchor=north, below=-0.3cm] {\scriptsize{\color{black} $\widehat{2}$ }};
			\draw [green,very thick] (-0.5,7) -- (-0.5,5.5);
			\draw (-0.7,6) node[anchor=north, below=-0.3cm] {\scriptsize{\color{black} $\widehat{1}$ }};
			\end{tikzpicture}
			\caption{Illustration of Problem 1 on a reconstructed tree $T^*$ with 5 taxa, starting at some time $t_{\circ}^{1}$ and evolving until time $t$. Internal branches are labelled $k=1,2,3,4$, following speciation order, and external branches are labelled $\widehat m=\widehat{1},\widehat{2},\widehat{3},\widehat{4},\widehat{5}$. Birth times of the internal and external branches are $t_{\circ}^{1},t_{\circ}^{2},t_{\circ}^{3},t_{\circ}^4$ and $t_{\circ}^{\widehat{1}},t_{\circ}^{\widehat{2}},t_{\circ}^{\widehat{3}},t_{\circ}^{\widehat{4}},t_{\circ}^{\widehat{5}}$, respectively. Times at which each internal branch ends with known speciation events, are $t_{\bullet}^{1},t_{\bullet}^{2},t_{\bullet}^{3},t_{\bullet}^{4}$. Phase $\widetilde \varphi(k)$ is observed just before speciation at time $t_{\bullet}^k$ on internal branch $k$. Here, we assume the BiSSE model, so the set of phases is $\{0,1\}$. Parent branches of external branches: $\mathcal{P}^{\widehat{1}}=\mathcal{P}^{\widehat{2}}=3,\mathcal{P}^{\widehat{3}}=2,\mathcal{P}^{\widehat{4}}=\mathcal{P}^{\widehat{5}}=4$. Parent branches of internal branches: $\mathcal{P}^{2}=\mathcal{P}^{3}=1,\mathcal{P}^{4}=2$.}
			\label{fig:Problem1}
		\end{center}
	\end{figure}
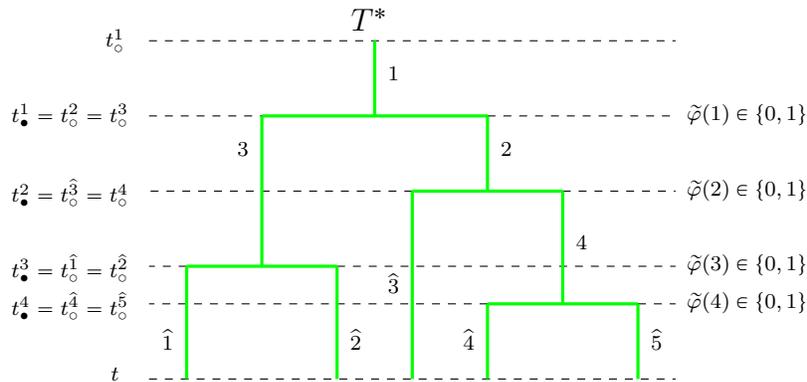

	Denote by $t_{\circ}^1,\ldots,t_{\circ}^K$ the birth times of the internal branches labelled $k=1,\ldots,K$, and by $t_{\bullet}^1,\ldots,t_{\bullet}^K$ the times at which they end with \underline{known} speciation events. Let $b^k=t_{\bullet}^k-t_{\circ}^k$ be the length of internal branch $k$ and $x^k=t-t_{\bullet}^k$ be the distance from its start to the tip of the tree, for $k=1,\ldots,K$. Similarly, denote by $ {t}_{\circ}^{\widehat 1},\ldots,{t}_{\circ}^{\widehat M}$ the birth times of the external branches labelled $\widehat m=\widehat 1,\ldots ,\widehat M$, which all end at time $t$. Let $  {b}^{\widehat  m}=t- {t}_{\circ}^{\widehat  m}$ be the length of external branch $\widehat m$, equal to the distance from its end of the tip of the tree, for $\widehat m=\widehat 1,\ldots,\widehat M$. Further, let $\widetilde\varphi(k)$ be the phase observed just before speciation on the $k$-th internal branch, and $\mathcal{P}^k, \mathcal{P}^{\widehat m}\in\{1,\ldots,K\}$ be the parent of the $k$-th internal and the $m$-th external branch respectively. We illustrate these notations in a tree example in Figure~\ref{fig:Problem1}.	
	
	Further, denote by ${\bf f}(T^*;1;t_{\circ})=[f(T^*;1;t_{\circ})_i]$ and ${\bf f}(T^*;2;t_{\circ})=[f(T^*;2;t_{\circ})_i]$ the likelihood vectors of two subtrees that start at some time $t_{\circ}<t$ at which speciation occurs at the end of some internal branch on the tree $T^*$. Denote by ${\bf f}(T^*;t_{\circ}^k)=[f(T^*;t_{\circ}^k)_i]$ the likelihood vector of an internal branch $k$ that starts at some time $t_{\circ}^k<t$ on the tree $T^*$, and by $f(T^*;t_{\circ}^{\widehat m})=[f(T^*;t_{\circ}^{\widehat m})_i]$ the likelihood vector of an external branch ${\widehat m}$ that starts at some time $t_{\circ}^{\widehat m}<t$ on the tree $T^*$. We note that since the extinction events are not known, speciation events that eventually ended with extinction may have occurred at various points of any internal branches or any external branches. Therefore, the probability density of observing the reconstructed species tree $T^*$ is given by the recursive formula,
	\begin{eqnarray}\label{eq:speciesknowlittle2}
		f(T^*)&=&
		\balpha {\bf G}(b^1,x^1)
		{\bf B}
		\left(\ 
		{\bf f}(T^*;1;t_{\bullet}^1)
		\otimes
		{\bf f}(T^*;2;t_{\bullet}^1)
		+
		{\bf f}(T^*;2;t_{\bullet}^1)
		\otimes
		{\bf f}(T^*;1;t_{\bullet}^1)
		\ \right)
		,\nonumber\\
		{\bf f}(T^*;t_{\circ}^k)
		&=&
		{\bf G}(b^k,x^k)
		{\bf B}
		\left(\ 
		{\bf f}(T^*;1;t_{\bullet}^{k})
		\otimes
		{\bf f}(T^*;2;t_{\bullet}^{k})
		+
		{\bf f}(T^*;2;t_{\bullet}^{k})
		\otimes
		{\bf f}(T^*;1;t_{\bullet}^{k})
		\ \right)
		,
		\quad \mbox{ for all } k,
		\nonumber
		\\
		{\bf f}(T^*;t_{\circ}^{\widehat m})
		&=&
		{\bf D}^{(1)}(b^{\widehat m})
		,
		\quad \mbox{ for all } \widehat m,
		\nonumber\\
	\end{eqnarray}
	where, for any $b^k>0$, $x^k>0$, the entry $[{\bf G}(b^k,x^k)]_{ij}$ is the probability that an internal branch $k$ that started at time~$t_{\circ}^k=t-x^k-b^k$ in phase $i$ and was observed at time $t_{\bullet}^k=t-x^k$ in phase $j$, is such that all potential speciation events that occurred on this branch within the open time interval $(t_{\circ}^k,t_{\bullet}^k)$, have resulted in branches that became extinct by time $t$.

 \begin{Remark}
We note that the likelihood expression in~\eqref{eq:speciesknowlittle2} does not require information about observed phases. However, if the phase $\varphi$ at the tip of some external branch $\widehat{m}$ in the reconstructed tree is known, then to modify~\eqref{eq:speciesknowlittle2}, replace the column vector ${\bf D}^{(1)}(b^{\widehat m})$ with the column vector given by
\begin{eqnarray}
\mathbf{D}_{\varphi}^{(1)}(t)&=&[\widetilde D_{i\varphi}^{(1)}(t)], 
\end{eqnarray}
where $\widetilde D_{i\varphi}^{(1)}(t)$ defined in~\eqref{defD1t} is the probability that a lineage on the tree that started at time $0$ in phase $i$ consists of one individual only at time $t$, who is in phase $\varphi$. In some applications, this may be useful, since the phase may represent a phenotype that can be measured in the current species.
  \end{Remark}

 \begin{Remark}
We note that although the expression in~\eqref{eq:speciesknowlittle2} and related Lemmas~\ref{lemDt}-\ref{lemgwzt} are stated in terms of quantities corresponding to time, they can also be applied to reconstructed non-clock trees in which instead of time, some other continuous quantity is used to represent the branch lengths. That is, the analysis here can be applied to the non-clock trees as well. Simply, write the expressions for distances between various points as sums of suitable branch lengths. As example, given branch lengths $b^k$ and $b^{\widehat m}$ (represented via any continuous quantity) of the tree in Figure~\ref{fig:Problem1}, let $t_{\circ}^1=0$ and then use $t_{\bullet}^1=b^1$, $t_{\bullet}^2=b^1+b^2$, $t_{\bullet}^3=b^1+b^3$, $t_{\bullet}^4=b^1+b^2+b^4$, and for each external branch $\widehat{m}$ which in non-clock trees may terminate at a different point than other branches, replace ``current'' time $t$  with $t_{\circ}^{\widehat m}+b^{\hat m}$.
 \end{Remark}

Below, we outline our method for computing ${\bf G}(b^k,x^k)$ required in~\eqref{eq:speciesknowlittle2}.  
	\begin{Lemma}\label{lemgwzt}
		For any $z,x>0$, matrix ${\bf G}(z,x)$ is a solution of the differential equation 
		\begin{equation}\label{eq:DE_Gt}
			\frac{d {\bf G}(z,x)}{dz}
			= {\bf D}_0 {\bf G}(z,x)		
			+
			{\bf B} 
			\left(
			{\bf E}(z+x)\otimes {\bf G}(z,x)
			\right)
			+
			{\bf B} 
			\left(
			{\bf G}(z,x) \otimes {\bf E}(z+x)
			\right)
		\end{equation}
		with the initial condition ${\bf G}(0,x)={\bf I}$.

		Furthermore, 	
		\begin{eqnarray}\label{eq:int_Gt}
			{\bf G}(z,x) = 
			e^{\mathbf{D}_{0}(z)}
			+
			\int_{u=0}^t
			e^{\mathbf{D}_{0}(z-u)}
			\mathbf{B}
			\left(
			\mathbf{E}(u+x) \otimes {\bf G}(u,x)
			+
			{\bf G}(u,x) \otimes \mathbf{E}(u+x)
			\right)
			du.
		\end{eqnarray} 
	\end{Lemma}

	\noindent{\bf Proof:} The result follows by a slight modification of the proof of Lemma~\ref{lemDt} so that in place of the quantities ${\bf E}(t)$ and ${\bf D}^{(1)}(t)$ we apply ${\bf E}(z+x)$ and ${\bf G}(z,x)$, respectively. We present the details below, for completeness.
	
	Consider the evolution of the process in the time interval $[0,z]$, where $0$ and $z$ corresponds to the start and the end of the branch of length $z$, respectively, and the following alternatives.
	\begin{enumerate}
		\item No speciation or extinction events occur in the time interval $(0,z)$. The probability of this alternative is recorded in the matrix
		\begin{eqnarray}\label{eq:Gzxalt1}
			e^{\mathbf{D}_{0}z}.
		\end{eqnarray}
		\item The first speciation event occurs at time $(z-u)$ for some $u\in(0,z)$, according to the probability density matrix
		$$e^{\mathbf{D}_{0}(z-u)}
		\mathbf{B}.$$ 
		Next, one of the two new branches becomes extinct by time $t$ (or equivalently, by the remaining time $u+x$); while the other ends with one individual by time $z$ (or equivalently, by the remaining time $u$), according to the probability vector
		$$\mathbf{E}(u+x) \otimes \mathbf{D}^{(1)}(u)
		+
		\mathbf{D}^{(1)}(u) \otimes \mathbf{E}(u+x).$$ 
		Therefore, by conditioning on $u\in(0,z)$, the probability vector of the second alternative is,
		\begin{eqnarray}\label{eq:Gzxalt2}
			\int_{u=0}^z
			e^{\mathbf{D}_{0}(z-u)}
			\mathbf{B}
			\left(
			\mathbf{E}(u+x) \otimes \mathbf{D}^{(1)}(u)
			+
			\mathbf{D}^{(1)}(u) \otimes \mathbf{E}(u+x)
			\right)
			du.
		\end{eqnarray}
	\end{enumerate}

	By adding the probabilities~\eqref{eq:Gzxalt1}-\eqref{eq:Gzxalt2} of the two alternatives, we obtain~\eqref{eq:int_Gt}. Next, by taking derivatives with respect to $z$ in~\eqref{eq:int_Gt}, we have,
	\begin{eqnarray*}
		\frac{\partial {\bf G}(z,x)}{\partial z}
		&=&
		\mathbf{D}_{0}
		\left(
		e^{\mathbf{D}_{0}(z)}
		+
		\int_{u=0}^t
		e^{\mathbf{D}_{0}(z-u)}
		\mathbf{B}
		\left(
		\mathbf{E}(u+x) \otimes {\bf G}(u,x)
		+
		{\bf G}(u,x) \otimes \mathbf{E}(u+x)
		\right)
		du
		\right)
		\\
		&&
		+
		{\bf B} 
		\left(
		{\bf E}(z+x)\otimes {\bf G}(z,x)
		\right)
		+
		{\bf B} 
		\left(
		{\bf G}(z,x) \otimes {\bf E}(z+x)
		\right),
	\end{eqnarray*} 
	and so~\eqref{eq:DE_Gt} follows. \hfill\rule{9pt}{9pt}

	\begin{Example}\label{ex:Gorecki}
		In order to illustrate our methodology, we compute the likelihood of the species trees in Figure~\ref{fig:Example2}, based on G{\'o}recki~\cite[Figure 1]{gorecki2014drml}, also discussed later in Section~\ref{sec:RecEx} (Figure~\ref{E1E17Tree}). 
		\begin{figure}[h!]
			\begin{center}
				\begin{tikzpicture}[>=stealth,redarr/.style={->}]	
				\draw (3.5,10.4) node[anchor=north, below=-0.17cm] {{\color{black} $T^*$ }};
				
				\draw [dashed] (-1.0,10) -- (6,10);
				\draw (-1.5,10) node[anchor=north, below=-0.3cm] {\scriptsize{\color{black} $t_{\circ}^{1}=0$ }};

				\draw [dashed] (-1.0,9) -- (6,9);
				\draw (-2.1,9) node[anchor=north, below=-0.3cm] {\scriptsize{\color{black} $t_{\bullet}^{1}=t_{\circ}^{2}=t_{\circ}^{\widehat 4}=2$ }};

				\draw [dashed] (-1.0,8) -- (6,8);
				\draw (-2.1,8) node[anchor=north, below=-0.3cm] {\scriptsize{\color{black} $t_{\bullet}^{2}=t_{\circ}^{{3}}=t_{\circ}^{\widehat 3}=3$ }};

				\draw [dashed] (-1.0,7) -- (6,7);
				\draw (-2.2,7) node[anchor=north, below=-0.3cm] {\scriptsize{\color{black} $t_{\bullet}^{3}=t_{\circ}^{\widehat{1}}=t_{\circ}^{\widehat{2}}=11$ }};

				\draw [dashed] (-1.0,5.5) -- (6,5.5);
				\draw (-1.5,5.5) node[anchor=north, below=-0.3cm] {\scriptsize{\color{black} $t=20$ }};

				\draw [green,very thick] (3.5,10) -- (3.5,9);
				\draw [green,very thick] (1.5,9) -- (5.5,9);
				\draw (3.3,9.5) node[anchor=north, below=-0.3cm] {\scriptsize{\color{black} $1$ }};
				
				\draw [green,very thick] (5.5,9) -- (5.5,5.5);
				\draw (5.3,8.5) node[anchor=north, below=-0.3cm] {\scriptsize{\color{black} $\widehat 4$ }};
				
				\draw [green,very thick] (1.5,9) -- (1.5,8);
				\draw (1.3,8.5) node[anchor=north, below=-0.3cm] {\scriptsize{\color{black} $2$ }};
				
				\draw [green,very thick] (0.5,8) -- (2.5,8);
				\draw (0.3,7.5) node[anchor=north, below=-0.3cm] {\scriptsize{\color{black} $3$ }};
				
				\draw [green,very thick] (2.5,8) -- (2.5,5.5);
				\draw (2.3,7.5) node[anchor=north, below=-0.3cm] {\scriptsize{\color{black} $\widehat 3$ }};
				
				\draw [green,very thick] (0.5,8) -- (0.5,7);

				\draw [green,very thick] (-0.5,7) -- (1.5,7);
				\draw (-0.7,6.5) node[anchor=north, below=-0.3cm] {\scriptsize{\color{black} $\widehat 1$ }};

				\draw [green,very thick] (-0.5,7) -- (-0.5,5.5);

				\draw [green,very thick] (1.5,7) -- (1.5,5.5);
				\draw (1.3,6.5) node[anchor=north, below=-0.3cm] {\scriptsize{\color{black} $\widehat 2$ }};
				
				\draw (6.5,9) node[anchor=north, below=-0.3cm] {\scriptsize{\color{black} $\widetilde\varphi(1)$ }};
				
				\draw (6.5,8) node[anchor=north, below=-0.3cm] {\scriptsize{\color{black} $\widetilde\varphi(2)$ }};
				
				\draw (6.5,7) node[anchor=north, below=-0.3cm] {\scriptsize{\color{black} $\widetilde\varphi(3)$ }};
				
				\draw [decorate,
				decoration = {brace}] (7,7) --  (7,5.5);
				\draw (7.5,6.5) node[anchor=north, below=-0.17cm] {{\color{black} $x^3$ }};
				
				\draw [decorate,
				decoration = {brace}] (8,8) --  (8,5.5);
				\draw (8.5,7) node[anchor=north, below=-0.17cm] {{\color{black} $x^2$ }};
				
				\draw [decorate,
				decoration = {brace}] (9,9) --  (9,5.5);
				\draw (9.5,7.5) node[anchor=north, below=-0.17cm] {{\color{black} $x^1$ }};

				\end{tikzpicture}
				\caption{Species tree based on~\cite[Figure 1]{gorecki2014drml}, showing topology and times (not in true scale). Internal branches are labelled $k=1,2,3$. External branches are labelled $\widehat m=\widehat 1,\ldots,\widehat 4$. Phase $\widetilde \varphi(k)$ is observed just before speciation at time $t_{\bullet}^k$ on internal branch $k=1,2,3$ with $x^k=t-t_{\bullet}^k$. 
				}
				\label{fig:Example2}
			\end{center}
		\end{figure}
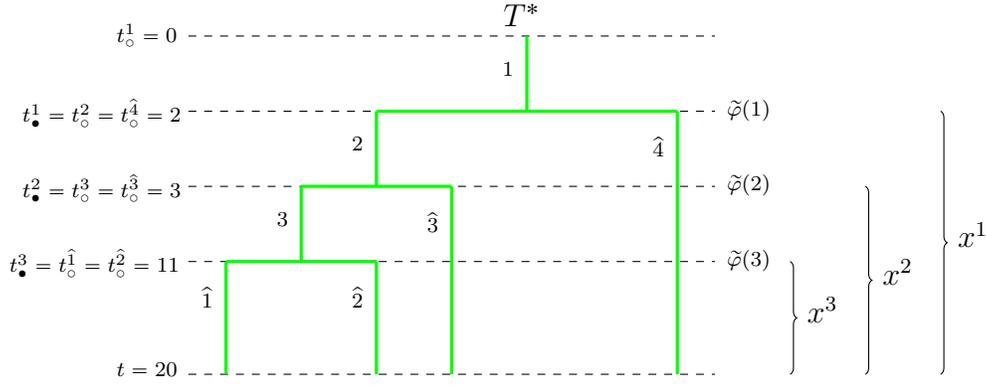

		Without loss of generality, we may assume that $t_{\circ}^1=0$, which then gives $t_{\bullet}^1=t_{\circ}^2=t_{\circ}^{\widehat 4}=2$, and $t_{\bullet}^2=t_{\circ}^3=t_{\circ}^{\widehat 3}=2+1=3$, and $t_{\bullet}^3=t_{\circ}^{\widehat 1}=t_{\circ}^{\widehat 2}=2+1+8=11$, and $t=2+1+8+9=20$. Then we have, $b^1=2$, $b^2=1$, $b^3=8$, $x^3=\widehat b^1=\widehat b^2=9$, $x^2=\widehat b^3=17$, $x^1=\widehat b^4=18$, and so by~\eqref{eq:speciesknowlittle2}  applied to the BiSSE model, since here
		$${\bf B}
		\left(\ 
		{\bf f}(T^*;1;t_{\bullet}^{k})
		\otimes
		{\bf f}(T^*;2;t_{\bullet}^{k})
		+
		{\bf f}(T^*;2;t_{\bullet}^{k})
		\otimes
		{\bf f}(T^*;1;t_{\bullet}^{k})
		\ \right)
		,
		\quad \mbox{ for all } k$$
		simplifies to
		$${\bf D}_1
		\left(\ 
		2\times{\bf f}(T^*;1;t_{\bullet}^{k})
		\odot
		{\bf f}(T^*;2;t_{\bullet}^{k})
		\ \right)
		,
		\quad \mbox{ for all } k,$$
		we have,
		\begin{eqnarray*}
			f(T^*)
			&=&
			2^3\times\balpha {\bf G}(b^1,x^1)
			{\bf D}_1\times
			\Big\{
			\\
			&&
			{\bf G}(b^2,x^2)
			{\bf D}_1\times
			\left(
			\quad 
			\left\{
			{\bf G}(b^3,x^3){\bf D}_1\times
			\left[
			{\bf D}^{(1)}(b^{\widehat 1})
			\odot
			{\bf D}^{(1)}(b^{\widehat 2})
			\right]
			\right\}
			\odot
			{\bf D}^{(1)}(b^{\widehat 3})
			\quad 
			\right)
			\\
			&&
			\quad \odot \quad 
			{\bf D}^{(1)}(b^{\widehat 4})
			\quad \Big\}
			.
		\end{eqnarray*}

		We apply the BiSSE model discussed in Example~\ref{ex:BiSSE} with $\balpha=[0.5\ 0.5]$ and: (a) Trait dependent speciation rates with: $\mu_0=0.1$, $\mu_1=0.1$, $q_{01}=0.9$, $q_{10}=0.001$, $\lambda_0=1$, $\lambda_1=0.099$; (b) Trait dependent speciation rates with: $\mu_0=0.1$, $\mu_1=0.1$, $q_{01}=0.9$, $q_{10}=0.001$, $\lambda_0=0.3$, $\lambda_1=0.099$; (c) Trait dependent extinction rates with: $\mu_0=0.999$, $\mu_1=0.099$, $q_{01}=0.2$, $q_{10}=0.001$, $\lambda_0=1$, $\lambda_1=1$; (d) Trait dependent extinction rates with: $\mu_0=0.2$, $\mu_1=0.099$, $q_{01}=0.2$, $q_{10}=0.001$, $\lambda_0=\lambda_1=1$. The results are summarised in Table~\ref{likely_species} below. We note that the most likely scenario is case (a), which by Remark~\ref{rem_balance} below corresponds to the most unbalanced tree.
		
		\begin{table}[H]
			\centering
			\begin{tabular}{l l l l l l l l}
				\hline
				Case &$\mu_0$&$\mu_1$&$q_{01}$&$q_{10}$&$\lambda_0$&$\lambda_1$&$\log{f(T^*)}$\\
				\hline
				(a) Trait-dependent speciation rates&$0.1$&$0.1$&$0.9$&$0.001$&$1$&$0.099$&$-12.2485$\\
				(b) Trait-dependent speciation rates&$0.1$&$0.1$&$0.9$&$0.001$&$0.3$&$0.099$&$-12.3811$\\
				(c) Trait-dependent extinction rates&$0.999$&$0.099$&$0.2$&$0.001$&$1$&$1$&$-51.8785$\\
				(d) Trait-dependent extinction rates&$0.2$&$0.099$&$0.2$&$0.001$&$1$&$1$&$-55.7678$\\
				\hline
			\end{tabular}
			\caption{Log likelihood of observing the species tree considered in Example~\ref{ex:Gorecki} for various parameters.}
			\label{likely_species}
		\end{table}
		
	\end{Example}

	\begin{Remark}\label{rem_balance}
		In the absence of a model, the probability $D_{n_L|n}(t)$ that a tree with $n$ tips has $n_L$ tips on the left subtree may be estimated from data by using statistical methods and the formula~\cite{1996A},
		\begin{eqnarray}
			D_{n_L|n}(t)\approx q_{n}(n_L,\beta) = \frac{1}{\alpha_{n}(\beta)}\frac{\Gamma(\beta+n_L+1)\Gamma(\beta+n-n_L+1)}{\Gamma(n_L+1)\Gamma(n-n_L+1)},\quad 1\leq n_L \leq n-1 ,
		\end{eqnarray}
		where $\alpha_{n}(\beta)$ is a normalising constant. We compare this estimate with $D_{n_L|n}(t)$ for different choice of parameters in Figure~\ref{fig:Dintapprox}.
		\begin{figure}[H]
			\centering
			\includegraphics[scale=0.5]{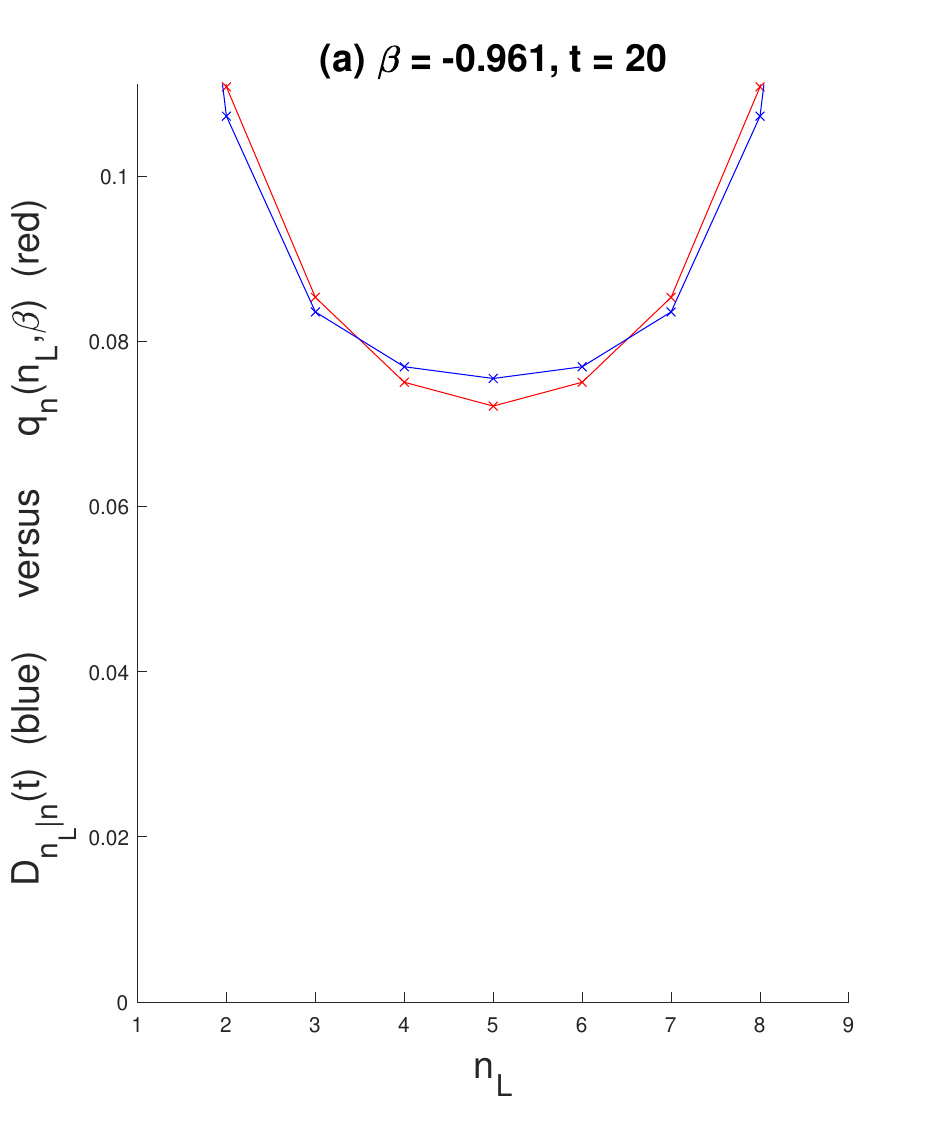}
			\quad
			\includegraphics[scale=0.5]{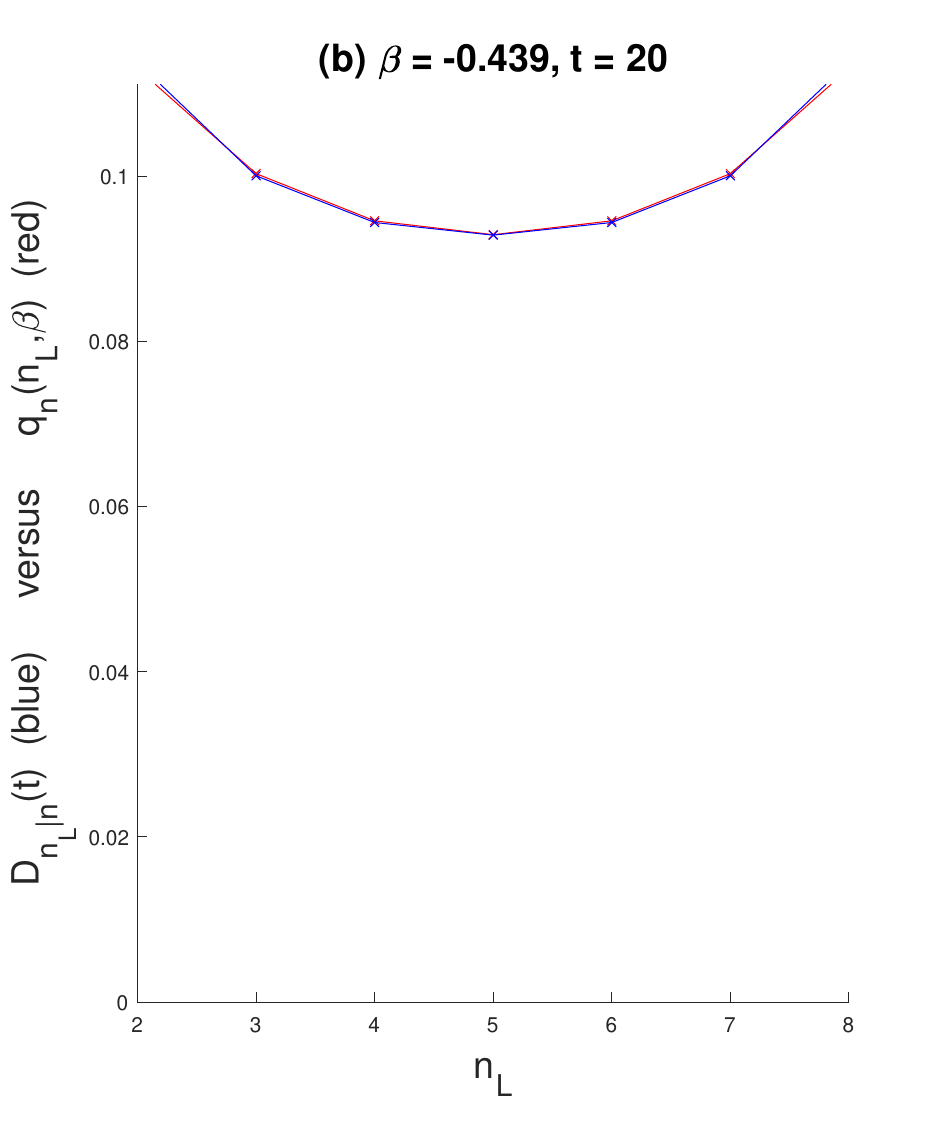}
			\\
			\includegraphics[scale=0.5]{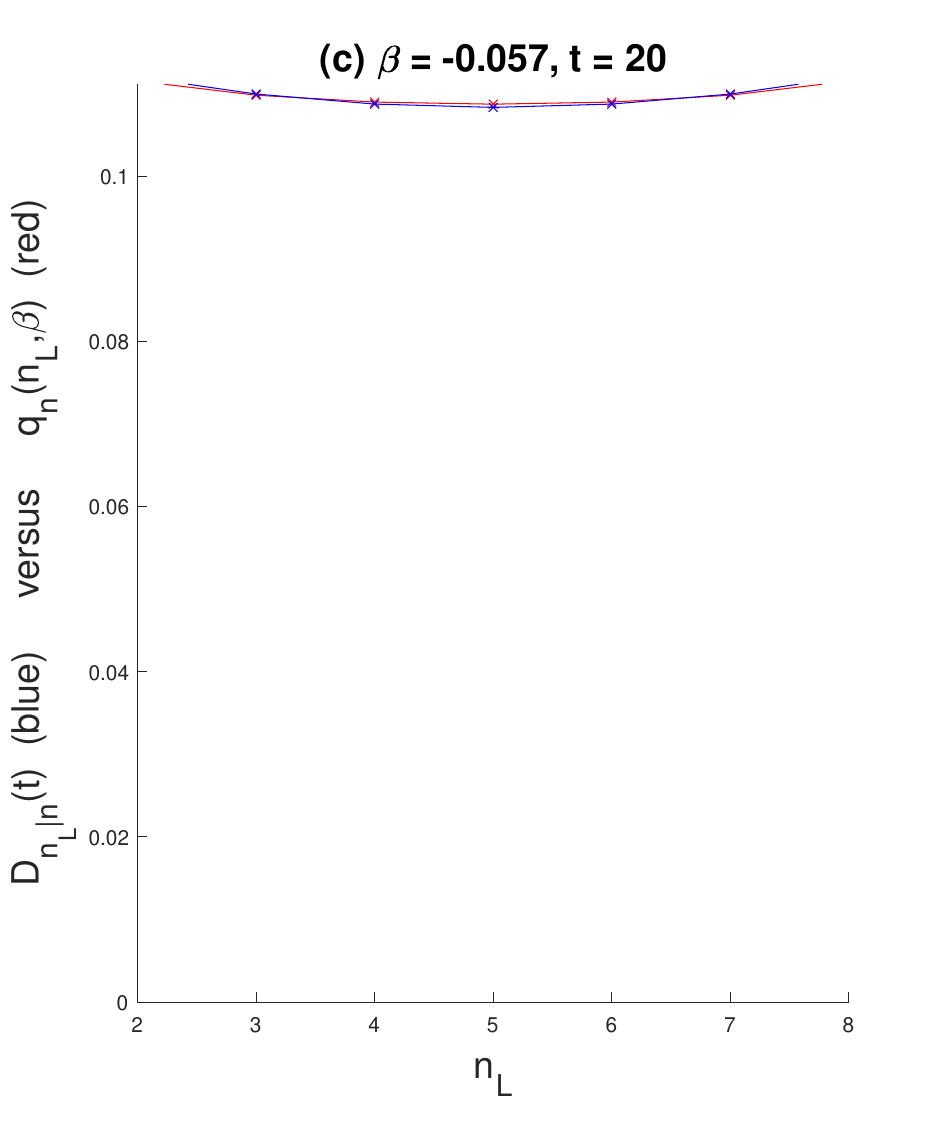}
			\quad
			\includegraphics[scale=0.5]{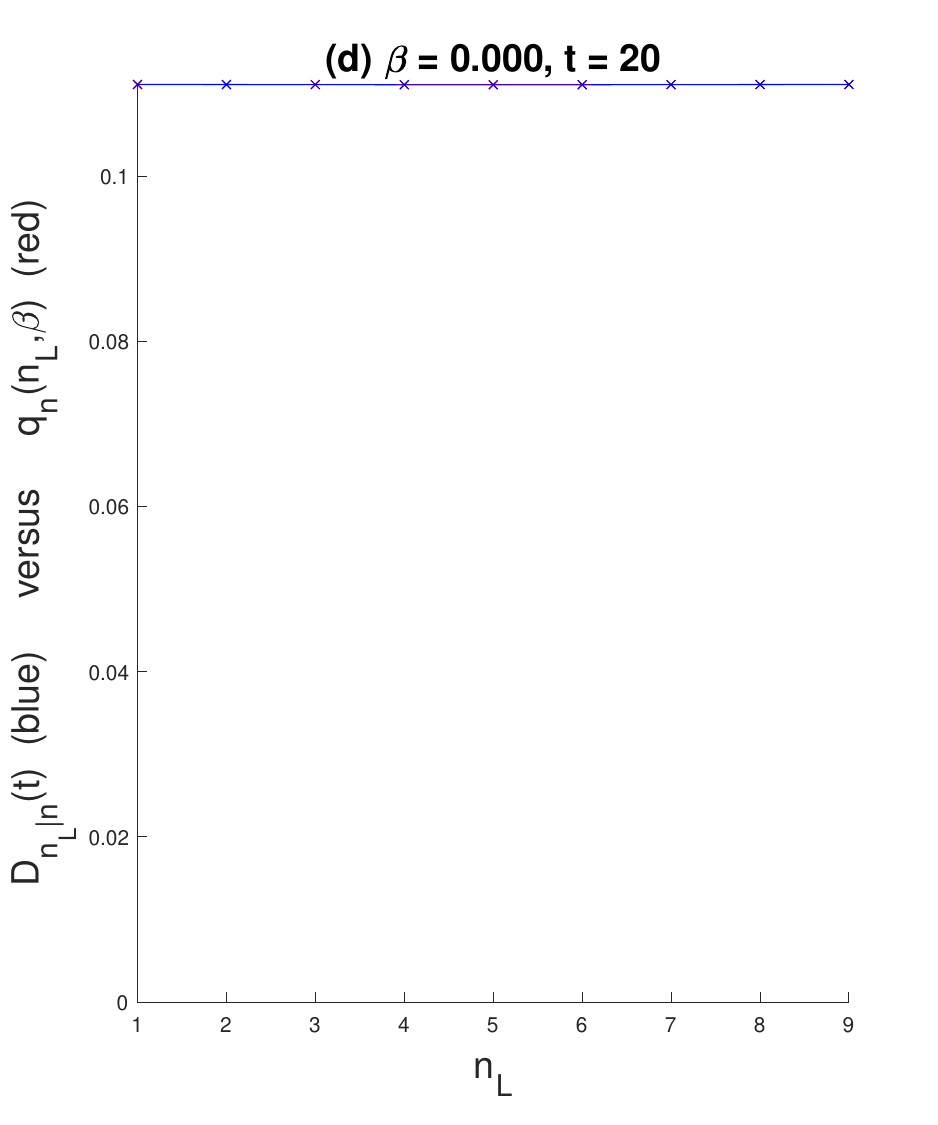}
			\caption{$D_{n_L|n}(t)$ (blue) derived for the BiSSE model discussed in Example~\ref{ex:BiSSE} versus $q_{n}(n_L,\beta)$ (red) obtained Kullback-Leibler divergence metric~\cite{kullback1951information} in order to get the best fit. (a) Trait dependent speciation rates with: $\mu_0=0.1$, $\mu_1=0.1$, $q_{01}=0.9$, $q_{10}=0.001$, $\lambda_0=1$, $\lambda_1=0.099$; the extinction probabilities for both states: $E_{0}=0.8283$, $E_{1}=0.9673$; (b) Trait dependent speciation rates with: $\mu_0=0.1$, $\mu_1=0.1$, $q_{01}=0.9$, $q_{10}=0.001$, $\lambda_0=0.3$, $\lambda_1=0.099$ with $E_{0}=E_{1}=1$; (c) Trait dependent extinction rates with: $\mu_0=0.999$, $\mu_1=0.099$, $q_{01}=0.2$, $q_{10}=0.001$, $\lambda_0=1$, $\lambda_1=1$ with $E_{0}=0.6636$, $E_{1}=0.0996$; (d) Trait dependent extinction rates with: $\mu_0=0.2$, $\mu_1=0.099$, $q_{01}=0.2$, $q_{10}=0.001$, $\lambda_0=\lambda_1=1$ with $E_{0}=0.1802$, $E_{1}=0.0991$.}
			\label{fig:Dintapprox}
		\end{figure}
	\end{Remark}
	
	We observe from Figure~\ref{fig:Dintapprox} that our model tends to generate unbalanced trees (with negative $\beta$ values), whenever we use trait-dependent speciation rates (Figure~\ref{fig:Dintapprox}a,b), or trait-dependent extinction rates (Figure~\ref{fig:Dintapprox}c,d). This implies that, under such conditions, the model produces trees that are neither uniform on ranked trees (URT) studied in~\cite{lambert2013birth}, nor the Yule trees introduced in~\cite{yule1925ii}. Instead, our model behaves similar to the model class 4, in which speciation rates depend on heritable traits with constant extinction rates; or the model class 5, in which extinction rates depend on heritable traits with constant speciation rates, as described in~\cite{lambert2013birth}.
	
	This behaviour may be explained as follows. In Figure~\ref{fig:Dintapprox}a, the process spends more time in $1$ than in $0$ (due to $q_{10}<q_{01}$) and when in $1$, the chance of speciation is lower ($\lambda_1<\lambda_0$). So the process visits $0$ very rarely, but when it does, the speciation is very likely to occur ($\lambda_0>>\mu_0$). On the other hand, when in $1$, the extinction is more likely to be observed ($\lambda_1<\mu_1$). Therefore, this results in a very unbalanced tree. In comparison, in Figure~\ref{fig:Dintapprox}b, the value of $\lambda_0$ is lower than in~(a), and so the effect of $\lambda_0>>\mu_0$ is reduced, which results in a less unbalanced tree than in (a).
	
	In Figure~\ref{fig:Dintapprox}c, the process spends more time in $1$ ($q_{10}<q_{01}$) and when it visits $1$, speciation is very likely to occur ($\lambda_1>>\mu_1$). Again, the visits to $0$ are rare, but when in $0$, the speciation and extinction rates are similar ($\lambda_0\approx \mu_0$). So the tree is unbalanced due to $\lambda_1>>\mu_1$, but less than in (a)-(b) due to $\lambda_0\approx \mu_0$. In comparison, in Figure~\ref{fig:Dintapprox}d, $\lambda_0=\lambda_1>> \mu_0$, and so speciation events are more likely to occur on either subtree, which results in $\beta\approx 0$. 
	
	Moreover, we see from Figure~\ref{fig:Dintapprox} that the variance in $\beta$ values is more easily affected by the speciation process rather than by the extinction process. A similar observation was made in~\cite{hagen2015age}, where the authors applied an age-dependent speciation model, and in~\cite{2020SOH}, which studied a model with phase-type distributed times to speciation.

	We note that the empirical data tends to produce $\beta$  values close to $-1$, as stated in~\cite{1996A}. Therefore, our model is a good candidate for fitting to the empirical data, since it is able to produce $\beta$ values similar to the empirical value $\beta=-1$, by using the intuitions described in Figure~\ref{fig:Dintapprox}a-b above.
	
	Finally, we note that since the MBT model can we represented as a quasi-birth-and-death process (QBD)~\cite{Kontoleon}, it follows that various quantities of interest can be computed using ideas similar to the theory of QBDs~\cite{2011H}. In fact, for models with a small dimension of the state, it is possible to compute the quantities of interest by directly representing the MBT as a QBD and then applying the expressions from the theory of QBDs. We illustrate this idea below.

\begin{figure}[H]
		\centering
		\includegraphics[scale=0.4]{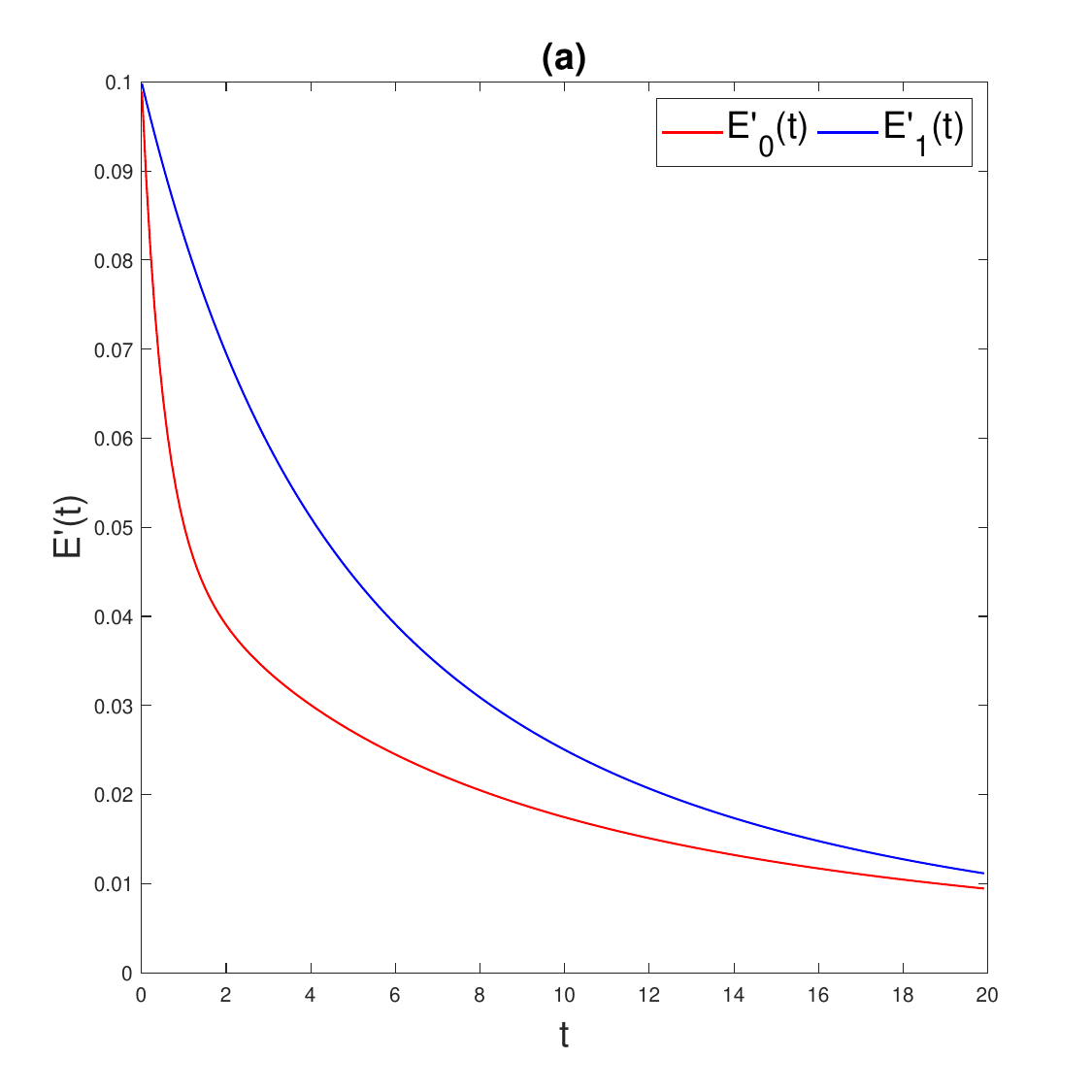}
		\quad
		\includegraphics[scale=0.4]{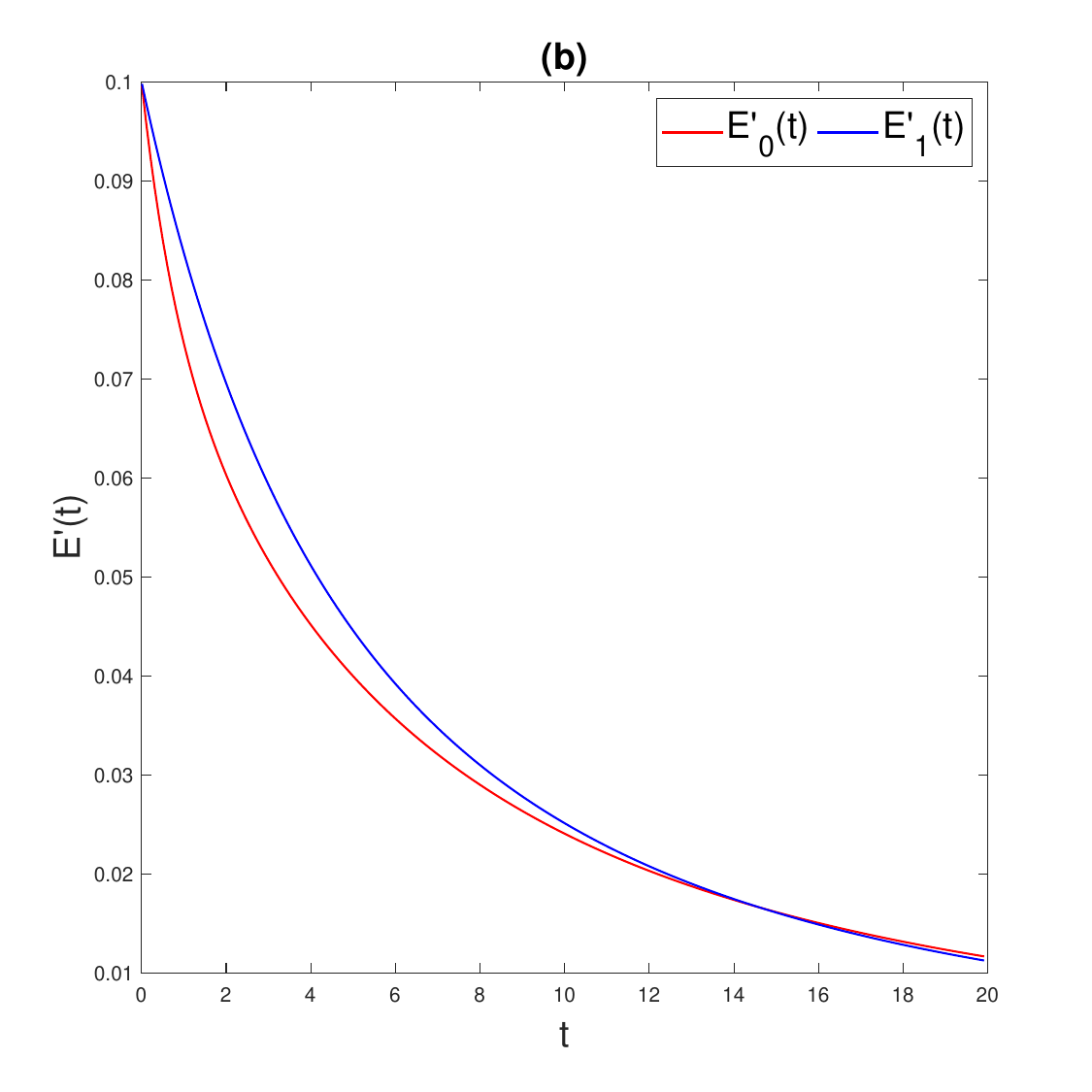}
		\\
		\includegraphics[scale=0.4]{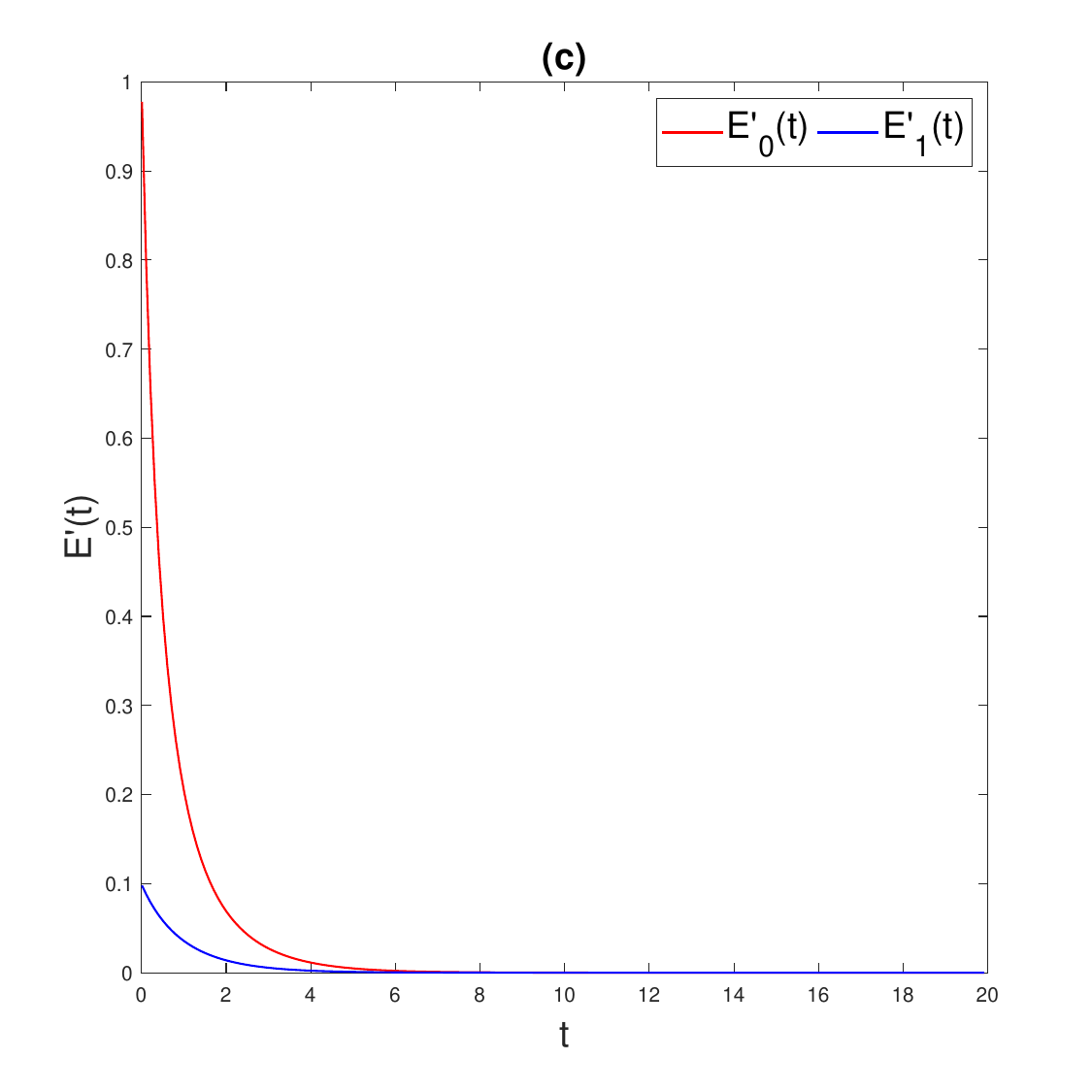}
		\quad
		\includegraphics[scale=0.4]{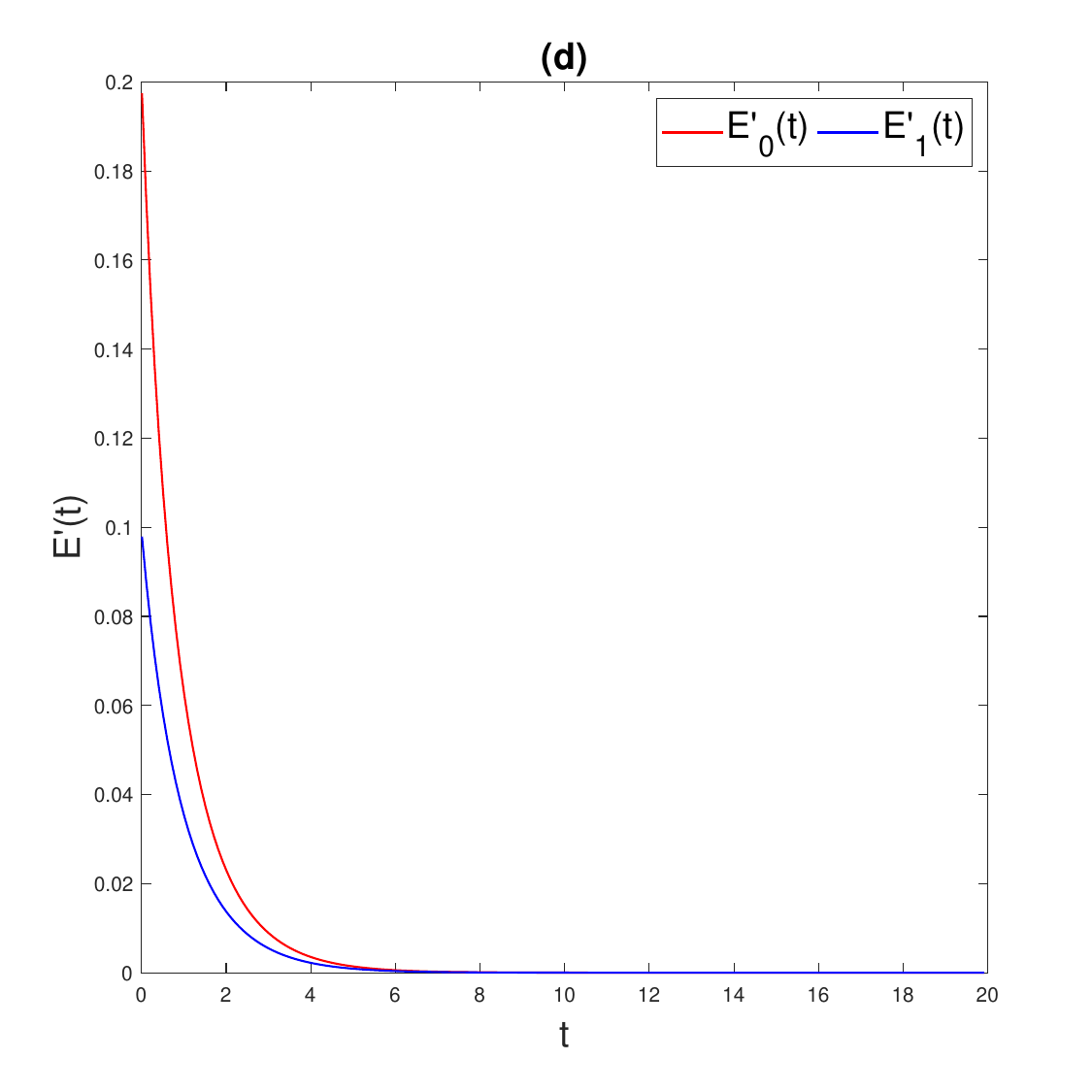}
		\caption{${\bf E}'(t)=[E^{'}_i(t)]$ is the vector recording probability densities $E^{'}_i(t)$ that a tree becomes extinct at time $t$ given it started at time $0$ in phase $i=0,1$.We assume the parameter values as in Figure~\ref{fig:Dintapprox}(a)-(d).}
		\label{fig:extinctdens}
	\end{figure}

	\begin{Example}
		\label{exBisseQBD}
		We may represent the MuSSE model in~\cite{2012F} ($\equiv$ the BiSSE model in~\cite{2007MMO} when $\ell=2$) as a level-dependent quasi-birth-and-death process (LD-QBD) $\{(Y(t),\varphi(t)):t\geq 0\}$ with state space
		\begin{eqnarray}
			\mathcal{S}&=&\{(n,\varphi): n=0,1,2,\ldots; \varphi=(k_1,\ldots,k_{\ell-1});\sum_{m=1}^{\ell-1} k_m \leq n; k_m=0,1,\ldots,n\},
		\end{eqnarray}
		where the level variable $Y(t)\in\{0,1,2,\ldots\}$ records the number of species, the $(\ell-1)$-dimensional phase variable $\varphi(t)=(k_1(t),\ldots,k_{\ell-1}(t))$ records the number of species $k_i(t)$ that are in phase $i$, $i=1,\ldots,\ell-1$, with $k_\ell=n-\sum_{m=1}^{\ell-1} k_m$, and generator ${\bf Q}=[q_{(n,k_1,\ldots,k_\ell)(n^{'},k_1^{'},\ldots,k_\ell^{'})}]_{(n,k_1,\ldots,k_\ell),(n^{'},k_1^{'},\ldots,k_\ell^{'})\in\mathcal{S}}$, is made of block matrices ${\bf Q}^{[n,n^{'}]}$, given by, 
		\begin{equation}\label{Qnnbasic}
			{\bf Q} = 
			[{\bf Q}^{[n,n^{'}]}]_{n,n^{'}=0,1,2\ldots}
			=
			\begin{bmatrix}
				{\bf 0} & {\bf 0} & {\bf 0} & {\bf 0} & \cdots\\
				{\bf Q}^{[1,0]} & {\bf Q}^{[1,1]} & {\bf Q}^{[1,2]} & {\bf 0} & \cdots\\
				{\bf 0} & {\bf Q}^{[2,1]} & {\bf Q}^{[2,2]} & {\bf Q}^{[2,3]} &  \cdots\\
				{\bf 0} & {\bf 0} & {\bf Q}^{[3,2]} & {\bf Q}^{[3,3]} & \cdots \\
				\vdots & \vdots & \vdots & \vdots & \ddots
			\end{bmatrix},
		\end{equation}
		whose nonzero off-diagonals are
		\begin{eqnarray}
			q_{(n,k_1,\ldots,k_\ell)(n^{'},k_1^{'},\ldots,k_\ell^{'})}&=&
			\left\{
			\begin{array}{ll}
				\lambda_i k_i& \quad n^{'}=n+1;k_i^{'}=k_i+1; k_m^{'}=k_m \mbox{ for }m\not= i;\\
				\mu_i k_i& \quad n^{'}=n-1;k_i^{'}=k_i-1; k_m^{'}=k_m \mbox{ for }m\not= i;\\
				q_{ij} k_i& \quad n^{'}=n;k_i^{'}=k_i-1;k_j^{'}=k_j+1; k_m^{'}=k_m \mbox{ for }m\not= i,j;\\
			\end{array}
			\right.
		\end{eqnarray}
		and the on-diagonals are given by $q_{(n,k_1,\ldots,k_\ell)(n,k_1,\ldots,k_\ell)}=-\sum_{(n,k_1,\ldots,k_\ell)\not= (n^{'},k_1^{'},\ldots,k_\ell^{'})}q_{(n,k_1,\ldots,k_\ell)(n^{'},k_1^{'},\ldots,k_\ell^{'})}$.

		The size of the block matrix ${\bf Q}^{[n,n^{'}]}$ is $|\mathcal{S}_{n}|\times |\mathcal{S}_{n^{'}}|$ with $|\mathcal{S}_{n}|={n + \ell-1 \choose \ell-1}$, where $\mathcal{S}_{n}=\{(k_1,\ldots,k_\ell): \sum_{m=1}^\ell k_m =n; k_m=0,1,\ldots,n\}$ is the set of possible phases at level $n$ ($\equiv$  the total number of nonnegative solutions $(k_1,\ldots,k_\ell)$ to the equation $\sum_{m=1}^\ell k_m =n$).

		In particular when $\ell=2$ (BiSSE model), we obtain $|\mathcal{S}_{n}|=n+1$, which is a moderate value for computations that involve blocks ${\bf Q}^{[n,n]}$, and so the quantities of interest in the analysis may be computed with ease. In this case, following the notation in~\cite{2007MMO} in which phases $1$ and $0$ are used (rather than $1$ and $2$ as in~\cite{2012F} when $\ell=2$), in the context of the BiSSE model we interpret $k_1$ as the total number of $1$s and $(n-k_1)$ as the total number of $0$s on the tips of the tree. 
		
		Denote ${\bf E}^{'}(t)=\frac{d {\bf E}(t)}{dt}$ and define the Laplace-Stieltjes transform
		\begin{eqnarray}
			\widehat {\bf E}(s)&=&\int_{t=0}^{\infty} e^{-st} {\bf E}^{'}(t) dt,
		\end{eqnarray}
		and note that the quantities $\widehat {\bf E}(s)$, ${\bf E}=\widehat {\bf E}(0)$ and ${\bf E}^{'}(t)$ in the MBT model are equivalent to the quantities $\widehat {\bf G}(s){\bf 1}=\int_{t=0}^{\infty} e^{-st} {\bf g}(t) dt{\bf 1}$, ${\bf G}{\bf 1}=\widehat {\bf G}(0){\bf 1}$ and ${\bf g}(t){\bf 1}$ in the LD-QBD model respectively, where $[{\bf g}(t)]_{ij}$ is the probability density that the LD-QBD first visits level $0$ at time $t$ and does so in phase $j$, given start from level $1$ in phase $i$.  $\widehat {\bf G}(s)$ can be computed using an iterative scheme in Phung-Duc et al. in~\cite{phung2010simple}, summarised in Algorithm~\ref{GMatrix_algorithm}. Next, by inverting $\widehat {\bf G}(s)$ using numerical inversion techniques by Abate and Whitt~\cite{abate1995numerical} we may obtain ${\bf g}(t)$, and then ${\bf E}^{'}(t)={\bf g}(t){\bf 1}$. We illustrate this in Figure~\ref{fig:extinctdens}. \hfill$\Box$
	\end{Example}

	\begin{algorithm}
		\caption{Approximate 
			$\{\widehat{\bf G}^{(n)}(s)\}_{n=1,\hdots,N}$ 
			\hfill (adapted from Phung-Duc et al.~\cite{phung2010simple})}
		\label{GMatrix_algorithm}
		\begin{algorithmic}[1] %
			\Procedure{GMatrices}{$s,N,{\bf Q},\epsilon$} 
			\State 
			Compute ${\bf G}^{(N)}_{k}$ for $k=1$ using
			\begin{align} 
				{\bf G}^{(N)}_{k} &= G_N \circ G_{N+1} \circ \hdots \circ G_{N+k-1}({\bf 0}) , \label{eq_Algorithm1G} \\
				G_n({\bf X}) &= (-({\bf Q}^{[n,n]}-s{\bf I})-{\bf Q}^{[n,n+1]}{\bf X})^{-1}{\bf Q}^{[n,n-1]}  
				\label{eq_Algorithm2G}  
			\end{align}
			where we denote $f\circ h(\cdot)=f(h(\cdot))$.
			\While{$||{\bf G}^{(N)}_k - {\bf G}^{(N)}_{k-1}||_\infty > \epsilon$} 
			\State Let $k=k + 1$.
			\State Compute ${\bf G}^{(N)}_{k}$ using \eqref{eq_Algorithm1G} and \eqref{eq_Algorithm2G}.
			\EndWhile
			\State Let $\hat {\bf G}^{(N)} = {\bf G}^{(N)}_{k}$.
			\For{$n = N-1, \hdots, 1$}
			\State Compute $\hat {\bf G}^{(n)} = G_{n}(\hat {\bf G}^{(n+1)})$ using \eqref{eq_Algorithm1G} and \eqref{eq_Algorithm2G}. 
			\EndFor
			\State Let $\widehat {\bf G}(s)\approx \hat {\bf G}^{(1)}$.
			\EndProcedure
		\end{algorithmic}
	\end{algorithm}

	\section{Gene trees}\label{Sec:GenTree}
	
	For the gene family tree of a {\em single species} we apply a class of models in which the evolution of each branch may depend on other branches, due to the interactions between the genes which share various functions. We assume no extinction of the gene family due to the protective mechanisms, and so essentially, we model the evolution of a gene family that has survived.
	
	We note that, in contrast to the the simulation model described in Diao et al.~\cite{2020DSLOH}, here we do not track the exact functional redundancies, but instead approximate them using the phase variable of the QBD. Consequently, we are able to apply many useful computational techniques from the theory of QBDs for the analysis of stationary and transient quantities.

	Consider a continuous-time level-dependent quasi-and-birth process (LD-QBD) $\{(Y(t),\varphi(t)):t\geq 0\}$ which is a CTMC with state space
	\begin{equation}
		\mathcal{S}=\{(n,k): n=1,2,\ldots; k=1,\ldots,K_n\},
	\end{equation}
	where the level variable $Y(t)\in\{1,2,\ldots\}$ records the number of genes in the family, the phase variable $\varphi(t)\in\{1,\ldots,K_n\}$ records some information about the gene family and/or an underlying environment that affects the evolution when $Y(t)=n$.
	
	We assume the initial distribution $\balpha=[\alpha_k]_{k=1,\ldots K_1}$ such that $\alpha_k=\mathbb{P}(Y(0)=1,\varphi(0)=k)$ for $k=1,\ldots K_1$, and generator ${\bf Q}=[q_{(n,k)(n^{'},k^{'})}]_{(n,k),(n^{'},k^{'})\in\mathcal{S}}$ made of block matrices ${\bf Q}^{[n,n^{'}]}$, given by,  
	\begin{equation}
		{\bf Q} = 
		[{\bf Q}^{[n,n^{'}]}]_{n,n^{'}=1,2,3,\ldots}
		=
		\begin{bmatrix}
			{\bf Q}^{[1,1]} & {\bf Q}^{[1,2]} & {\bf 0} & {\bf 0} & \cdots\\
			{\bf Q}^{[2,1]} & {\bf Q}^{[2,2]} & {\bf Q}^{[2,3]} & {\bf 0} & \cdots\\
			{\bf 0} & {\bf Q}^{[3,2]} & {\bf Q}^{[3,3]} & {\bf Q}^{[3,4]} &  \cdots\\
			{\bf 0} & {\bf 0} & {\bf Q}^{[4,3]} & {\bf Q}^{[4,4]} & \cdots \\
			\vdots & \vdots & \vdots & \vdots & \ddots
		\end{bmatrix},
	\end{equation}
	where, for a given $n$ and $n^{'}$, block ${\bf Q}^{[n,n^{'}]}$ records the transition rates $q_{(n,k)(n^{'},k^{'})}$ from states $(n,k)$ to $(n^{'},k^{'})$, $n^{'}\in\{n-1,n,n+1\}$, with
	\begin{eqnarray}
		{\bf Q}^{[n,n^{'}]}&=&\left[q_{(n,k)(n^{'},k^{'})}\right]_{k=1,\ldots,K_n,k^{'}=1,\ldots,K_{n^{'}}}.
	\end{eqnarray}

	Below, we describe LD-QBDs models for the evolution of a gene family based on the work by Diao et al. in~\cite{2020DSLOH}, which incorporate the various complex processes that affect the evolution of genes. These models are mechanistic in the sense that they incorporate the four key Poisson rates that drive the evolution, where
	\begin{itemize}
		\item $u_c$ is per gene rate of null mutation in the coding region of a gene;
		\item $u_r$ is per region rate of null mutation in regulatory region of a gene;
		\item $u_d$ is per gene rate of duplication of a gene; and
		\item $u_f$ is per gene rate of obtaining a new function of a gene.
	\end{itemize}
	So, in practice, the assumptions are made about the values of $u_c$, $u_r$, $u_d$ and $u_f$, and the remaining parameters in the generator matrix ${\bf Q}$ are obtained through the simulation model in Diao et al.~\cite{2020DSLOH}, as we illustrate in Example~\ref{ex2bJ}. The advantage of the QBD model over the simulation model is that we can apply the theory of matrix-analytic methods to compute a range of useful metrics in an efficient manner.

	\begin{Example}\label{ex1J}
		Consider a QBD $\{(Y(t),\varphi(t)):t\geq 0\}$ based on a model proposed by Diao et al. in~\cite{2020DSLOH}, where we modify the state space so as to record the number of branches in each subtree, with
		\begin{equation}
			\mathcal{S}=\{(n,m,n_{(L)},m_{(L)},k): n=1,\ldots; m=0,\ldots,n; n_{(L)}=1,\ldots,n-1; m_{(L)}=0,\ldots,m; k=1,\ldots,K\},
		\end{equation}
		where $n$ is the number of genes of family, $m$ is the number of redundant genes, $n_{(L)}$ is the number of genes on the left branch, $m_{(L)}$ is the number of the redundant genes on the left branch, and $k$ is some information about the gene family used the model the total number of functions in the family.

		The transition rates of such QBD summarised in Table~\ref{table:offdiag}, expressed in terms of parameters $u_c$, $u_d$, $u_r$ and $u_f$, follow by appropriately modifying~\cite[Table 1]{2020DSLOH}, where
		\begin{eqnarray}
			{p}_{k,\ell,\ell_{(L)}}=
			\begin{cases}
				p_{k,\ell}\times p_{\ell_{(L)},\ell}& \textit{if}\quad n_{(L)}^{'}= n_{(L)}-1\\
				p_{k,\ell}\times \widehat{p}_{\ell_{(L)},\ell}& \textit{if}\quad n_{(L)}^{'}= n_{(L)}
			\end{cases}
		\end{eqnarray}
		is the probability that $\ell$ previously redundant genes each become non-redundant and $\ell_{(L)}$ genes are lost on the left branch of the tree when a gene loses its coding region; and
		\begin{eqnarray}
			\widehat{p}_{k,\ell,\ell_{(L)}}=
			\begin{cases}
				\widehat{p}_{k,\ell}\times p_{\ell_{(L),\ell}}&\textit{if}\quad n_{(L)}^{'}= n_{(L)}-1,\\
				\widehat{p}_{k,\ell}\times \widehat{p}_{\ell_{(L),\ell}}& \textit{if}\quad n_{(L)}^{'}= n_{(L)};
			\end{cases}
		\end{eqnarray}
		is the probability that $\ell$ previously redundant genes each become non-redundant and $\ell_{(L)}$ genes are lost on the left branch of the tree when a gene loses its last function; where
		\begin{eqnarray}
			p_{\ell_{(L)},\ell}=\frac{{m_{(L)}-1 \choose \ell_{(L)}-1}{m-m_{(L)} \choose \ell-\ell_{(L)} } }{{m-1 \choose \ell-1}}
		\end{eqnarray}
		is the conditional probability that a gene is lost on the left branch and $\ell$ redundant genes become non-redundant include $\ell_{(L)}$ genes are on the left branch of the gene tree, with $\sum_{\ell=1}^{m_{(L)}}p_{\ell_{L},\ell}=1$; and
		\begin{eqnarray}
			\widehat{p}_{\ell_{(L)},\ell}=\frac{{m-m_{(L)}-1 \choose \ell-\ell_{(L)}-1}{m_{(L)} \choose \ell_{(L)} } }{{m-1 \choose \ell-1}}
		\end{eqnarray}
		is the conditional probability that a gene is lost on the right branch and $\ell$ redundant genes become non-redundant include $\ell_{(L)}$ genes are on the left branch of the gene tree, with $\sum_{\ell=1}^{m_{(L)}}\widehat{p}_{\ell_{(L)},\ell}=1$. $p_{k,\ell}$ and $\widehat{p}_{k,\ell}$ are the probability of $\ell$ redundant genes becoming non-redundant under the current deterioration $k$ when a gene loses its coding region or loses its last function.

		Further,
		\begin{eqnarray}
			\alpha_r = P(E_k),\
			\delta=P(E_m|E_n),\
			\gamma = P(E_n),
		\end{eqnarray}
		where $E_n,E_m$ and $E_k$ are events, respectively, that $n$ stays the same, $m$ stays the same and $k$ stays the same, given one of the genes loses a function.

		Moreover,  $p_k$ is the proportion of functions, per gene, that are permitted to be lost, given current deterioration $k$, which decreases as $k$ increases. A suitable form of funcion $p_k$ is
		\begin{eqnarray}\label{eq:pkdef}
			p_k=\frac{K-k}{K}+\epsilon,
		\end{eqnarray}
		for some $\epsilon>0$. Note that then $nz_0p_k$ is the estimated number of functions that are permitted to be lost in the gene family.\hfill$\Box$
		
	\end{Example}

	\begin{Example}\label{ex2J}
		Consider another QBD $\{(Y(t),\varphi(t)):t\geq 0\}$ also based on a model proposed by Diao et al. in~\cite{2020DSLOH}, obtained by assuming $u_f=0$ in Example~\ref{ex1J}, which implies that the number of functions in the gene family may not increase. Denote by $z_0$ a positive integer which records the average number of functions in the regulatory regions of the genes in the gene family. We derive the expressions for the transition rates of the model by modifying \cite[Table 4]{2020DSLOH}, see Table~\ref{table:offdiag1}.
		
		By an extension of the analysis in~\cite{2020DOH,2020DSLOH}, when the condition $u_c+u_r>u_d$ is met, the process is stable. In such case, we define the stationary distribution vector $\bm{\pi}=[\bm{\pi}_n]_{n=1,2,\ldots}$ such that, for all $n=1,2,\ldots$, vectors
		\begin{eqnarray}
			\bm{\pi}_n &=&
			[
			\pi_{n,m,n_{(L)},m_{(L)},k}]_{m=0,1,\dots,n;n_{(L)}=1,\dots,n-1;m_{(L)}=0,1,\dots,\min\{m,n_{(L)}\};k=1,\dots,K} 
		\end{eqnarray}
		are given by
		\begin{eqnarray}
			\pi_{n,m^{'},n_{(L)}^{'},m_{(L)}^{'},k^{'}}&=&
			\lim_{t\to\infty}\mathcal{P}
			\left(
			(Y(t),\varphi(t))=(n,m^{'},n_{(L)}^{'},m_{(L)}^{'},k^{'})
			\right),
		\end{eqnarray}
		where the limits exist and do not depend on the initial state by the stability of the process. 
		
		We also define we the probability $q_n(i)$ that a tree with $n$ tips has $i$ tips on the left, as
		\begin{eqnarray}\label{qniEq}
			q_n(i)=\sum_{m=0}^{n}\sum_{m_{(L)}=0}^{\min{(m,i)}}\sum_{k=1}^{K}\pi_{n,m,i,m_{(L)},k},
		\end{eqnarray}
		which is a key metric of interest in the phylogenetic analysis. \hfill$\Box$

		\begin{table}[hp]
			\scriptsize
			\begin{displaymath}
				\begin{array}{|l|l|c|}
					\hline
					\vspace{-0.4cm}
					& & \\\vspace{-0.4cm}
					q_{(n,m,n_{(L)},m_{(L)},k)(n^{'},m^{'},n_{(L)}^{'},m_{(L)}^{'},k^{'})} & (n^{'},m^{'},n_{(L)}^{'},m_{(L)}^{'},k^{'}) & \mbox{under condition}\\
					& & \\\hline 
					\vspace{-0.4cm}
					& & \\
					(n_{(L)}-m_{(L)}) u_d & (n+1,m+2,n_{(L)}+1,m_{(L)}+2,k) & k=1\\
					\big((n-n_{(L)})-(m-m_{(L)})\big)u_d&(n+1,m+2,n_{(L)},m_{(L)},k)&k=1\\
					m_{(L)} u_d & (n+1,m+1,n_{(L)}+1,m_{(L)}+1,k) & k=1\\
					(m-m_{(L)})u_d & (n+1,m+1,n_{(L)},m_{(L)},k) & k=1\\
					\hline
					\vspace{-0.4cm}
					& & \\
					\alpha_d\times(n_{(L)}-m_{(L)}) u_d & (n+1,m+2,n_{(L)}+1,m_{(L)}+2,k) & k>1\\
					\alpha_d\times\big((n-n_{(L)})-(m-m_{(L)})\big) u_d & (n+1,m+2,n_{(L)},m_{(L)},k) & k>1\\
					(1-\alpha_d)\times(n_{(L)}-m_{(L)}) u_d & (n+1,m+2,n_{(L)}+1,m_{(L)}+2,k-1) & k>1\\
					(1-\alpha_d)\times\big((n-n_{(L)})-(m-m_{(L)})\big) u_d & (n+1,m+2,n_{(L)},m_{(L)},k-1) & k>1\\
					\alpha_d\times m_{(L)} u_d & (n+1,m+1,n_{(L)}+1,m_{(L)}+1,k) & k>1\\
					\alpha_d\times(m-m_{(L)}) u_d&
					(n+1,m+1,n_{(L)},m_{(L)},k) & k>1\\
					(1-\alpha_d)\times m_{(L)} u_d & (n+1,m+1,n_{(L)}+1,m_{(L)}+1,k-1) & k>1\\
					(1-\alpha_d)\times (m-m_{(L)}) u_d & (n+1,m+1,n_{(L)},m_{(L)},k-1) & k>1\\\hline
					\vspace{-0.4cm}
					& & \\
					m_{(L)} u_c p_{k,\ell,\ell_{(L)}}+(1-\gamma)\alpha_r\times n_{(L)}z_0p_ku_r\widehat p_{k,\ell,\ell_{(L)}} 
					& \makecell{(n-1,m-\ell,n_{(L)}-1,m_{(L)}-\ell_{(L)},k),\\\ell=1,\ldots,m, \ell_{(L)}=1,\ldots,m_{(L)}} & m>0,n_{(L)}>1,k<K\\
					(m-m_{(L)})u_c p_{k,\ell,\ell_{(L)}}+(1-\gamma)\alpha_r (n-n_{(L)})z_0p_ku_r\widehat p_{k,\ell,\ell_{(L)}}&\makecell{ (n-1,m-\ell,n_{(L)},m_{(L)}-\ell_{(L)},k),\\ \ell=1,\ldots,m,\ell_{L}=0,\ldots,\ell-1}&m>0,n_{(L)}<n-1,k<K\\
					(1-\gamma)(1-\alpha_r)\times n_{(L)}z_0p_ku_r\widehat p_{k,\ell,\ell_{(L)}}  & \makecell{(n-1,m-\ell,n_{(L)}-1,m_{(L)}-\ell_{(L)},k+1),\\ \ell=1,\ldots,m,\ell_{(L)}=1,\ldots,m_{(L)}} & m>0,n_{(L)}>1,k<K\\
					(1-\gamma)(1-\alpha_r)\times (n-n_{(L)})z_0p_ku_r\widehat p_{k,\ell,\ell_{(L)}}  & \makecell{(n-1,m-\ell,n_{(L)},m_{(L)}-\ell_{(L)},k+1),\\ \ell=1,\ldots,m,\ell_{(L)}=0,\ldots,\ell-1} & m>0,n_{(L)}<n-1,k<K\\
					\gamma\beta(1-\alpha_r)\times nz_0p_k u_r  & (n,m,n_{(L)},m_{(L)},k+1) & m>0,k<K\\
					\gamma(1-\beta)\alpha_r\times  n_{(L)}z_0p_k u_r+m_{(L)} u_f & (n,m-1,n_{(L)},m_{(L)}-1,k) & m>0,k<K\\
					\gamma(1-\beta)\alpha_r\times  (n-n_{(L)})z_0p_k u_r+(m-m_{(L)}) u_f & (n,m-1,n_{(L)},m_{(L)},k) & m>0,k<K\\
					\gamma(1-\beta)(1-\alpha_r)\times  n_{(L)}z_0p_k u_r  & (n,m-1,n_{(L)},m_{(L)}-1,k+1) & m>0,k<K\\
					\gamma(1-\beta)(1-\alpha_r)\times  (n-n_{(L)})z_0p_k u_r  & (n,m-1,n_{(L)},m_{(L)},k+1) & m>0,k<K\\\hline
					\vspace{-0.4cm}
					& & \\
					m_{(L)}u_c p_{k,\ell,\ell_{(L)}}+(1-\gamma)\times n_{(L)}z_0p_ku_r\widehat p_{k,\ell,\ell_{(L)}}  & \makecell{ (n-1,m-\ell,n_{(L)}-1,m_{(L)}-\ell_{(L)},k),\\ \ell=1,\ldots,m,\ell_{L}=1,\ldots,m_{(L)}} & m>0,n_{(L)}>1,k=K\\
					(m-m_{(L)})u_c p_{k,\ell,\ell_{(L)}}+(1-\gamma) (n-n_{(L)})z_0p_ku_r\widehat p_{k,\ell,\ell_{(L)}}&\makecell{ (n-1,m-\ell,n_{(L)},m_{(L)}-\ell_{(L)},k),\\ \ell=1,\ldots,m,\ell_{(L)}=0,\ldots,\ell-1}&m>0,n_{(L)}<n-1,k=K\\
					\gamma(1-\beta)\times n_{(L)}z_0p_k u_r +m_{(L)}u_f & (n,m-1,n_{(L)},m_{(L)}-1,k) & m>0,k=K\\
					\gamma(1-\beta)\times (n-n_{(L)})z_0p_k u_r +(m-m_{(L)})u_f & (n,m-1,n_{(L)},m_{(L)},k) & m>0,k=K\\\hline
					\vspace{-0.4cm}
					& & \\
					(1-\alpha_r)\times nz_0p_k u_r & (n,m,n_{(L)},m_{(L)},k+1) & m=0,k<K\\\hline 
				\end{array}
			\end{displaymath}
			\caption{Non-zero off-diagonals $q_{(n,m,n_{(L)},m_{(L)},k)(n^{'},m^{'},n_{(L)}^{'},m_{(L)}^{'},k^{'})}$ in the generator ${\bf Q}$ in Example~\ref{ex1J}.}
			\label{table:offdiag}
		\end{table}
		\begin{table}[hp]
			\scriptsize
			\begin{displaymath}
				\begin{array}{|l|l|c|}
					\hline
					\vspace{-0.4cm}
					& & \\\vspace{-0.4cm}
					q_{(n,m,n_{(L)},m_{(L)},k)(n^{'},m^{'},n_{(L)}^{'},m_{(L)}^{'},k^{'})} & (n^{'},m^{'},n_{(L)}^{'},m_{(L)}^{'},k^{'}) & \mbox{under condition}\\
					& & \\\hline 
					\vspace{-0.4cm}
					& & \\
					(n_{(L)}-m_{(L)}) u_d & (n+1,m+2,n_{(L)}+1,m_{(L)}+2,k) & k=1\\
					\big((n-n_{(L)})-(m-m_{(L)})\big)u_d&(n+1,m+2,n_{(L)},m_{(L)},k)&k=1\\
					m_{(L)} u_d & (n+1,m+1,n_{(L)}+1,m_{(L)}+1,k) & k=1\\
					(m-m_{(L)})u_d & (n+1,m+1,n_{(L)},m_{(L)},k) & k=1\\
					\hline
					\vspace{-0.4cm}
					& & \\
					\alpha_d\times(n_{(L)}-m_{(L)}) u_d & (n+1,m+2,n_{(L)}+1,m_{(L)}+2,k) & k>1\\
					\alpha_d\times\big((n-n_{(L)})-(m-m_{(L)})\big) u_d & (n+1,m+2,n_{(L)},m_{(L)},k) & k>1\\
					(1-\alpha_d)\times(n_{(L)}-m_{(L)}) u_d & (n+1,m+2,n_{(L)}+1,m_{(L)}+2,k-1) & k>1\\
					(1-\alpha_d)\times\big((n-n_{(L)})-(m-m_{(L)})\big) u_d & (n+1,m+2,n_{(L)},m_{(L)},k-1) & k>1\\
					\alpha_d\times m_{(L)} u_d & (n+1,m+1,n_{(L)}+1,m_{(L)}+1,k) & k>1\\
					\alpha_d\times(m-m_{(L)}) u_d&
					(n+1,m+1,n_{(L)},m_{(L)},k) & k>1\\
					(1-\alpha_d)\times m_{(L)} u_d & (n+1,m+1,n_{(L)}+1,m_{(L)}+1,k-1) & k>1\\
					(1-\alpha_d)\times (m-m_{(L)}) u_d & (n+1,m+1,n_{(L)},m_{(L)},k-1) & k>1\\\hline
					\vspace{-0.4cm}
					& & \\
					\makecell{m_{(L)} u_c p_{k,\ell^{max},\ell_{(L)}}+\\(1-\gamma)\alpha_r\times n_{(L)}z_0p_ku_r\widehat p_{k,\ell,\ell_{(L)}}}  
					& \makecell{(n-1,m-\ell,n_{(L)}-1,m_{(L)}-\ell_{(L)},k),\\\ell=1,\ldots,z_0+1-(n-m), \ell_{(L)}=1,\ldots,m_{(L)}} & m>0,n_{(L)}>1,k<K\\
					\makecell{(m-m_{(L)})u_c p_{k,\ell,\ell_{(L)}}+\\(1-\gamma)\alpha_r\times (n-n_{(L)})z_0p_ku_r\widehat p_{k,\ell,\ell_{(L)}}}&\makecell{ (n-1,m-\ell,n_{(L)},m_{(L)}-\ell_{(L)},k),\\ \ell=1,\ldots,z_0+1-(n-m),\ell_{L}=0,\ldots,\ell-1}&m>0,n_{(L)}<n-1,k<K\\
					(1-\gamma)(1-\alpha_r)\times n_{(L)}z_0p_ku_r\widehat p_{k,\ell,\ell_{(L)}}  & \makecell{(n-1,m-\ell,n_{(L)}-1,m_{(L)}-\ell_{(L)},k+1),\\ \ell=1,\ldots,z_0+1-(n-m),\ell_{(L)}=1,\ldots,m_{(L)}} & m>0,n_{(L)}>1,k<K\\
					(1-\gamma)(1-\alpha_r)\times (n-n_{(L)})z_0p_ku_r\widehat p_{k,\ell,\ell_{(L)}}  & \makecell{(n-1,m-\ell,n_{(L)}-1,m_{(L)}-\ell_{(L)},k+1),\\ \ell=1,\ldots,z_0+1-(n-m),\ell_{(L)}=0,\ldots,\ell-1} & m>0,n_{(L)}<n-1,k<K\\
					\gamma\beta(1-\alpha_r)\times nz_0p_k u_r  & (n,m,n_{(L)},m_{(L)},k+1) & m>0,k<K\\
					\gamma(1-\beta)\alpha_r\times  n_{(L)}z_0p_k u_r & (n,m-1,n_{(L)},m_{(L)}-1,k) & m>0,k<K\\
					\gamma(1-\beta)\alpha_r\times  (n-n_{(L)})z_0p_k u_r& (n,m-1,n_{(L)},m_{(L)},k) & m>0,k<K\\
					\gamma(1-\beta)(1-\alpha_r)\times  n_{(L)}z_0p_k u_r  & (n,m-1,n_{(L)},m_{(L)}-1,k+1) & m>0,k<K\\
					\gamma(1-\beta)(1-\alpha_r)\times  (n-n_{(L)})z_0p_k u_r  & (n,m-1,n_{(L)},m_{(L)},k+1) & m>0,k<K\\\hline
					\vspace{-0.4cm}
					& & \\
					\makecell{m_{(L)}u_c p_{k,\ell,\ell_{(L)}}+\\(1-\gamma)\times n_{(L)}z_0p_ku_r\widehat p_{k,\ell,\ell_{(L)}}}  & \makecell{ (n-1,m-\ell,n_{(L)}-1,m_{(L)}-\ell_{(L)},k),\\ \ell=1,\ldots,z_0+1-(n-m),\ell_{L}=1,\ldots,m_{(L)}} & m>0,n_{(L)}>1,k=K\\
					\makecell{(m-m_{(L)})u_c p_{k,\ell,\ell_{(L)}}+\\(1-\gamma)\times (n-n_{(L)})z_0p_ku_r\widehat p_{k,\ell,\ell_{(L)}}}&\makecell{ (n-1,m-\ell,n_{(L)},m_{(L)}-\ell_{(L)},k),\\ \ell=1,\ldots,z_0+1-(n-m),\ell_{L}=0,\ldots,\ell-1}&m>0,n_{(L)}<n-1,k=K\\
					\gamma(1-\beta)\times n_{(L)}z_0p_k u_r & (n,m-1,n_{(L)},m_{(L)}-1,k) & m>0,k=K\\
					\gamma(1-\beta)\times (n-n_{(L)})z_0p_k u_r  & (n,m-1,n_{(L)},m_{(L)},k) & m>0,k=K\\\hline
					\vspace{-0.4cm}
					& & \\
					(1-\alpha_r)\times nz_0p_k u_r & (n,m,n_{(L)},m_{(L)},k+1) & m=0,k<K\\\hline 
					& &
					\vspace{-0.4cm}\\
					mu_c+(1-\gamma)\times nz_0p_ku_r&(1,0,1,0,4)&n=2,m=1,2\\
					u_d&(2,2,1,1,1)&n=1\\\hline
				\end{array}
			\end{displaymath}
			\caption{Non-zero off-diagonals $q_{(n,m,n_{(L)},m_{(L)},k)(n^{'},m^{'},n_{(L)}^{'},m_{(L)}^{'},k^{'})}$ in the generator ${\bf Q}$ under assumption $u_f=0$ in Example~\ref{ex2J}. For $\ell^{max}=z_0+1-(n-m)$ we use $p_{k,\ell^{max},\ell{(L)}}=\sum_{\ell\geq \ell^{max}} p_{k,\ell,\ell^{(L)}}$ and $\widehat{p}_{k,\ell^{max},\ell^{(L)}}=\sum_{\ell\geq \ell^{max}} \widehat{p}_{k,\ell,\ell^{(L)}}$.}
			\label{table:offdiag1}
		\end{table}

	\end{Example}

	\subsection{Computing metrics of a gene tree}
	
	Below we discuss tree metrics for the transient (time-dependent) as well stationary behaviour of the process. Metrics based on the stationary behaviour are useful when the process is stable (stationary distribution exists) and has been evolving for a sufficiently long time (so that the stationary distribution is a reasonable approximation of the current distribution). Metrics based on the transient behaviour are applicable regardless of whether the stationary distribution exists. We give examples of applications of these metrics for various models.
	
	Define the stationary distribution vector $\bm{\pi}=[\bm{\pi}_n]_{n=1,2,\ldots}$, $\bm{\pi}_n=[\pi_{n,k}]_{k=1,\dots,K_{n}}$, given by
	\begin{eqnarray}
		\pi_{n,k}=\lim_{t\to\infty}\mathbb{P}\left(Y(t)=n,\varphi(t)=k\right),
	\end{eqnarray}
	whenever the limits exist and do not depend on the initial state. In order to compute $\bm{\pi}_n$ for $n=1,2,\ldots$, we apply an iterative scheme in Phung-Duc et al. in~\cite{phung2010simple}, 
	\begin{enumerate}
		\item Find ${\bf x}_1$ such that ${\bf x}_1({\bf Q}^{[1,1]}+{\bf R}^{(2)}{\bf Q}^{[2,1]})=0$ and let $\Sigma_1={\bf x}_1{\bf 1}$;
		\item For $n=2$ to $N$, let ${\bf x}_n={\bf x}_{n-1}{\bf R}^{(n)}$ and $\Sigma_n=\Sigma_{n-1}+a_n$, where $a_n={\bf x}_n{\bf 1}$;
		\item For $n=1$ to $N$, let $\bm{\pi}_n={\bf x}_n/\Sigma_N$;  
	\end{enumerate}
	where matrices ${\bf R}^{(n)}\equiv {\bf R}^{(n)}(s)|_{s=0}$ are computed using Algorithm~\ref{rateMatrices_algorithm} below, based an algorithm in~\cite{phung2010simple}, with a slight extension to ${\bf R}^{(n)}(s)$.

	The iterative scheme for  is $\bm{\pi}_n$ in~\cite{phung2010simple} is based on a recursive expression
	\begin{eqnarray}
		\bm{\pi}_{n}&=&\bm{\pi}_{n-1}{\bf R}^{(n)},
	\end{eqnarray}
	where matrix ${\bf R}^{(n)}=[R^{(n)}_{ij}]_{i=1,\dots,K_{n-1},j=1,\dots,K_{n}}$ is such that
	\begin{equation}
		R^{(n)}_{i,j} = [{\bf Q}^{[n-1,n-1]}]_{ii}\mathbb{E}_{n-1,i}\left[\int_0^{\tau_{n}} {\bf 1}(Y(t) = n,\varphi(t)=j)dt\right],
	\end{equation}
	records the expected sojourn time in state $(n,j)$ per unit sojourn time in $(n-1,i)$, before returning to level $n-1$ given the process starts in state $(n-1,i)$, and ${\bf R}^{(n)}(s) = [R^{(n)}_{i,j}(s)]_{i,j \in \mathcal{S}}$ is the corresponding Laplace-Stieltjes transform given by,
	\begin{equation}
		R^{(n)}_{i,j}(s) = [{\bf Q}^{[n-1,n-1]}]_{ii}\mathbb{E}_{n-1,i}\left[\int_0^{\tau_{n}}e^{-st} {\bf 1}(Y(t) = n,\varphi(t)=j)dt\right].
	\end{equation}
	The algorithm for ${\bf R}^{(n)}(s)$ is based on the approximation 
	\begin{equation}
		{\bf R}^{(n)}(s) = \lim_{k\to \infty} {\bf R}^{(n)}_k(s),
	\end{equation}
	where ${\bf R}^{(n)}_k(s)$ has interpretation similar to ${\bf R}^{(n)}(s)$, except that there is a taboo on visiting level $n-1+k$.

	Let ${\bf f}(t)=[{\bf f}_n(t)]_{n=0,1,\infty}$ be the distribution of the process at time $t$, such that
	\begin{equation}
		[{\bf f}_n(t)]_i = \mathbb{P}_{\boldsymbol \alpha}(Y(t) = n, \varphi(t) = i),
	\end{equation}
	is the probability of observing state $(n,i)$ at time $t$ given the initial distribution $\boldsymbol \alpha$, and let ${\bf F}_n(s)$ be the corresponding Laplace-Stieltjes transform given by,
	\begin{equation}
		{\bf F}_n(s) = \int_{t=0}^\infty e^{-st}{\bf f}_n(t)dt.
	\end{equation}
	Then by Joyner and Fralix~\cite{joyner2016new}, we have a recursive formula for $ {\bf F}_n(s)$ for $n\geq 1$, 
	\begin{eqnarray} \label{eq_transdist1}
		F_1(s) &=& -\boldsymbol \alpha({\bf Q}^{[1,1]} - s{\bf I} + {\bf R}^{(2)}(s){\bf Q}^{[2,1]})^{-1},
		\\
		\label{eq_transdist2}
		F_n(s) &=& F_{n-1}(s){\bf R}^{(n)}(s).
	\end{eqnarray}
	and so the quantities ${\bf f}_n(t)$ can be computed with the numerical inversion techniques in Den Iseger~\cite{DenIseger_2006}, or Horv{\'a}th et al.~\cite{horvath2020numerical}.

	\begin{algorithm}
		\caption{Approximate $\{{\bf R}^{(n)}(s)\}_{n=2,\hdots,N}$ 
		\hfill (adapted from Phung-Duc et al.~\cite{phung2010simple})}
		\label{rateMatrices_algorithm}
		\begin{algorithmic}[1] %
			\Procedure{rateMatrices}{$s,N,{\bf Q},\epsilon$} 
			\State Compute ${\bf R}^{(N)}_{k}$ for $k=1$ using
			\begin{align} 
				{\bf R}^{(N)}_{k} &= R_N \circ R_{N+1} \circ \hdots \circ R_{N+k-1}({\bf 0}) , \label{eq_Algorithm1} \\
				R_n({\bf X}) &= {\bf Q}^{[n-1,n]}(-({\bf Q}^{[n,n]}-s{\bf I})-{\bf X}{\bf Q}^{[n+1,n]})^{-1},    \label{eq_Algorithm2}   
			\end{align}
			where we denote $f\circ h(\cdot)=f(h(\cdot))$.
			\While{$||{\bf R}^{(N)}_k - {\bf R}^{(N)}_{k-1}||_\infty > \epsilon$} 
			\State Let $k=k + 1$.
			\State Compute ${\bf R}^{(N)}_{k}$ using \eqref{eq_Algorithm1} and \eqref{eq_Algorithm2}.
			\EndWhile
			\State Let $\hat {\bf R}^{(N)} = {\bf R}^{(N)}_{k}$. 
			\For{$n = N-1, \hdots, 2$}
			\State Compute $\hat {\bf R}^{(n)} = R_{n}(\hat {\bf R}^{(n+1)})$ using \eqref{eq_Algorithm1} and \eqref{eq_Algorithm2}. 
			\EndFor
			\State Let $\{{\bf R}^{(n)}(s)\}_{n=2,\hdots,N}\approx \{\hat {\bf R}^{(n)}\}_{n=2,\hdots,N}$.
			\EndProcedure
		\end{algorithmic}
	\end{algorithm}

	\begin{Example}\label{ex2bJ}
		Consider Example~\ref{ex2J}. Let $K=4$ and $\epsilon=0$ so that $p_k$ function defined in~\eqref{eq:pkdef} is given by $p_k=\frac{K-k}{K}$ for $k=1,2,3,4$.
		
		Assuming $u_d=u_c=1, u_r=0.3, u_f=0$ and $z_0=5$, we estimated the values of $\alpha_d$, $\alpha_r$, $\delta$ and $\gamma$ using the average proportion of time the corresponding events were observed in the simulation model in Diao et al.~\cite{2020DSLOH}. We have 
		\begin{eqnarray}\label{para}
			\alpha_d\approx 0.9076,\quad
			\alpha_r\approx 0.7557,\quad
			\delta\approx 0.4642,\quad
			\gamma\approx 0.3172.
		\end{eqnarray}
		
		We also evaluated $p_{k,\ell}$ and $\widehat{p}_{k,\ell}$ using the proportion of time $\ell$ redundant genes were lost when $n^{'}=n-1$ and deterioration was $k$ (Table~\ref{pkl}), and the proportion of time $\ell$ redundant genes were lost when $n^{'}=n$ and deterioration was $k$ (Table~\ref{hatpkl}).
		\begin{table}[H]
			\centering
			\begin{tabular}{l l l l l}
				\hline
				$p_{k,\ell}$&$k=1$&$k=2$&$k=3$&k=4\\
				\hline
				$\ell=1$&$0.9075$&$0.6599$&$0.4671$&$0.2408$\\
				$\ell=2$&$0.0920$&$0.3334$&$0.5269$&$0.7561$\\
				$\ell=3$&$5.3578\times 10^{-4}$&$0.0067$&$0.0060$&$0.0031$\\
				$\ell=4$&$8.2682\times 10^{-7}$&$3.8669\times 10^{-5}$&$2.1421\times 10^{-5}$&$0$\\
				\hline
			\end{tabular}
			\caption{Probabilities $p_{k,\ell}$, for $k=1,2,3,4$, $\ell=1,2,3,4$ when $m\geq 4$. When $m=1,2,3$, we use $p_{k,m}=\sum_{\ell\geq m}p_{k,\ell}$ due to the requirement that $\ell\leq m$.}
			\label{pkl}
		\end{table}
		\begin{table}[H]
			\centering
			\begin{tabular}{l l l l l}
				\hline
				$\widehat{p}_{k,\ell}$&$k=1$&$k=2$&$k=3$&k=4\\
				\hline
				$\ell=1$&$0.9313$&$0.7768$&$0.6419$&$0.2056$\\
				$\ell=2$&$0.0687$&$0.2232$&$0.3581$&$0.7944$\\
				\hline
			\end{tabular}
			\caption{Probabilities $\widehat{p}_{k,\ell}$, for $k=1,2,3,4$, $\ell=1,2$ when $m\geq 2$. When m=1, we use $\widehat{p}_{k,m}=\sum_{\ell\geq m}\widehat p_{k,\ell}=\sum_{\ell\geq 1}\widehat p_{k,\ell}=1$ due to the requirement that $\ell\leq m$.}
			\label{hatpkl}
		\end{table}

		Next, we considered how the values of $\delta$ and $\gamma$ may affect the $q_n(i)$ function. We found that $\delta$ has limited affect on the value of $q_n(i)$, see  Figure~\ref{QBDbetagamma}(a)-(b). This makes sense since $\delta$ is the probability that $m$ stays the same given one of the genes loses a function, which is a type of event that does not directly change the balance of tree, since we assumed that $m>n-z_0$. 
		
		On the other hand, we found that decreasing $\gamma$ results in larger $\beta$ (more balanced trees when compared to the original value $\gamma=0.3172$) and increasing $\gamma$ has in the opposite effect (less balanced trees when compared to the original value $\gamma=0.3172$), see Figure~\ref{QBDbetagamma}(c)-(d). This makes sense since $(1-\gamma)$ is the probability that a gene is lost due to losing its last function and so $n\to (n-1)$, which is a type of event that directly changes the balance of the tree. Specifically, smaller $\gamma$ results in a higher total rate of gene loss, which leads to more balanced trees, matching the result in Diao et al. \cite{2020DOH}.
		\begin{figure}[hp]
			\centering
			\subfloat[]{\includegraphics[width=7.5 cm,height=7.5 cm]{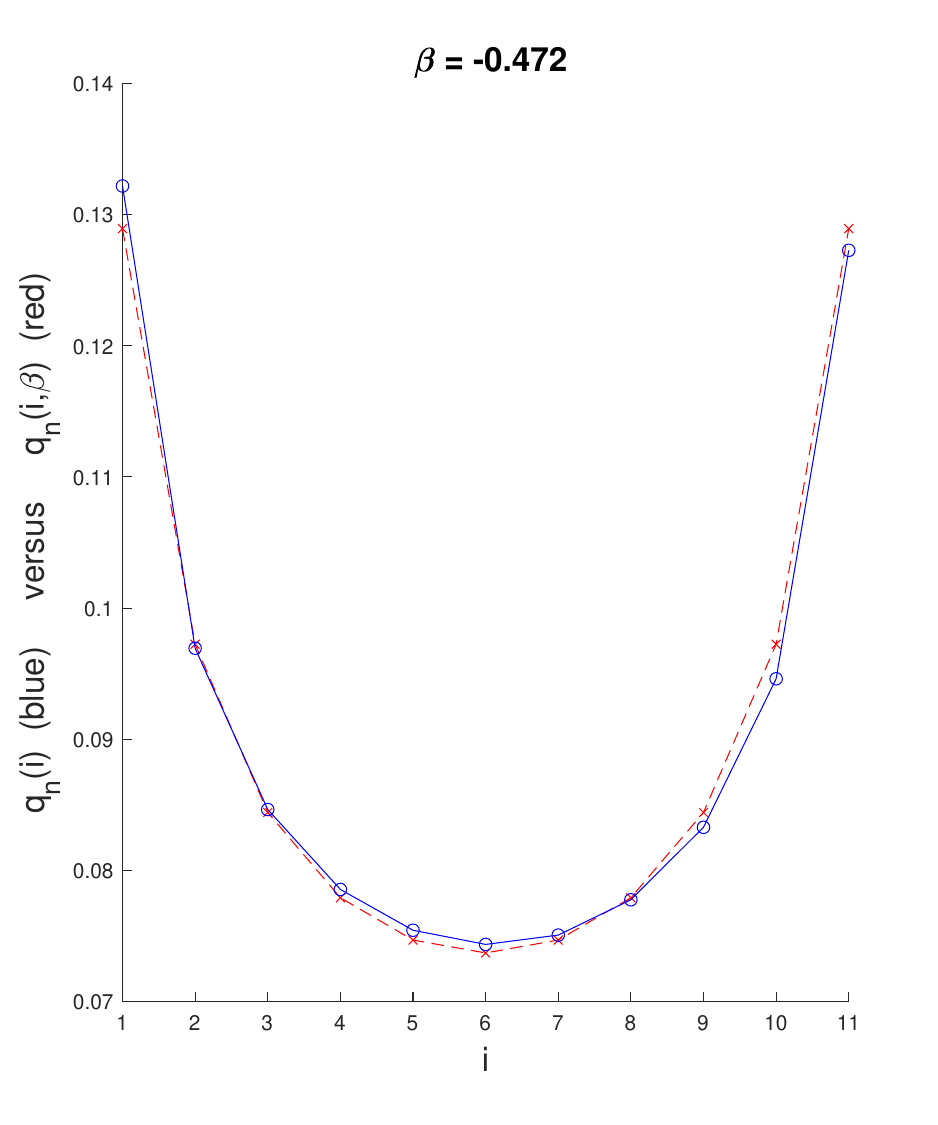}}
			\quad
			\subfloat[]{\includegraphics[width=7.5 cm,height=7.5 cm]{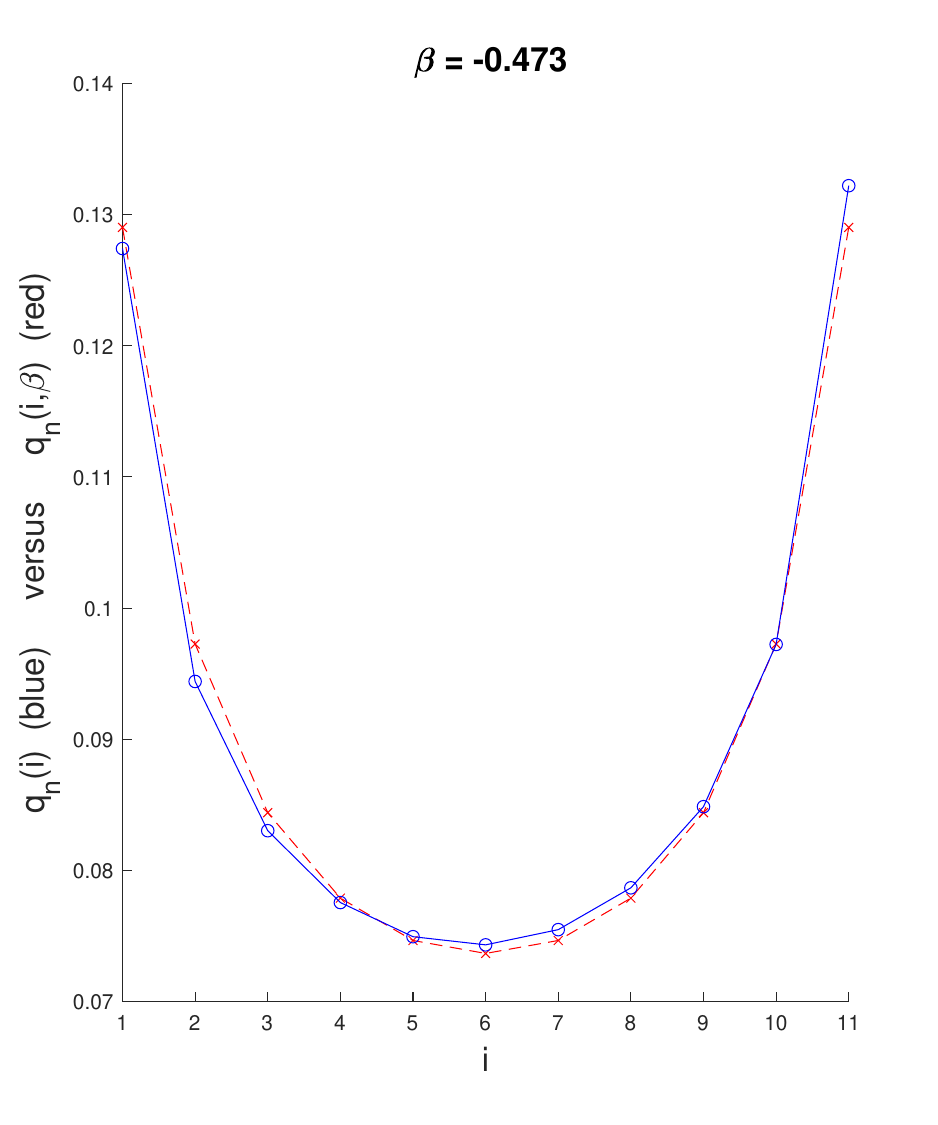}}
			\\
			\subfloat[]{\includegraphics[width=7.5 cm,height=7.5 cm]{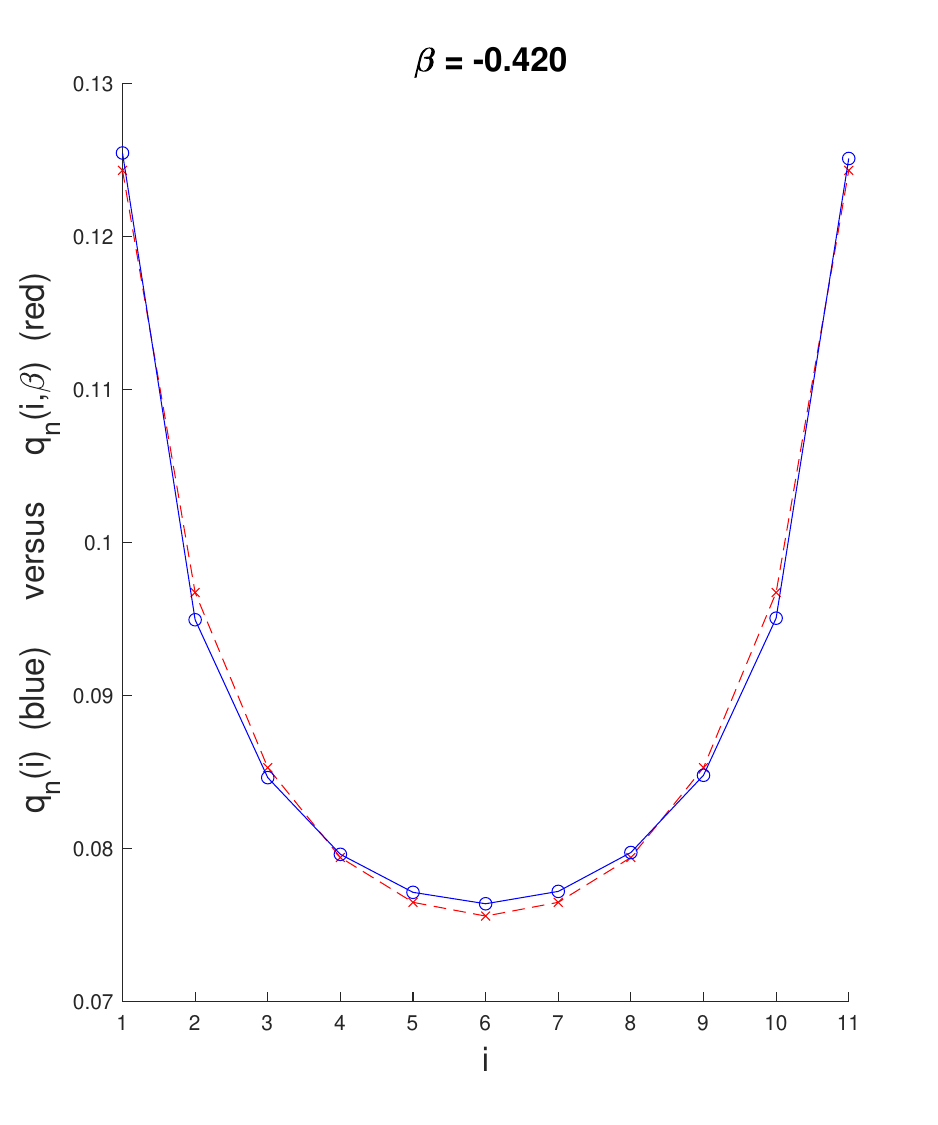}}
			\quad
			\subfloat[]{\includegraphics[width=7.5 cm,height=7.5 cm]{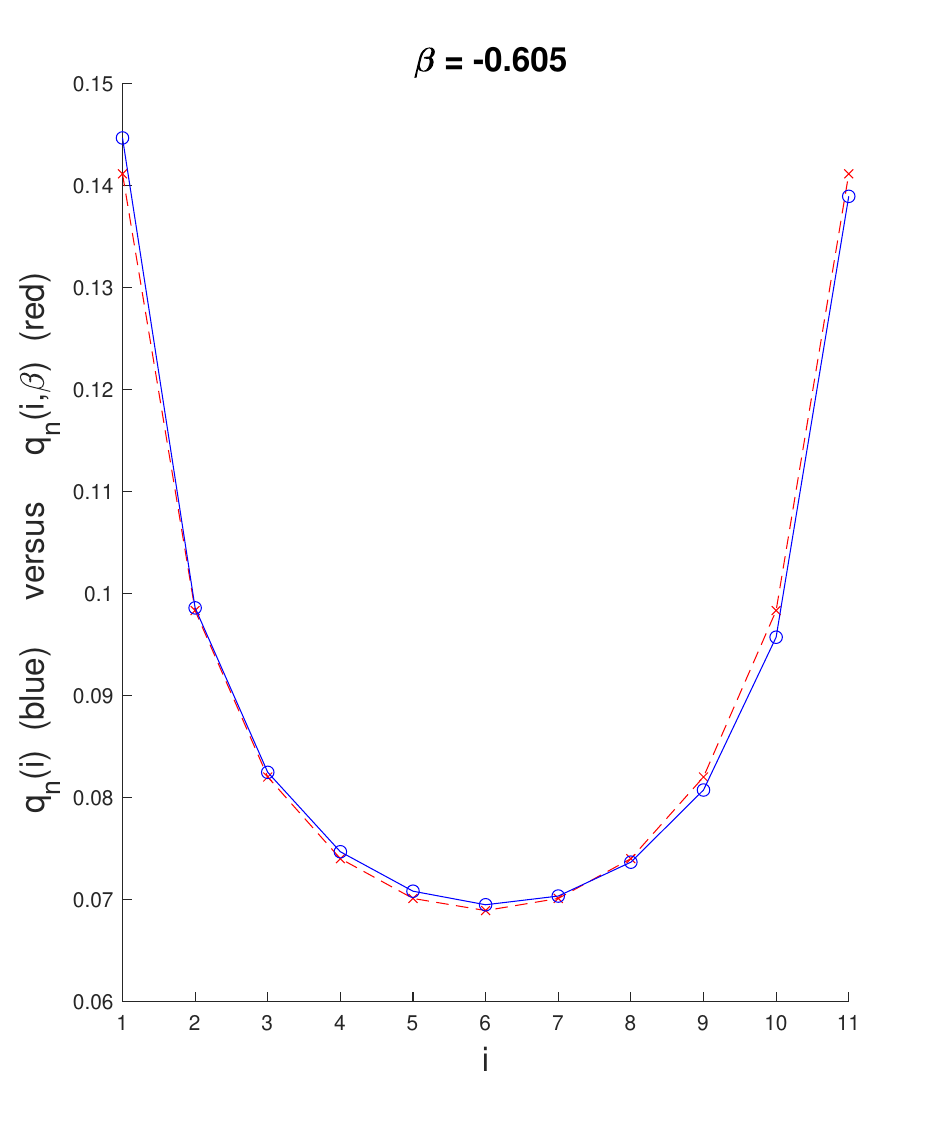}}	
			\caption{$q_n(i)$ (blue) derived for the QBD model discussed in Example~\ref{ex2bJ} versus $q_n(i,\beta)$ (red) obtained from Kullback-Leibler divergence in order to get the best fit $\beta$. We assume the parameters $u_d=u_c=1, u_r=0.3, u_f=0$ and $z_0=5$; and (a) $\delta=0.2$; (b) $\delta=0.8$; (c) $\gamma=0.2$; (d) $\gamma=0.8$. }
			\label{QBDbetagamma}
		\end{figure}

		We also examined the effect of $u_r$ on the tree balance and found that increasing $u_r$ results in smaller~$\beta$, see Figure~\ref{QBDucur0806}. This makes sense since a larger value of $u_r$ means that there is a higher chance of gene losing a function.
		\begin{figure}[hp]
			\centering
			\includegraphics[width=10 cm,height=10 cm]{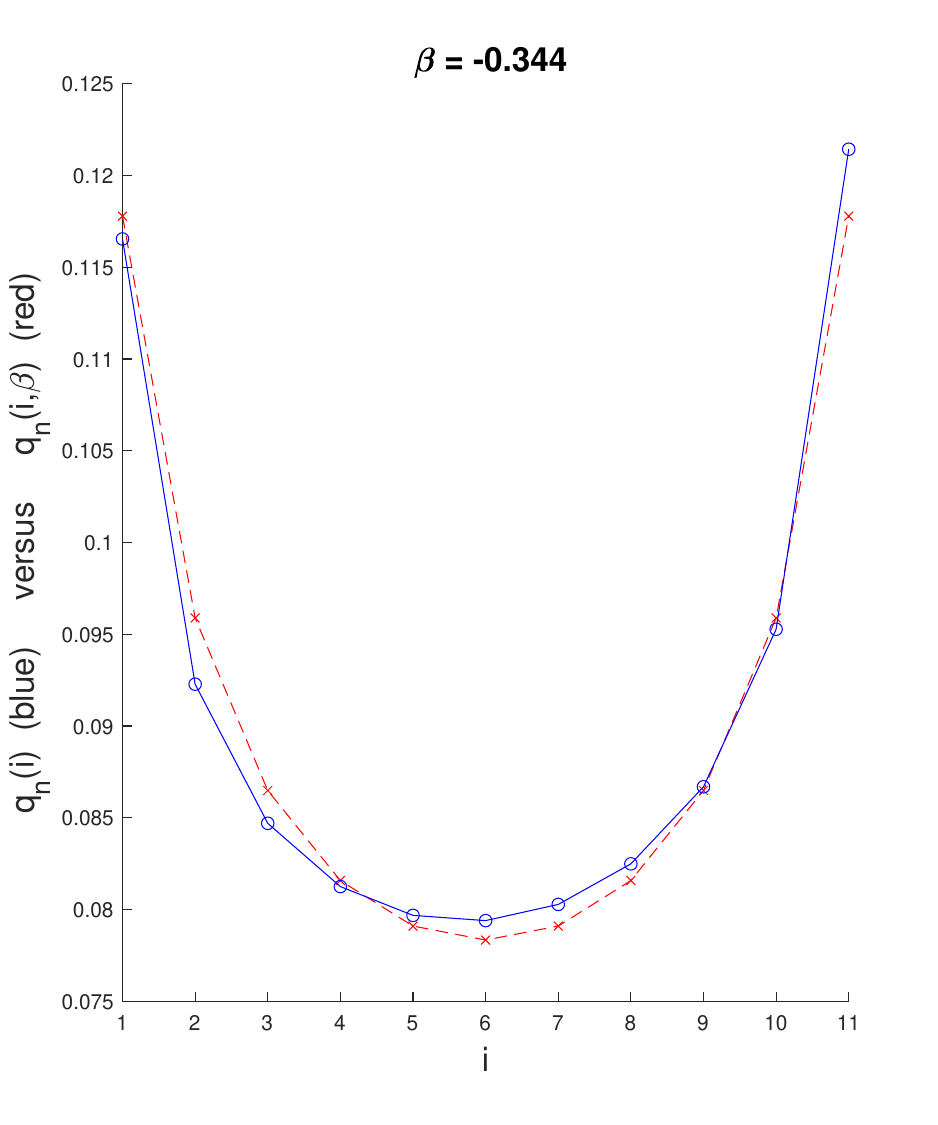}
			\caption{$q_n(i)$ (blue) derived for the QBD model discussed in Example~\ref{ex2bJ} versus $q_n(i,\beta)$ (red) obtained from Kullback-Leibler divergence in order to get the best fit $\beta$. We assume the parameters $u_d=1=u_c=1, u_r=0.4, u_f=0$ and $z_0=5$; and $\delta=0.8$.}
			\label{QBDucur0806}
		\end{figure}
		
		Finally, instead of $\epsilon=0$ and so $p_k=\frac{K-k}{K}$, we applied $\epsilon=\frac{1}{2K}$ resulting in the formula $p_k=\frac{2K-2k+1}{2K}$, see Figure~\ref{QBDgamma0208K}. There is a limited influence of the change of $p_k$ on $\beta$, since $p_k$ is the proportion of functions, per gene, that are permitted to be lost. It makes sense that $p_k$ does not directly change the tree balance but it does slightly affect the probability of the loss of the last function in a gene. This leads to a slightly unbalanced tree when compared to the tree under a model with $p_k=\frac{K-k}{K}$.
		\begin{figure}[hp]
			\centering
			\subfloat[]{\includegraphics[width=7.5 cm,height=7.5 cm]{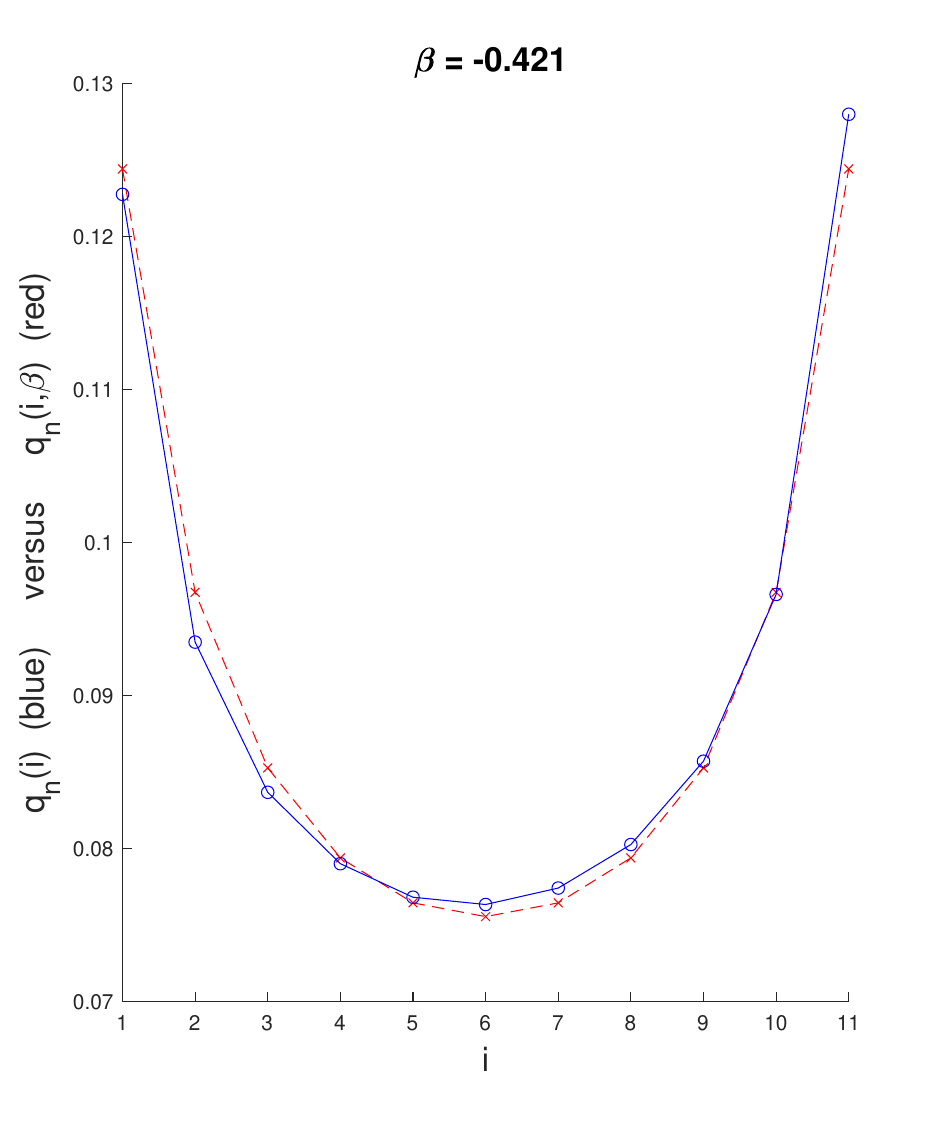}}
			\quad
			\subfloat[]{\includegraphics[width=7.5 cm,height=7.5 cm]{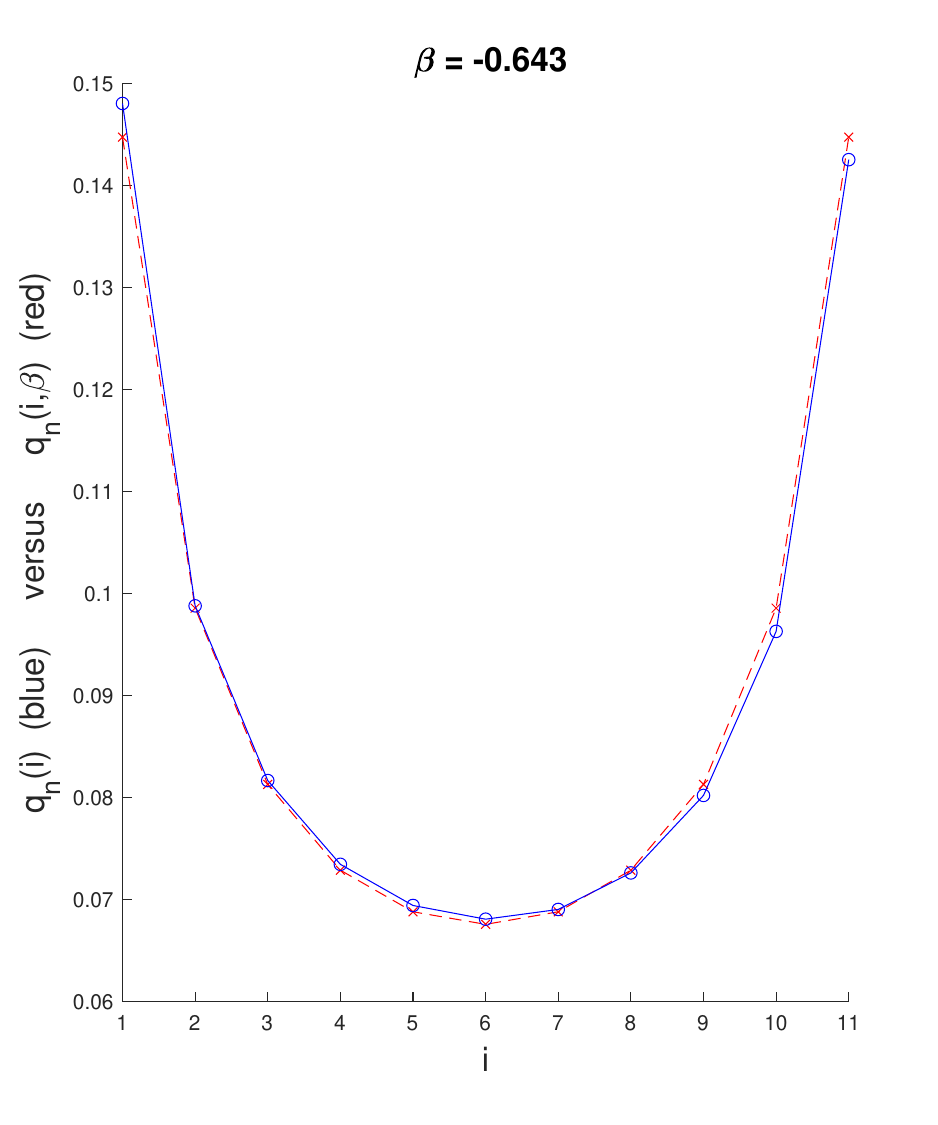}}
			\caption{$q_n(i)$ (blue) derived for the QBD model discussed in Example \ref{ex2bJ} versus $q_n(i,\beta)$ (red) obtained from Kullback-Leibler divergence in order to get the best fit $\beta$. We assume $p_k=\frac{2K-2k+1}{2K}$ and the parameters $u_d=u_c=1, u_r=0.3, u_f=0$ and $z_0=5$; and (a) $\gamma=0.2$ ; (b) $\gamma=0.8$. }
			\label{QBDgamma0208K}
		\end{figure}

	\end{Example}

	\section{Reconciliation}\label{sec:reconciliation}
	
\begin{figure}[H]
			\begin{center}
				\begin{tikzpicture}[>=stealth,redarr/.style={->}]	
				\draw (3.5,10.4) node[anchor=north, below=-0.17cm] {{\color{black} $T^*$ }};

				\draw [dashed] (-1.0,10) -- (6,10);
				\draw (-1.5,10) node[anchor=north, below=-0.3cm] {\scriptsize{\color{black} $t_0$ }};

				\draw [dashed] (-1.0,9) -- (6,9);
				\draw (-1.5,9) node[anchor=north, below=-0.3cm] {\scriptsize{\color{black} $t_1$ }};

				\draw [dashed] (-1.0,8) -- (6,8);
				\draw (-1.5,8) node[anchor=north, below=-0.3cm] {\scriptsize{\color{black} $t_2$ }};

				\draw [dashed] (-1.0,5.5) -- (6,5.5);
				\draw (-1.5,5.5) node[anchor=north, below=-0.3cm] {\scriptsize{\color{black} $t$ }};

				\draw [green,very thick] (3.5,10) -- (3.5,9);
				\draw [green,very thick] (1.5,9) -- (5.5,9);

				\draw [green,very thick] (5.5,9) -- (5.5,5.5);
			\draw (5.5,5.15) node[anchor=north, below=-0.17cm] {{\color{black} c }};
				
				\draw [green,very thick] (1.5,9) -- (1.5,8);

				\draw [green,very thick] (0.5,8) -- (2.5,8);

				\draw [green,very thick] (2.5,8) -- (2.5,5.5);
	\draw (2.5,5.25) node[anchor=north, below=-0.17cm] {{\color{black} b }};
				
			\draw [green,very thick] (0.5,8) -- (0.5,5.5);
			\draw (0.5,5.15) node[anchor=north, below=-0.17cm] {{\color{black} a }};

				\draw [decorate,
				decoration = {brace}] (6.5,8) --  (6.5,5.5);
				\draw (7,7) node[anchor=north, below=-0.17cm] {{\color{black} $2$ }};
				
				\draw [decorate,
				decoration = {brace}] (6.5,9) --  (6.5,8.2);
				\draw (7,8.7) node[anchor=north, below=-0.17cm] {{\color{black} $1$ }};

\draw [decorate,
decoration = {brace}] (6.5,10) --  (6.5,9.2);
\draw (7,9.7) node[anchor=north, below=-0.17cm] {{\color{black} $0$ }};
				
\draw [solid, very thick] (10,9) -- (13,9);	
\draw [solid, very thick] (10,9) -- (10,8);
\draw [solid, very thick] (13,9) -- (13,8);
\draw [solid, very thick] (9,8) -- (11,8);	
\draw [solid, very thick] (12,8) -- (14,8);
\draw [solid, very thick] (9,8) -- (9,7);
\draw [solid, very thick] (11,8) -- (11,7);
\draw [solid, very thick] (12,8) -- (12,7);
\draw [solid, very thick] (14,8) -- (14,7);	

\draw (9,6.7) node[anchor=north, below=-0.17cm] {{\color{black} a }};
\draw (11,6.75) node[anchor=north, below=-0.17cm] {{\color{black} b }};
\draw (12,6.75) node[anchor=north, below=-0.17cm] {{\color{black} b }};
\draw (14,6.7) node[anchor=north, below=-0.17cm] {{\color{black} c }};

\filldraw (10,8) circle (5pt);
\filldraw (12.85,7.85) rectangle (13.15,8.15);	
				
				\end{tikzpicture}
				\caption{On the left: Species tree with one gene in species a, two genes in species b, and one gene in species c. Speciation events have occurred at times $t_1$ and $t_2$. On the right: Gene tree to reconcile with the species tree. Only information about topology is given. Two events (circle, square) need to be allocated to the species tree as either speciation events at times $t_1,t_2$ or duplication events in time intervals $(t_0,t_1)$, $(t_1,t_2)$, $(t_2,t)$. Once allocated, additional gene loss events may need to be included to match the number of genes in each species. The likelihood of the resulting reconciliation can be computed using Algorithm~\ref{reconciliation_algorithm}. 
				}
				\label{fig:DPapproach}
			\end{center}
		\end{figure}
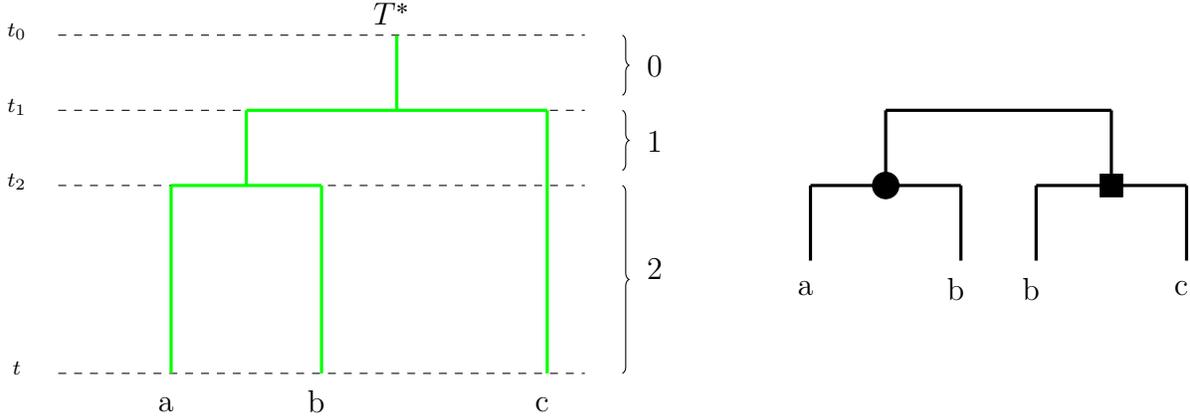

Below, we describe a method for fitting a gene tree to a given species tree, see Figure~\ref{fig:DPapproach}. We first consider a gene family tree of a {\em single species} 
and derive the likelihood of its various segments. We will use these expressions as building blocks of an iterative algorithm for the likelihood of reconciliation in a multiple-species tree. The key idea in the algorithm is to compute the likelihood recursively, by taking suitable products of the likelihood of the segments of the tree.	

	\begin{Definition}
		Let $T_G$ be a gene tree of one species that consists of internal branches ending with duplication events $D$, internal branches ending with loss of gene events $L$, and external branches which are yet to be absorbed into either $D$ or $L$.
	\end{Definition}

	Consider such defined gene tree $T_G$, modelled by the QBD $\{(Y(t),\varphi(t)):t\geq 0\}$  of Section~\ref{Sec:GenTree}. Let $\tau_0$ be the time at which the tree started from one branch and $(\tau-\tau_0)$ be the age of the tree. Suppose that $W$ events (duplication or loss of gene) in total have occurred on the tree and denote the consecutive times of events $\tau_1,\ldots,\tau_W$, with $\tau_0<\tau_1<\ldots <\tau_W<\tau$. Let $Y(\tau_w)$ be the number of genes in the family and $\varphi(\tau_w)$ be the phase observed immediately after the event at time $\tau_w$, for all $w=1,\ldots,W$. Then the likelihood of observing such gene tree is given by,
	\begin{eqnarray}
		f(T_G(\tau_0,\tau))_{\varphi(\tau_0),\ldots,\varphi(\tau_W),\varphi(\tau)}&=& 
		[\balpha]_{\varphi(\tau_0)} 
		\left( 
		\prod_{w=1}^{W}
		[e^{{\bf Q}^{[Y(\tau_{w-1}),Y(\tau_{w-1})]}(\tau_w-\tau_{w-1})}
		{\bf Q}^{[Y(\tau_{w-1}),Y(\tau_w)]}
		]_{\varphi(\tau_{w-1}),\varphi(\tau_w)}
		\right)
		\nonumber\\
		&&\times
		[e^{{\bf Q}^{[Y(\tau_W),Y(\tau_W)]}(\tau-\tau_W)}]_{\varphi(\tau_W),\varphi(\tau)}
		.
	\end{eqnarray} 
	
	We are also interested in the likelihood of observing the segment of the gene tree in the time interval $[u,v]$ for any $\tau_0\leq u<v\leq \tau$, which we use in the derivation of the reconciliation likelihood below. Denote by $a(t)=\min\{w:\tau_w>t\}$ the first event on the gene tree $T_G$ after time $t$ and by $b(t)=\max\{w:\tau_w<t\}$ the last event on the gene tree $T_G$ before time $t$, for any $t\in [\tau_0,\tau]$.

	Then, for any time points $u$, $v$ on the gene tree such that $\tau_0<u<v<\tau$, the likelihood of observing the segment of the gene tree in the time interval $[u,v]$ that starts in phase $\varphi(u)$ and ends in phase $\varphi(v)$ is given by,
	\begin{eqnarray}
		F(T_G(u,v))_{\varphi(u),\varphi(v)}
		&=&
		\Big[
		e^{{\bf Q}^{[Y(u),Y(u)]}(\tau_{a(u)}-u)}
		\prod_{w=a(u)}^{b(v)}
		e^{{\bf Q}^{[Y(\tau_{w-1}),Y(\tau_{w-1})]}(\tau_w-\tau_{w-1})}
		{\bf Q}^{[Y(\tau_{w-1}),Y(\tau_w)]}
		\nonumber\\
		&&
		\times
		e^{{\bf Q}^{[Y(\tau_{b(v)}),Y(\tau_{b(v)})]}(v-\tau_{b(v)})}
		\Big]_{\varphi(u),\varphi(v)},
		\label{Fterm}
	\end{eqnarray} 
	which we collect in a matrix ${\bf F}(T_G(u,v))=[F(T_G(u,v))_{\varphi(u),\varphi(v)}]$. In the case the times of the events are not known, but the order of the events (and hence the levels) in the time interval $[u,v]$ are known, we may evaluate the corresponding Laplace transform, 
	\begin{eqnarray}
		\widetilde{\bf F}(s)&=&
		\int_{z=0}^{\infty}e^{-sz}{\bf F}(u,u+z)dz
		\nonumber\\
		&=&
		\left(
		s{\bf I}-{\bf Q}^{[Y(u),Y(u)]}
		\right)^{-1}
		\prod_{w=a(u)}^{b(v)}
		\left(
		s{\bf I}-{\bf Q}^{[Y(\tau_{w-1}),Y(\tau_{w-1})]}
		\right)^{-1}
		\prod_{w=a(u)}^{b(v)}
		{\bf Q}^{[Y(\tau_{w-1}),Y(\tau_w)]}
		\nonumber\\
		&&
		\times
		\left(
		s{\bf I}-{\bf Q}^{[Y(\tau_{b(v)}),Y(\tau_{b(v)})]}
		\right)^{-1},
		\label{Fterm2}
	\end{eqnarray}
	and then invert it at $z=v-u$ using the algorithm by Abate and Whitt~\cite{abate1995numerical}, Den Iseger~\cite{DenIseger_2006}, or Horv{\'a}th et al.~\cite{horvath2020numerical}, to obtain ${\bf F}(u,v)$, and use it in place of ${\bf F}(T_G(u,v))$.

	Next, the likelihood of observing the segment of the gene tree in the time interval $[\tau_0,u]$ that ends in phase $\varphi(u)$ is given by,
	\begin{eqnarray}
		\lefteqn{
			f(T_G(\tau_0,u))_{\varphi(u)}
		}
		\nonumber\\
		&=&
		\Big[
		\balpha 
		\prod_{w=1}^{b(u)}
		e^{{\bf Q}^{[Y(\tau_{w-1}),Y(\tau_{w-1})]}(\tau_w-\tau_{w-1})}
		{\bf Q}^{[Y(\tau_{w-1}),Y(\tau_w)]}
		e^{{\bf Q}^{[Y(\tau_{b(u)}),Y(\tau_{b(u)})]}(u-\tau_{b(u)})}
		\Big]_{\varphi(u)},
		\label{fterm}
	\end{eqnarray}
	which we collect in a row vector ${\bf f}(T_G(\tau_0,u))=[f(T_G(\tau_0,u))_{\varphi(u)}]$. In the case the times of the events are not known, but the order of the events (and hence the levels) in the time interval $[\tau_0,u]$ are known, we may evaluate the corresponding Laplace transform, 
	\begin{eqnarray}
		\widetilde{\bf f}(s)&=&
		\int_{z=0}^{\infty}e^{-sz}{\bf f}(\tau_0,\tau_0+z)dz
		\nonumber\\
		&=&
		\balpha 
		\prod_{w=1}^{b(u)}
		\left(s{\bf I}-  {\bf Q}^{[Y(\tau_{w-1}),Y(\tau_{w-1})]}\right)^{-1}
		{\bf Q}^{[Y(\tau_{w-1}),Y(\tau_{w})]}
		\left(s{\bf I}-  {\bf Q}^{[Y(\tau_{b(u)}),Y(\tau_{b(u)})]}\right)^{-1},
		\label{fterm2}
	\end{eqnarray}
	and then invert it at $z=u-\tau_0$ using the algorithm by Abate and Whitt~\cite{abate1995numerical}, Den Iseger~\cite{DenIseger_2006}, or Horv{\'a}th et al.~\cite{horvath2020numerical}, to obtain ${\bf f}(\tau_0,u)$, and use it in place of ${\bf f}(T_G(\tau_0,u))$. In particular, we may apply 
	\begin{eqnarray}
		\left(s{\bf I}-{\bf Q}^{[1,1]} \right)^{-1}&=&
		\left(s+u_d \right)^{-1},
		\label{fterm3}
	\end{eqnarray}
	since since $\int_{u=0}^{\infty}e^{-su}u_d e^{-u_d u}=\left(s+u_d \right)^{-1}$ where $u_d$ is the duplication rate.

	Furthermore, the likelihood of observing the segment of the gene tree in the time interval $[v,\tau]$ that starts in phase $\varphi(v)$ is given by,
	\begin{eqnarray}
		\widehat{f}(T_G(v,\tau))_{\varphi(v)}
		&=&
		\Big[
		e^{{\bf Q}^{[Y(v),Y(v)]}(\tau_{a(u)}-v)}
		\prod_{w=a(v)}^{W}
		e^{{\bf Q}^{[Y(\tau_{w-1}),Y(\tau_{w-1})]}(\tau_w-\tau_{w-1})}
		{\bf Q}^{[Y(\tau_{w-1}),Y(\tau_w)]}
		\nonumber\\
		&&\times 
		e^{{\bf Q}^{[Y(\tau_W),Y(\tau_W)]}(\tau-\tau_W)}
		{\bf 1}
		\Big]_{\varphi(v)},
		\label{fhatterm}
	\end{eqnarray}
	which we collect in a column vector $\widehat{\bf f}(T_G(v,\tau))=[\widehat{f}(T_G(v,\tau))_{\varphi(v)}]$. In the case the times of the events are not known, but the order of the events (and hence the levels) in the time interval $[v,\tau]$ are known, we may evaluate the corresponding Laplace transform, 
	\begin{eqnarray}
		\widetilde{\widehat{\bf f}}(s)&=&
		\int_{z=0}^{\infty}e^{-sz}\widehat{\bf f}(v,v+z)dz
		\nonumber\\
		&=&
		\left(
		s{\bf I}-{\bf Q}^{[Y(v),Y(v)]}
		\right)^{-1}
		\prod_{w=a(v)}^{W}
		\left(
		s{\bf I}-{\bf Q}^{[Y(\tau_{w-1}),Y(\tau_{w-1})]}
		\right)^{-1}
		{\bf Q}^{[Y(\tau_{w-1}),Y(\tau_w)]}
		\nonumber\\
		&&\times
		\left(
		s{\bf I}-{\bf Q}^{[Y(\tau_W),Y(\tau_W)]}
		\right)^{-1}
		{\bf 1}
		,
		\label{fhatterm2}
	\end{eqnarray}
	and then invert it at $z=\tau-v$ using the algorithm by Abate and Whitt~\cite{abate1995numerical}, Den Iseger~\cite{DenIseger_2006}, or Horv{\'a}th et al.~\cite{horvath2020numerical} to obtain $\widehat{\bf f}(v,\tau)$, and use it in place of $\widehat{\bf f}(T_G(v,\tau))$.

	\subsection{The likelihood of a reconciliation}\label{sec:likform}
	
	Now, consider the likelihood of a species-gene tree with multiple species. Let $T^*$ be a reconstructed species tree, modelled by the MBT $\{(M(t),\varphi(t)):t\geq 0\}$  of Section~\ref{sec:speciestree}. Let $t_0$ be the time at which the tree started from one branch and $(t-t_0)$ be the age of the tree. Suppose that $N$ speciation events in total have occurred on the tree and denote the consecutive times of events $t_1,\ldots,t_N$, with $t_0<t_1<\ldots <t_N<t$, and let $t=t_{N+1}$.

	Denote by ${T^*}_G^{(0)},\ldots,{T^*}_G^{(N)}$ the collection of reconstructed gene trees corresponding to the $N+1$ different species born at times $t_0,\ldots,t_N$, with ${T^*}_G^{(0)}$ being the reconstructed gene tree of the parent species, in some proposed reconciliation. 
	
	Further, let $T^*_G$ be some species-gene tree constructed by embedding gene trees ${T^*}_G^{(0)},\ldots,{T^*}_G^{(N)}$ into the species tree $T^*$. We divide the species-gene tree $T^*_G$ into subtrees as follows. Denote by ${T^*}_{G}^{(k,n)}$ the $(k,n)$-th subtree that starts at time $t_n$ and ends at time $t$, $n=0,\ldots,N$, counting $k=0,1,\ldots ,n$ from the left to the right on the species-gene tree. Then, ${\bf f}({T^*}_{G}^{(0,0)}(t_0,t_1))$ is the likelihood row vector of observing the segment of the $(0,0)$-th subtree in the time interval $[t_0,t_1]$, ${\bf F}({T^*}_{G}^{(k,n)}(t_n,t_{n+1}))$ is the likelihood matrix of observing the segment of the $(k,n)$-th subtree in the time interval $[t_n,t_{n+1}]$ for $n=1,\ldots,N-1$, and $\widehat{\bf f}({T^*}_{G}^{(k,N)}(t_N,t))$ is the likelihood column vector of observing the segment of the $(k,N)$-th subtree in the time interval $[t_N,t]$.

	Then the likelihood of such given reconciliation is given by the following recursive formula,
	\begin{eqnarray}
		\lefteqn{
			f(T^*;{T^*}_G^{(0)},\ldots,{T^*}_G^{(N)};
			(t_0,t_1,\ldots,t_N))
			=
			{\bf f}({T^*}_G^{(0)}(t_0,t_1))
		}
		\nonumber\\
		&&
		\times
		\left({\bf f}({T^*}_G^{(0,1)}(t_1,t)) \odot {\bf f}({T^*}_G^{(1,1)}(t_1,t))\right),
		\label{rec_all}
	\end{eqnarray}
	where, for $n=1,\ldots,N-1$, assuming that the speciation at time $t_n$ occurs on the $\ell_n$-th subtree ${T^*}_{G}^{(\ell_n,n-1)}$ for some $\ell_n=0,\ldots,n-1$,
	\begin{eqnarray}
		{\bf f}({T^*}_G^{(k,n)}(t_n,t))&=&
		\left\{
		\begin{array}{ll}
			{\bf F}({T^*}_G^{(k,n)}(t_n,t_{n+1}))
			{\bf f}({T^*}_G^{(k,n+1)}(t_{n+1},t))
			& \quad k< \ell_n,\bigskip
			\\
			\left(
			{\bf F}({T^*}_G^{(k,n)}(t_n,t_{n+1}))
			{\bf f}({T^*}_G^{(k,n+1)}(t_{n+1},t))
			\right)
			&\\
			\odot
			\left(
			{\bf F}({T^*}_G^{(k +1,n)}(t_n,t_{n+1}))
			{\bf f}({T^*}_G^{(k +1,n+1)}(t_{n+1},t))
			\right)
			& \quad k= \ell_n,\bigskip
			\\
			{\bf F}({T^*}_G^{(k,n)}(t_n,t_{n+1}))
			{\bf f}({T^*}_G^{(k+1,n+1)}(t_{n+1},t))
			& \quad k> \ell_n,
		\end{array}	
		\right.
		\label{rec_n}
	\end{eqnarray}
	and
	\begin{eqnarray}
		{\bf f}({T^*}_G^{(k,N)}(t_N,t))&=&
		\left\{
		\begin{array}{ll}
			\widehat{\bf f}({T^*}_{G}^{(k,N)}(t_N,t))
			& \quad k< \ell_N,\bigskip
			\\
			\widehat{\bf f}({T^*}_G^{(k,N)}(t_N,t))
			\odot
			\widehat{\bf f}({T^*}_G^{(k+1,N)}(t_N,t))
			& \quad k= \ell_N,\bigskip
			\\
			\widehat{\bf f}({T^*}_{G}^{(k,N)}(t_N,t))
			& \quad k> \ell_N,
		\end{array}	
		\right.
		\label{rec_N}
	\end{eqnarray}
	where the individual terms of the products above are given by~\eqref{Fterm}-\eqref{fhatterm2}.
	
	In practice, we compute the reconciliation likelihood  $f(T^*;{T^*}_G^{(0)},\ldots,{T^*}_G^{(N)};
	(t_0,t_1,\ldots,t_N))$ by applying the recursion in the direction from the tips to the root, using Algorithm~\ref{reconciliation_algorithm} below.

	\begin{algorithm}
		\caption{Compute $f(T^*;{T^*}_G^{(0)},\ldots,{T^*}_G^{(N)};
			(t_0,t_1,\ldots,t_N))$ }
		\label{reconciliation_algorithm}
		\begin{algorithmic}[1] %
			\For{$k = 0, \hdots, N$}
			\State Compute $f({T^*}_G^{(k,N)}(t_N,t))$ using~\eqref{rec_N}. 
			\EndFor
			\For{$n = N-1, \hdots, 1$}
			\For{$k = 0, \hdots, n$}
			\State Compute $f({T^*}_G^{(k,n)}(t_n,t))$ using~\eqref{rec_n}.
			\EndFor
			\EndFor
			\State Compute $f(T^*;{T^*}_G^{(0)},\ldots,{T^*}_G^{(N)};
			(t_0,t_1,\ldots,t_N))$ using~\eqref{rec_all}.
		\end{algorithmic}
	\end{algorithm}

	\subsection{Example}\label{sec:RecEx}

	\begin{figure}[H]
		\centering
		\includegraphics[width=9 cm,height=6 cm]{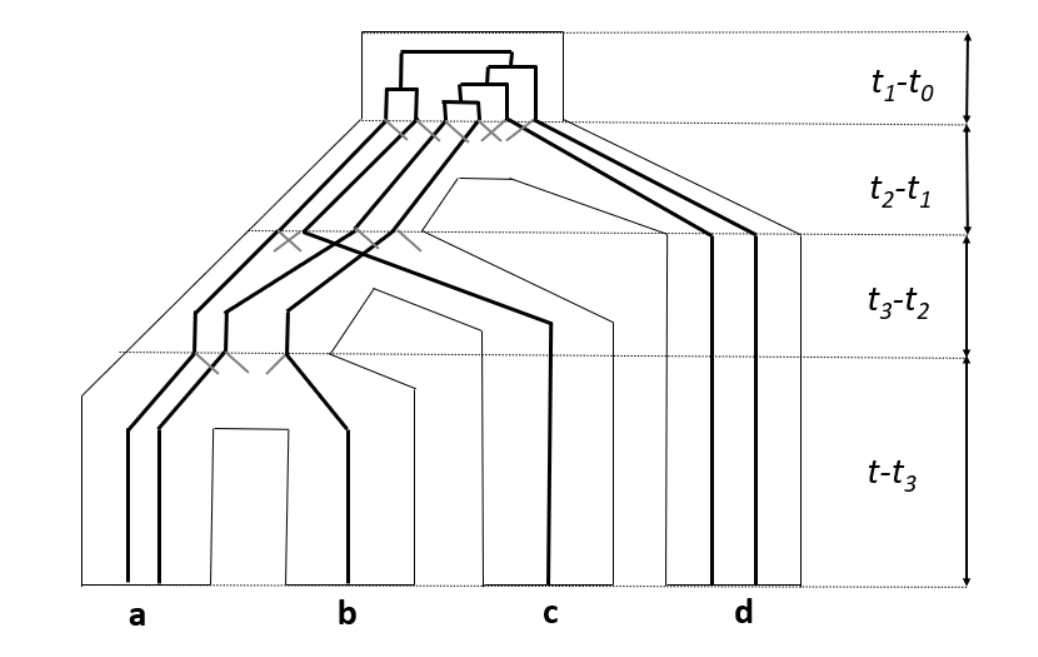}
		\includegraphics[width=9 cm,height=6 cm]{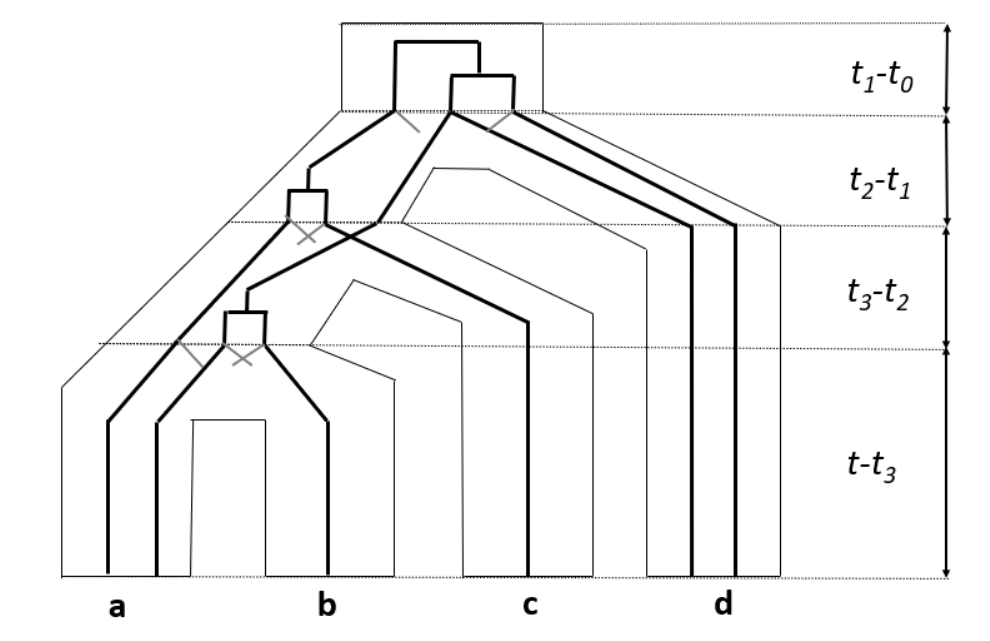}\\[3ex]
		\includegraphics[width=9 cm,height=6 cm]{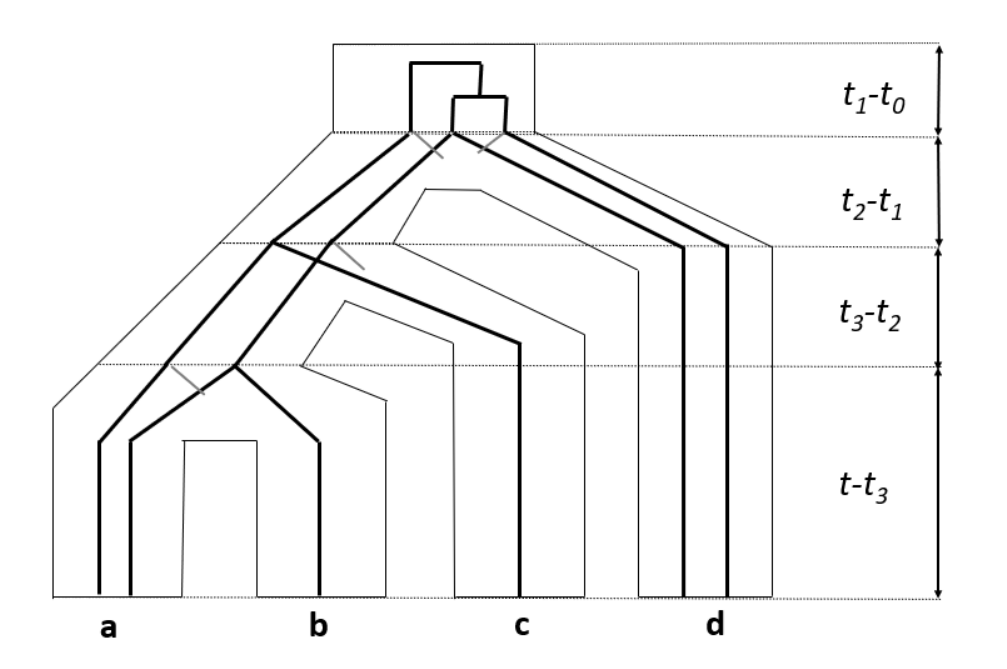}
		\quad\quad\quad
		\begin{tikzpicture}[every tree node/.style={font=\normalsize},
		level distance=1.2cm,sibling distance=.8cm, 
		edge from parent path={(\tikzparentnode.south) -- +(0, 0pt) -| (\tikzchildnode)}],
		frontier/.style={distance from root=350 pt} 
		
		\Tree 
		[
		\edge node[near end,left] {};
		[[
		\edge node[near end,left] {};
		[
		\edge node[near end,left] {};
		\edge node[] {}; [.a ]
		]]
		\edge node[near end,right] {};
		[
		\edge node[near end,right] {};
		[
		\edge node[near end,right] {};     
		\edge node[] {}; [.c ]
		]]]
		\edge node[near end,right] {};
		[   
		\edge node[near end, left] {}; 
		[
		\edge node[near end, left] {};
		[
		\edge node[] {}; [.a ]
		\edge node[] {}; [.b ]
		]
		[
		\edge node[near end,right] {};
		\edge node[] {}; [.d ]
		]]
		[
		\edge node[near end,right] {};
		[
		\edge node[near end,right] {};
		\edge node[] {}; [.d ]
		]]]]
		\node[draw] at (0,.5) {Gene tree $G$};
		\end{tikzpicture}
		\caption{Species-gene tree E1 (top left), E17 (top right), and E21 (bottom) based on~\cite[Figure~1]{gorecki2014drml}. Black lines represent evolution of genes due to gene duplication events (which may occur inside some time interval) or speciation events (occurring at the speciation times $t_1,t_2,t_3$). Grey lines represent gene loss. In E1, there are $5$ gene duplications in the time interval $[t_0,t_1]$, and later $13$ gene loss events in total. In E17, there are $2$ gene duplications in $[t_0,t_1]$, one in $[t_1, t_2]$ and one in $[t_2,t_3]$; as well as $7$ gene loss events in total. In E21, there are $2$ gene duplications in $[t_0,t_1]$, and $4$ gene loss events in total. }
		\label{E1E17Tree}
	\end{figure}
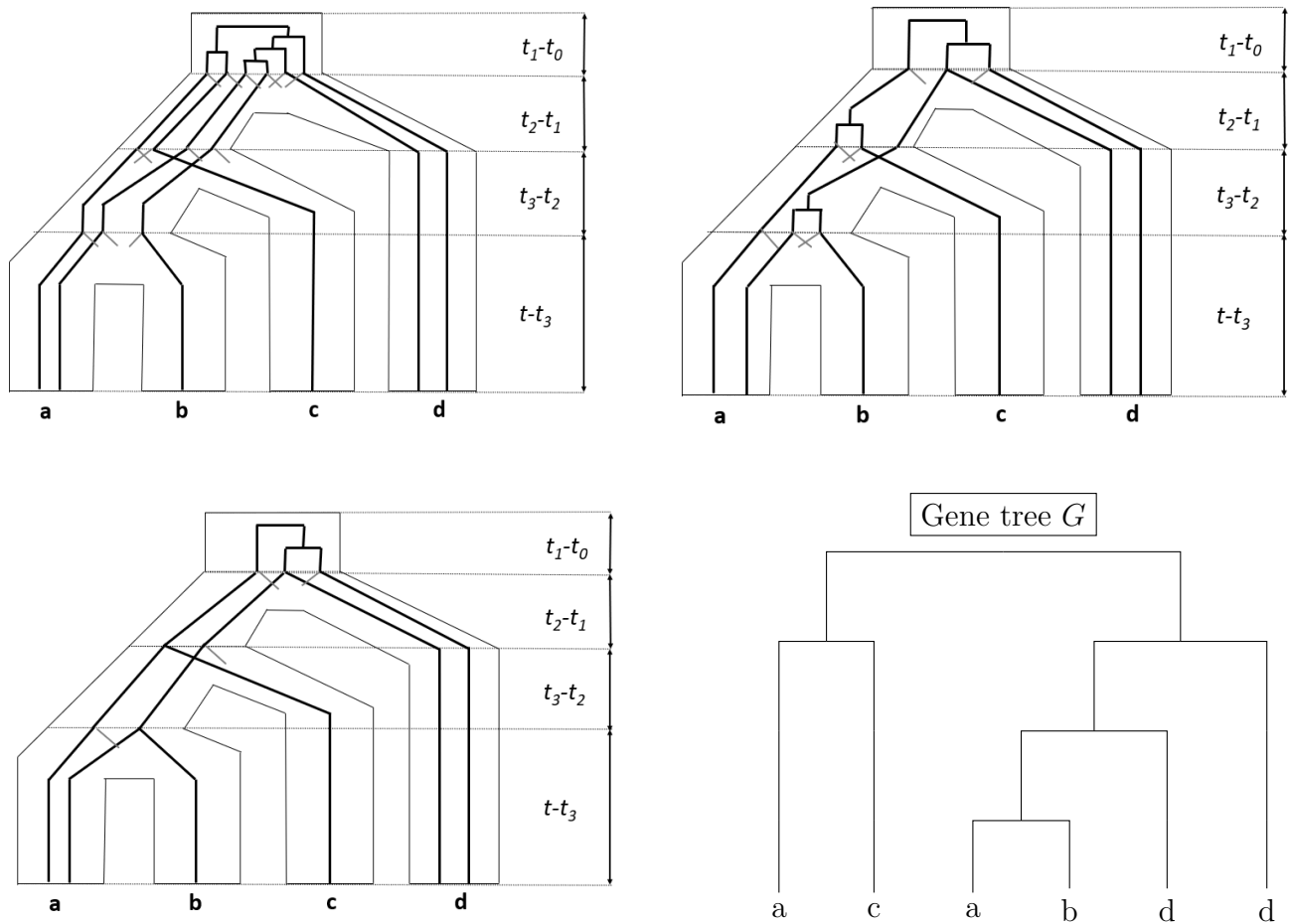

	In order to illustrate our methodology, we now compute the likelihoods of the three species-gene trees represented in Figure~\ref{E1E17Tree}, based on G{\'o}recki~\cite[E1, E17 and E21 in Figure 1]{gorecki2014drml}. The figure shows three examples of reconciliation such that species $a$ has two genes, species $b$ has one gene, species $c$ has one gene, and species $d$ has two genes. We assume that all genes are from the same family that started with one gene at time $t_0$ and then evolved until time $t$. Speciation events were at times $t_1$, $t_2$, and $t_3$.

	Since the species tree has the same shape in each of the three species-gene trees, we apply the following formula for the likelihood of each of them, using equations~\eqref{rec_all}, \eqref{rec_n}, and \eqref{rec_N}. We have, 
	\begin{eqnarray}\label{eq:likelihood}
		\lefteqn{
			f(T^{*};T^{*(0)}_G,T^{*(1)}_G,T^{*(2)}_G,T^{*(3)}_G;(t_0,t_1,t_2,t_3))=\textbf{f}(T^{*(0)}_G(t_0,t_1))
		}
		\nonumber\\ 
		&&\times
		\Big\{
		\Big[
		\textbf{F}(T^{*(0,1)}_G(t_1,t_2))
		\nonumber\\ 
		&&
		\quad
		\times 
		\Bigg(
		\left(
		\textbf{F}(T^{*(0,2)}_G(t_2,t_3))
		\times
		(\widehat{\textbf{f}}(T^{*(0,3)}_G(t_3,t))\odot\widehat{\textbf{f}}(T^{*(1,3)}_G(t_3,t)))
		\right)
		\nonumber\\ 
		&&
		\quad
		\odot 
		\left(
		\textbf{F}(T^{*(1,2)}_G(t_2,t_3)\times\widehat{\textbf{f}}(T^{*(2,3)}_G(t_3,t))
		\right)
		\Bigg)
		\Big]
		\nonumber\\ 
		&&
		\odot
		\left[\textbf{F}(T^{*(1,1)}_G(t_1,t_2))\times\textbf{F}(T^{*(2,2)}_G(t_2,t_3))\times\widehat{\textbf{f}}(T^{*(3,3)}_G(t_3,t))
		\right]
		\Big\},
	\end{eqnarray}
	where, by~\cite[Figure 1]{gorecki2014drml}, we have $t_1-t_0=2$, $t_2-t_1=1$, $t_3-t_2=8$ and $t-t_3=9$, while $t_0$ is unknown. Also recall that to evaluate the likelihood of the reconstructed species tree, we may apply the methodology discussed in Example~\ref{ex:Gorecki} in Section~\ref{sec:recspetree} (Figure~\ref{fig:Example2}).
	
	The log-likelihood value of species-gene trees E1, E17 and E21, computed using the expression in~\eqref{eq:likelihood}, is $\log \mathcal{L}(E1)=-73.96$, $\log \mathcal{L}(E17)=-58.63$, and $\log \mathcal{L}(E21)=-26.04$, respectively. That is, tree E21, which has the smallest number of duplication events (and the largest number of necessary extinction events in order to match the gene tree), has the highest likelihood, $\mathcal{L}(E21)=exp(-26.04)$.

\section{Conclusion}\label{sec:concl}

We developed theoretical results for the gene-tree species-tree reconciliation problem. We built our analysis on Markovian models and efficient techniques from the theory of matrix-analytic methods~\cite{latouche2012matrix,neuts1981matrix}. To model the reconstructed species tree, we applied a Markovian Binary Tree~\cite{2015S,2014HF,2011H,2009HLR,Kontoleon}, while to model subfunctionalisation in the evolution of the gene tree, we applied Quasi-Birth-and-Death process~\cite{asmussen2003applied,latouche2010level,latouche1999introduction,latouchet1998invariant}. We derived novel results for the application of these models to incomplete data and provided an algorithm for the computation of the likelihood of reconciliation.

Recently, a set of monocot gene families with speciation events, whole genome duplication events, and smaller scale genome duplication events annotated in the gene trees based upon a reference species tree has been described in~\cite{DLibetal}. Datasets like this are ideal for the application of novel methods like the one described here. In the future, we will apply these theoretical results for the analysis of such multi-species gene family data.

\section{Acknowledgements}

We would like to thank the Australian Research Council for funding this research through Discovery Project DP180100352.


Sections~\ref{sec:speciestree}-~\ref{sec:recspetree} have contributed to a chapter in the PhD thesis by Soewongsono~\cite{Soewongsono}. Sections~\ref{Sec:GenTree}-\ref{sec:reconciliation}  have contributed to a chapter in the PhD thesis by Diao~\cite{Diao}.

The following are the contributions of the authors, Albert C. Soewongsono (ACS), Barbara R. Holland (BRH), Ma\l gorzata M. O'Reilly (MMO), Jiahao Diao (JD), Tristan L. Stark (TLS), David A. Liberles (DAL), Amanda Wilson (AW): 
\begin{itemize}
	\item Sections~\ref{sec:speciestree}-\ref{sec:recspetree}, Problem formulation and methodology development: ACS, BRH, and MMO;
	\item Sections~\ref{sec:speciestree}-\ref{sec:recspetree}, Derivation and proofs of the new expressions for the MBTs: ACS and MMO;
	\item Section~\ref{sec:recspetree}, Coding: ACS, BRH, and MMO;
	\item Section~\ref{sec:recspetree}, Numerical analysis: ACS;
	\item Sections~\ref{Sec:GenTree}-\ref{sec:reconciliation}, Problem formulation and methodology development: JD, TS, BRH, and MMO;
	\item Sections-\ref{sec:reconciliation}-\eqref{sec:likform}, Derivation of the new expressions and an algorithm for the reconciliation: MMO;
	\item Sections~\ref{Sec:GenTree}-\ref{sec:reconciliation}, Coding: JD, BRH, and MMO;
	\item Sections~\ref{Sec:GenTree}-\ref{sec:reconciliation}, Numerical analysis: JD;
	\item 
	Conceptualisation, Biological background: DAL, AW;
	\item 
	Conceptualisation, Mathematical background: BRH, MMO.
\end{itemize}

	\bibliographystyle{abbrv}
	\bibliography{all_phylo}

\begin{thebibliography}{10}

\bibitem{abate1995numerical}
J.~Abate and W.~Whitt.
\newblock Numerical inversion of laplace transforms of probability
  distributions.
\newblock {\em ORSA Journal on computing}, 7(1):36--43, 1995.

\bibitem{1996A}
D.~Aldous.
\newblock Probability distributions on cladograms.
\newblock In D.~Aldous and R.~Pemantle, editors, {\em Random Discrete
  Structures}, pages 1--18, New York, NY, 1996. Springer New York.

\bibitem{anisimova2013state}
M.~Anisimova, D.~A. Liberles, H.~Philippe, J.~Provan, T.~Pupko, and A.~von
  Haeseler.
\newblock State-of the art methodologies dictate new standards for phylogenetic
  analysis.
\newblock {\em BMC evolutionary biology}, 13(1):1--8, 2013.

\bibitem{asmussen2003applied}
S.~Asmussen.
\newblock {\em Applied probability and queues}, volume~2.
\newblock Springer, 2003.

\bibitem{cavalli1967phylogenetic}
L.~L. Cavalli-Sforza and A.~W. Edwards.
\newblock Phylogenetic analysis. models and estimation procedures.
\newblock {\em American journal of human genetics}, 19(3 Pt 1):233, 1967.

\bibitem{DenIseger_2006}
P.~Den~Iseger.
\newblock Numerical transform inversion using gaussian quadrature.
\newblock {\em Probability in the Engineering and Informational Sciences},
  20(1):1--44, 2006.

\bibitem{Diao}
J.~Diao.
\newblock {\em Mechanistic {M}arkov Models for the Evolution of Gene Families}.
\newblock PhD thesis, The University of Tasmania, 2023.

\bibitem{2020DOH}
J.~Diao, M.~M. O’Reilly, and B.~Holland.
\newblock A subfunctionalisation model of gene family evolution predicts
  balanced tree shapes.
\newblock {\em Molecular Phylogenetics and Evolution}, 176:107566, 2022.

\bibitem{2020DSLOH}
J.~Diao, T.~L. Stark, D.~A. Liberles, M.~M. O'Reilly, and B.~R. Holland.
\newblock Level-dependent {QBD} models for the evolution of a family of gene
  duplicates.
\newblock {\em Stochastic Models}, 36(2):285--311, 2020.

\bibitem{duchen2021effect}
P.~Duchen, M.~L. Alfaro, J.~Rolland, N.~Salamin, and D.~Silvestro.
\newblock On the effect of asymmetrical trait inheritance on models of trait
  evolution.
\newblock {\em Systematic Biology}, 70(2):376--388, 2021.

\bibitem{felsenstein1973maximum}
J.~Felsenstein.
\newblock Maximum-likelihood estimation of evolutionary trees from continuous
  characters.
\newblock {\em American journal of human genetics}, 25(5):471, 1973.

\bibitem{fitzjohn2010quantitative}
R.~G. FitzJohn.
\newblock Quantitative traits and diversification.
\newblock {\em Systematic biology}, 59(6):619--633, 2010.

\bibitem{2012F}
R.~G. Fitz{J}ohn.
\newblock Diversitree: Comparative phylogenetic analyses of diversification in
  {R}.
\newblock {\em Methods in Ecology and Evolution}, 3(6):1084--1092, 2012.

\bibitem{gorecki2014drml}
P.~G{\'o}recki and O.~Eulenstein.
\newblock {DrML}: probabilistic modeling of gene duplications.
\newblock {\em Journal of Computational Biology}, 21(1):89--98, 2014.

\bibitem{10.1111/evo.12653}
K.~M. Gotanda, C.~Correa, M.~M. Turcotte, G.~Rolshausen, and A.~P. Hendry.
\newblock {Linking macrotrends and microrates: Re-evaluating microevolutionary
  support for Cope's rule}.
\newblock {\em Evolution}, 69(5):1345--1354, 05 2015.

\bibitem{grant1999ecology}
P.~R. Grant.
\newblock {\em Ecology and evolution of Darwin's finches}.
\newblock Princeton University Press, 1999.

\bibitem{Guo2015}
P.~C. Guo and S.~F. Xu.
\newblock The modified {N}ewton–{S}hamanskii method for the solution of a
  quadratic vector equation arising in markovian binary trees.
\newblock {\em Calcolo}, 52(3):317--325, 2015.

\bibitem{hagen2015age}
O.~Hagen, K.~Hartmann, M.~Steel, and T.~Stadler.
\newblock Age-dependent speciation can explain the shape of empirical
  phylogenies.
\newblock {\em Systematic biology}, 64(3):432--440, 2015.

\bibitem{2015S}
S.~Hautphenne.
\newblock A structured {M}arkov chain approach to branching processes.
\newblock {\em Stochastic Models}, 31(3):403--432, 2015.

\bibitem{2014HF}
S.~Hautphenne and M.~Fackrell.
\newblock An {EM} algorithm for the model fitting of {M}arkovian binary trees.
\newblock {\em Computational Statistics and Data Analysis}, 70:19--34, 2014.

\bibitem{2009HLR}
S.~Hautphenne, G.~Latouche, and M.-A. Remiche.
\newblock Transient features for {M}arkovian binary trees.
\newblock In {\em Proceedings of the Fourth International ICST Conference on
  Performance Evaluation Methodologies and Tools}, 2009.

\bibitem{2011H}
S.~Hautphenne, G.~Latouche, and M.-A. Remiche.
\newblock Algorithmic approach to the extinction probability of branching
  processes.
\newblock {\em Methodology and Computing in Applied Probability},
  13(1):171--192, 2011.

\bibitem{DLibetal}
C.~N. Henry, K.~Piper, A.~E. Wilson, J.~L. Miraszek, C.~S. Probst, Y.~Rong, and
  D.~A. Liberles.
\newblock {WGDT}ree: a phylogenetic software tool to examine conditional
  probabilities of retention following whole genome duplication events.
\newblock {\em BMC Bioinformatics}, 23(1):505, 2022.

\bibitem{hermansen2017adaptive}
R.~A. Hermansen, B.~P. Oswald, S.~Knight, S.~D. Shank, D.~Northover, K.~L.
  Korunes, S.~N. Michel, and D.~A. Liberles.
\newblock The adaptive evolution database ({TAED}): a new release of a database
  of phylogenetically indexed gene families from chordates.
\newblock {\em Journal of Molecular Evolution}, 85(1):46--56, 2017.

\bibitem{horvath2020numerical}
G.~Horv{\'a}th, I.~Horv{\'a}th, S.~A.-D. Almousa, and M.~Telek.
\newblock Numerical inverse {L}aplace transformation using concentrated matrix
  exponential distributions.
\newblock {\em Performance Evaluation}, 137:102067, 2020.

\bibitem{joyner2016new}
J.~Joyner and B.~Fralix.
\newblock A new look at {M}arkov processes of {G/M/1}-type.
\newblock {\em Stochastic Models}, 32(2):253--274, 2016.

\bibitem{Kontoleon}
N.~Kontoleon.
\newblock {\em The Markovian binary tree : A model of the macroevolutionary
  process}.
\newblock PhD thesis, The University of Adelaide,
  \url{http://hdl.handle.net/2440/22320}, 2006.

\bibitem{kullback1951information}
S.~Kullback and R.~A. Leibler.
\newblock On information and sufficiency.
\newblock {\em The annals of mathematical statistics}, 22(1):79--86, 1951.

\bibitem{lambert2013birth}
A.~Lambert and T.~Stadler.
\newblock Birth--death models and coalescent point processes: The shape and
  probability of reconstructed phylogenies.
\newblock {\em Theoretical population biology}, 90:113--128, 2013.

\bibitem{latouche2010level}
G.~Latouche.
\newblock Level-independent quasi-birth-and-death processes.
\newblock {\em Wiley encyclopedia of operations research and management
  science}, 2010.

\bibitem{latouche1999introduction}
G.~Latouche and V.~Ramaswami.
\newblock {\em Introduction to matrix analytic methods in stochastic modeling}.
\newblock SIAM, 1999.

\bibitem{latouche2012matrix}
G.~Latouche, V.~Ramaswami, J.~Sethuraman, K.~Sigman, M.~S. Squillante, and
  D.~Yao.
\newblock {\em Matrix-Analytic Methods in Stochastic Models}, volume~27.
\newblock Springer Science \& Business Media, 2012.

\bibitem{latouchet1998invariant}
G.~Latouchet, C.~E. Pearce, and P.~G. Taylor.
\newblock Invariant measures for quasi-birth-and-death processes.
\newblock {\em Stochastic Models}, 14(1-2):443--460, 1998.

\bibitem{2007MMO}
W.~P. Maddison, P.~E. Midford, and S.~P. Otto.
\newblock Estimating a binary character's effect on speciation and extinction.
\newblock {\em Systematic Biology}, 56(5):701--710, 2007.

\bibitem{mayr1982speciation}
E.~Mayr.
\newblock Speciation and macroevolution.
\newblock {\em Evolution}, 36(6):1119--1132, 1982.

\bibitem{mayr1999systematics}
E.~Mayr.
\newblock {\em Systematics and the origin of species, from the viewpoint of a
  zoologist}.
\newblock Harvard University Press, 1999.

\bibitem{neeetal}
S.~Nee, E.~Holmes, R.~May, and P.~Harvey.
\newblock {Extinction rates can be estimated from molecular phylogenies}.
\newblock {\em Philos Trans R Soc Lond B Biol Sci}, 344(1307):77--82, 04 1994.

\bibitem{neeetal2}
S.~Nee, A.~Mooers, and P.~Harvey.
\newblock {Tempo and mode of evolution revealed from molecular phylogenies}.
\newblock {\em Proc Natl Acad Sci U S A}, 89(17):8322--8326, 09 1992.

\bibitem{neuts1981matrix}
M.~Neuts.
\newblock Matrix geometric solutions in stochastic models: An algorithmic
  approach, johns hopkins university press.
\newblock {\em Baltimore, MD, USA}, 1981.

\bibitem{nosil2008speciation}
P.~Nosil.
\newblock Speciation with gene flow could be common.
\newblock {\em Molecular Ecology}, 17(9):2103--2106, 2008.

\bibitem{peng2009primate}
Z.~Peng, N.~Elango, D.~E. Wildman, and S.~V. Yi.
\newblock Primate phylogenomics: developing numerous nuclear non-coding,
  non-repetitive markers for ecological and phylogenetic applications and
  analysis of evolutionary rate variation.
\newblock {\em BMC genomics}, 10(1):1--11, 2009.

\bibitem{phung2010simple}
T.~Phung-Duc, H.~Masuyama, S.~Kasahara, and Y.~Takahashi.
\newblock A simple algorithm for the rate matrices of level-dependent {QBD}
  processes.
\newblock In {\em Proceedings of the 5th International Conference on Queueing
  Theory and Network Applications}, pages 46--52, 2010.

\bibitem{Ross2010}
S.~M. Ross.
\newblock {\em Introduction to Probability Models}.
\newblock Academic Press, San Diego, CA, USA, sixth edition, 2010.

\bibitem{roth2007evolution}
C.~Roth, S.~Rastogi, L.~Arvestad, K.~Dittmar, S.~Light, D.~Ekman, and D.~A.
  Liberles.
\newblock Evolution after gene duplication: models, mechanisms, sequences,
  systems, and organisms.
\newblock {\em Journal of Experimental Zoology Part B: Molecular and
  Developmental Evolution}, 308(1):58--73, 2007.

\bibitem{schilthuizen2003shape}
M.~Schilthuizen.
\newblock Shape matters: the evolution of insect genitalia.
\newblock In {\em Proceedings Of The Section Experimental And Applied
  Entomology-Netherlands Entomological Society}, volume~14, pages 9--16, 2003.

\bibitem{serrano2015decoupled}
M.~L. Serrano-Serrano, M.~Perret, M.~Guignard, A.~Chautems, D.~Silvestro, and
  N.~Salamin.
\newblock Decoupled evolution of floral traits and climatic preferences in a
  clade of neotropical gesneriaceae.
\newblock {\em BMC evolutionary biology}, 15(1):1--12, 2015.

\bibitem{singhal2022no}
S.~Singhal, G.~R. Colli, M.~R. Grundler, G.~C. Costa, I.~Prates, and D.~L.
  Rabosky.
\newblock No link between population isolation and speciation rate in squamate
  reptiles.
\newblock {\em Proceedings of the National Academy of Sciences},
  119(4):e2113388119, 2022.

\bibitem{Soewongsono}
A.~C. Soewongsono.
\newblock {\em Mathematical Models for the Evolution of Species}.
\newblock PhD thesis, The University of Tasmania, 2023.

\bibitem{2020SOH}
A.~C. Soewongsono, B.~R. Holland, and M.~M. O’Reilly.
\newblock The shape of phylogenies under phase-type distributed times to
  speciation and extinction.
\newblock {\em Bulletin of Mathematical Biology}, 84(10):1--45, 2022.

\bibitem{stark2021characterizing}
T.~L. Stark and D.~A. Liberles.
\newblock Characterizing amino acid substitution with complete linkage of sites
  on a lineage.
\newblock {\em Genome Biology and Evolution}, 13(10):evab225, 2021.

\bibitem{teufel2018using}
A.~I. Teufel, A.~M. Ritchie, C.~O. Wilke, and D.~A. Liberles.
\newblock Using the mutation-selection framework to characterize selection on
  protein sequences.
\newblock {\em Genes}, 9(8):409, 2018.

\bibitem{walker2022structural}
M.~E. Walker, J.~B. Simpson, and M.~R. Redinbo.
\newblock A structural metagenomics pipeline for examining the gut microbiome.
\newblock {\em Current Opinion in Structural Biology}, 75:102416, 2022.

\bibitem{wheeler2000species}
Q.~D. Wheeler and R.~Meier.
\newblock {\em Species concepts and phylogenetic theory: a debate}.
\newblock Columbia University Press, 2000.

\bibitem{yule1925ii}
G.~U. Yule.
\newblock {II.--} {A} mathematical theory of evolution, based on the
  conclusions of {D}r. {J. C. W}illis, {F. R. S.}
\newblock {\em Philosophical transactions of the Royal Society of London.
  Series B, containing papers of a biological character}, 213(402-410):21--87,
  1925.

\end{thebibliography}
	
\end{document}